\documentclass[11pt,a4paper]{article}
\usepackage{graphicx}
\usepackage{amsmath}
\usepackage{amssymb}
\usepackage{multirow}
\usepackage{color}

\newcommand{\nonu}{\nonumber}

\newcommand{\mrm}[1]{\mathrm{#1}}

\renewcommand{\d}{{\mathrm d}}

\newcommand{\p}{{\mathrm p}}
\newcommand{\q}{{\mathrm q}}

\renewcommand{\u}{{\mathrm u}}

\newcommand{\dbar}{\overline{\mathrm d}}

\newcommand{\pbar}{\overline{\mathrm p}}
\newcommand{\qbar}{\overline{\mathrm q}}

\newcommand{\ubar}{\overline{\mathrm u}}

\newcommand{\Pom}{\mathbb{P}}
\newcommand{\Reg}{\mathbb{R}}

\newcommand{\aem}{\alpha_{\mathrm{em}}}

\newcommand{\pT}{p_{\perp}}

{\end{list}}
\newcounter{enumct}

\newlength{\abstwidth}
\begin{document}
\sloppy
 
\pagestyle{empty}
 
\begin{flushright}
LU TP 18-06\\
MCnet-18-08\\
April 2018
\end{flushright}

\vspace{\fill}

\begin{center}
{\Huge\bf Models for Total, Elastic\\[2mm] 
and Diffractive Cross Sections}\\[6mm]
{\Large Christine O. Rasmussen and Torbj\"orn Sj\"ostrand}\\[3mm]
{\large\it Theoretical Particle Physics,
Department of Astronomy and Theoretical Physics,\\[1mm]
Lund University, S\"olvegatan 14A, SE-223 62 Lund, Sweden}
\end{center}

\vspace{\fill}

\begin{center}
\begin{minipage}{\abstwidth}
\begin{center}
{\bf Abstract}
\end{center}
The LHC has brought much new information on total, elastic and 
diffractive cross sections, which is not always in agreement with 
extrapolations from lower energies. The default framework in the 
\textsc{Pythia} event generator is one case in point. In this article 
we study and implement two recent models, as more realistic 
alternatives. Both describe total and
elastic cross sections, whereas one also includes single diffraction. 
Noting some issues at high energies, a variant of the latter is
proposed, and extended also to double and central diffraction.
Further, the experimental definition of diffraction is based on the
presence of rapidity gaps, which however also could be caused by 
colour reconnection in nondiffractive events, a phenomenon that is 
studied in the context of a specific model. Throughout comparisons 
with LHC and other data are presented.  
\end{minipage}
\end{center}

\vspace{\fill}

\phantom{dummy}

\clearpage

\pagestyle{plain}
\setcounter{page}{1}

\section{Introduction}

\begin{figure}[ht!]
\begin{minipage}[c]{0.3\linewidth}
\centering
\includegraphics[width=\linewidth]{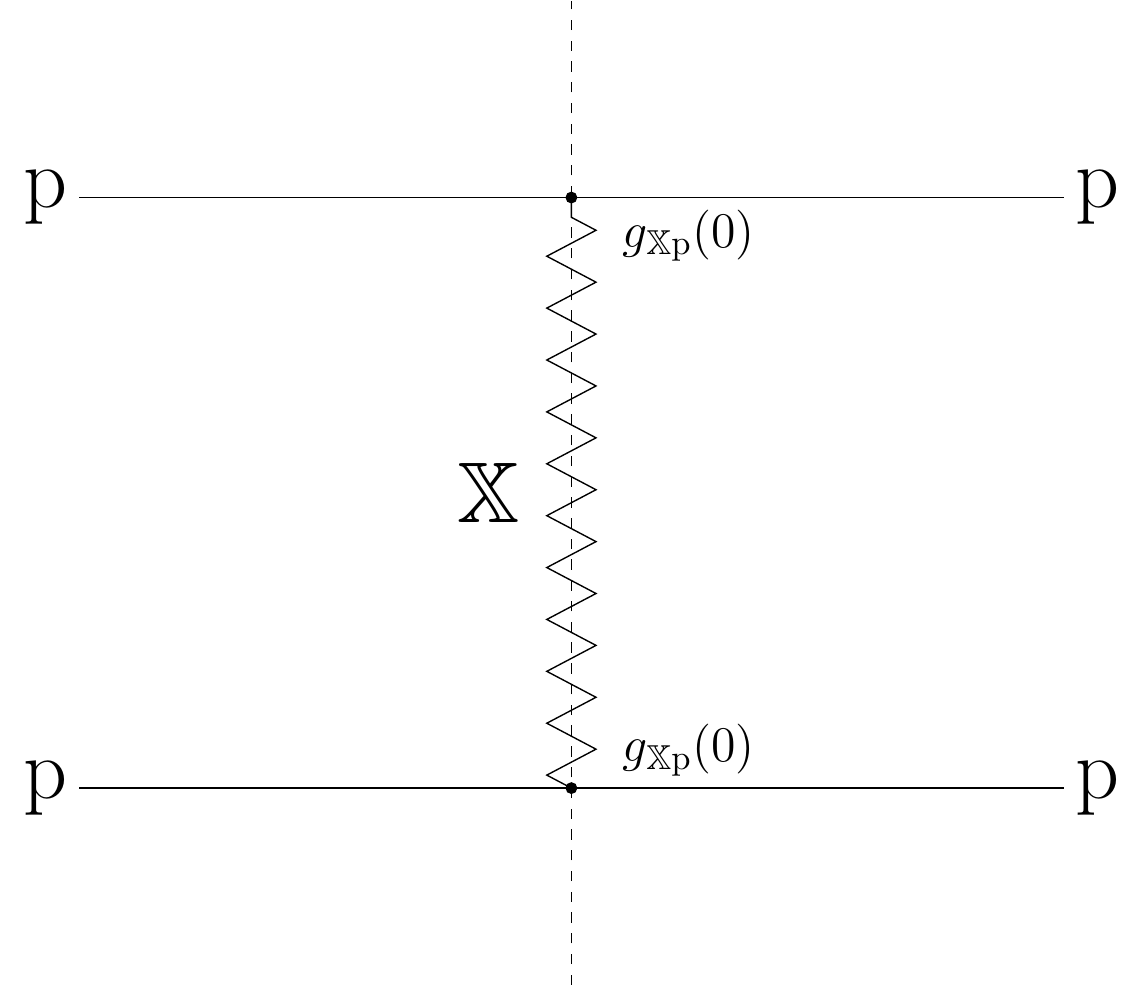}\\
(a)
\end{minipage}
\hfill
\begin{minipage}[c]{0.3\linewidth}
\centering
\includegraphics[width=\linewidth]{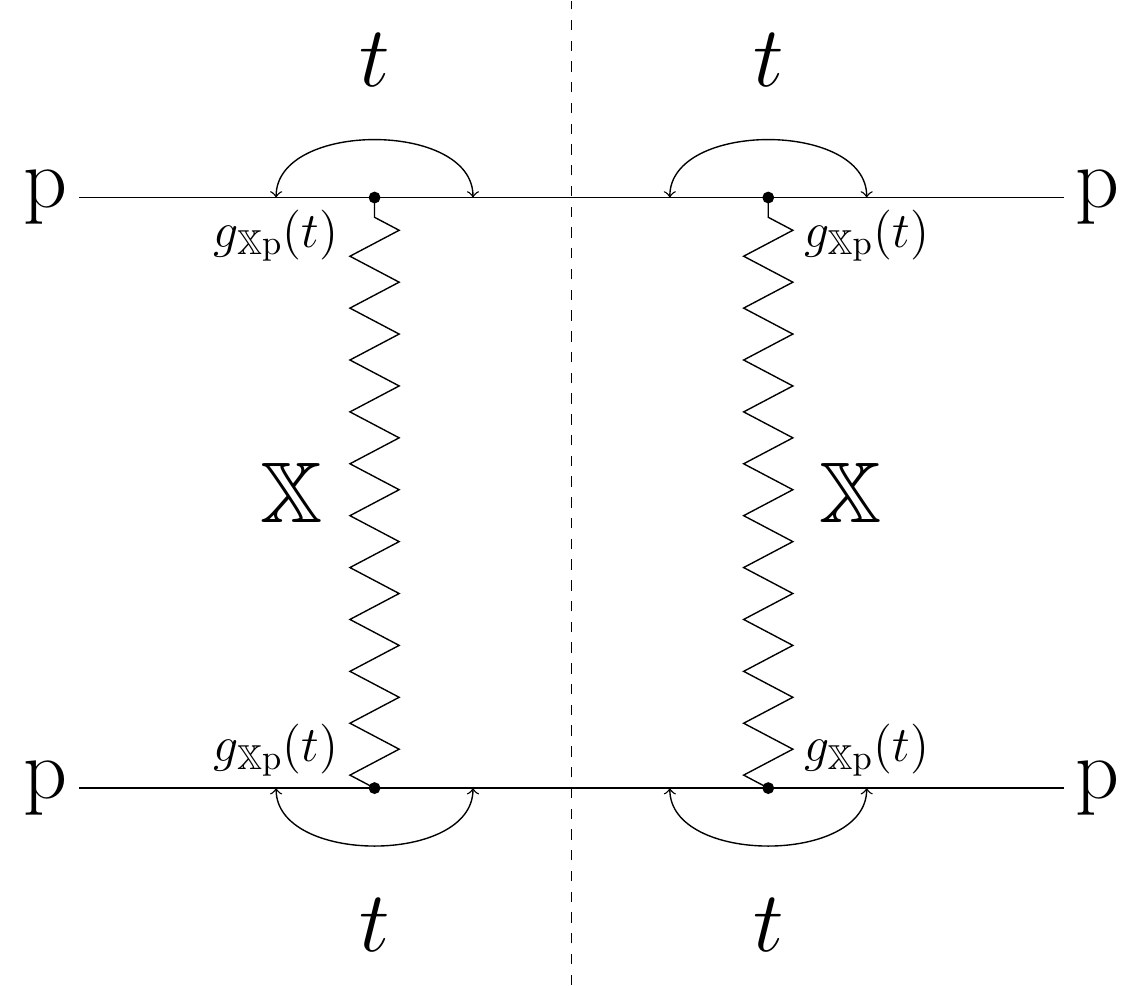}\\
(b)
\end{minipage}
\hfill
\begin{minipage}[c]{0.3\linewidth}
\centering
\includegraphics[width=\linewidth]{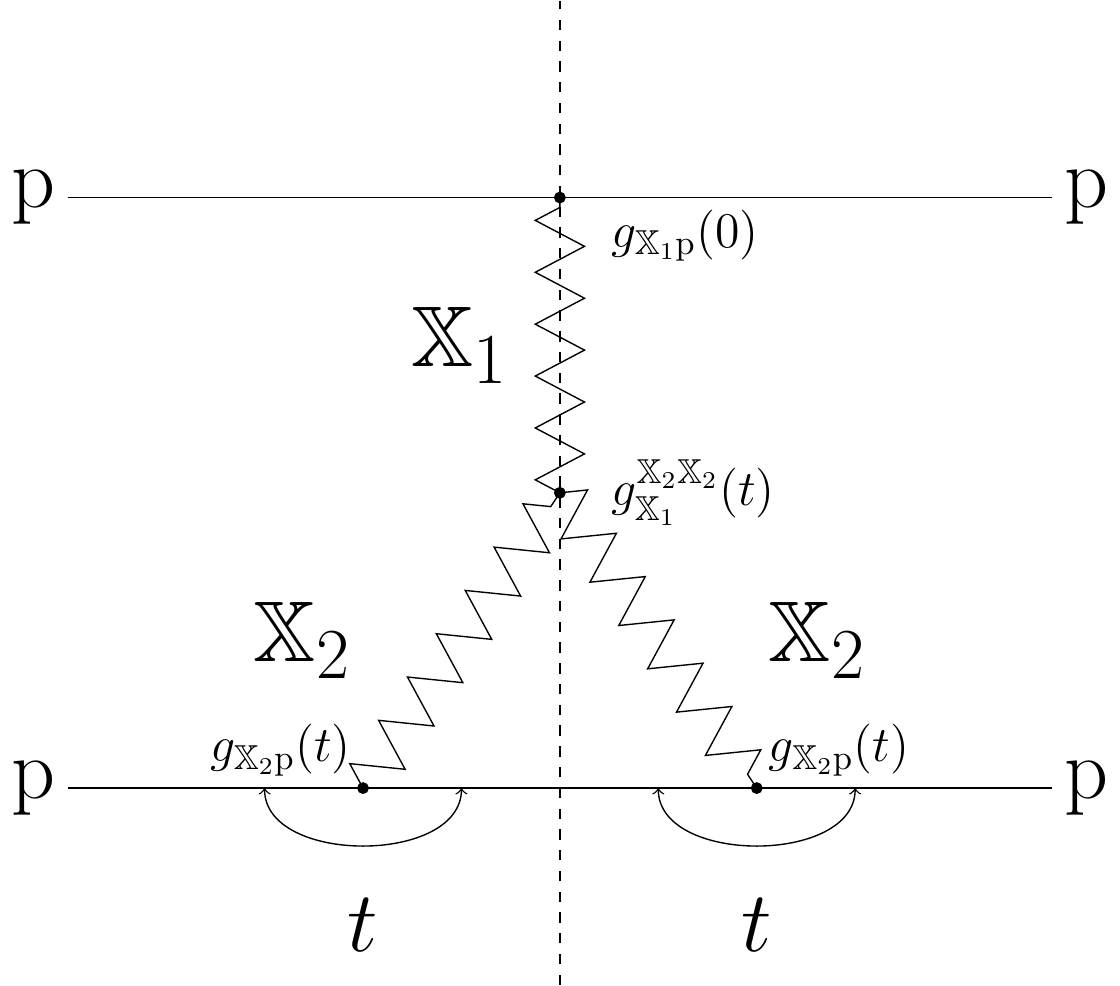}\\
(c)
\end{minipage}
\begin{minipage}[c]{0.3\linewidth}
\centering
\includegraphics[width=\linewidth]{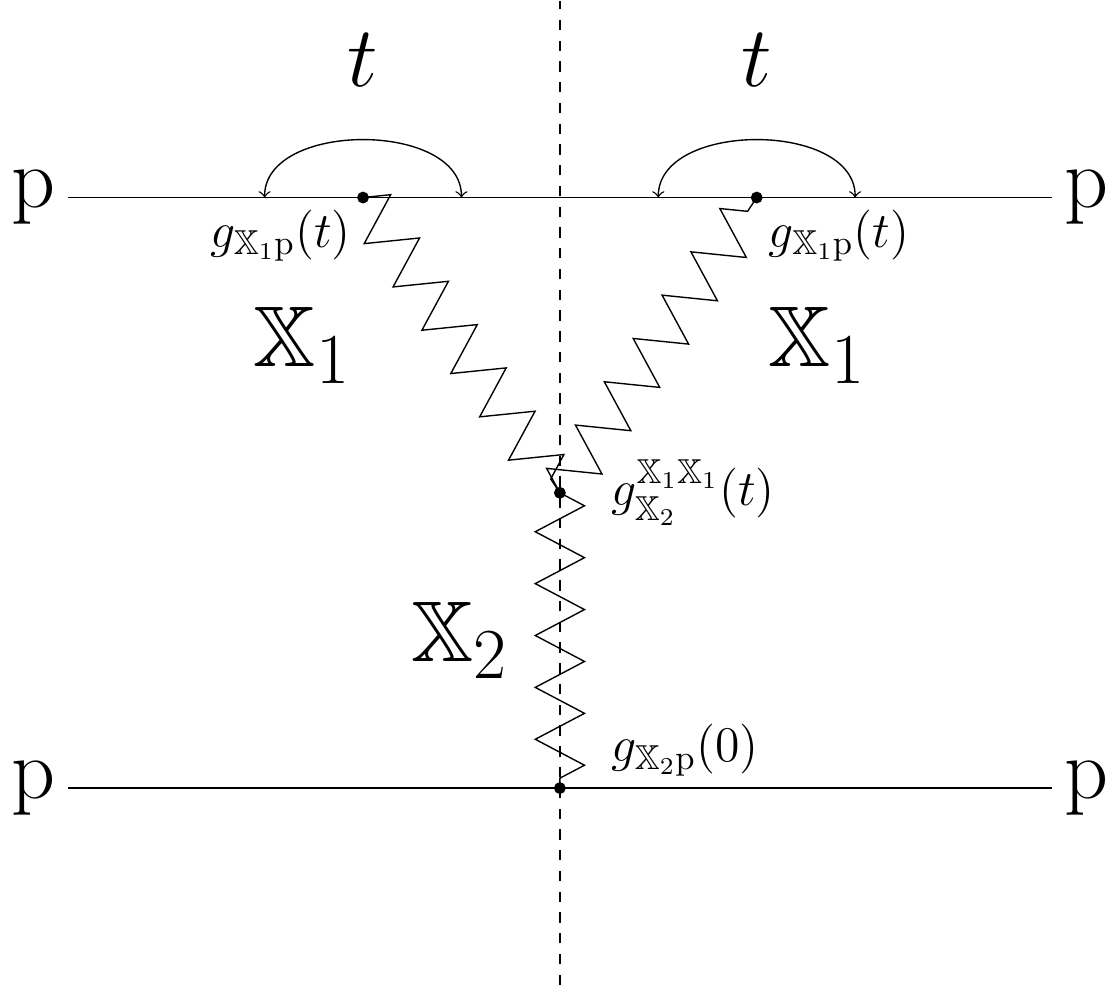}\\
(d)
\end{minipage}
\hfill
\begin{minipage}[c]{0.3\linewidth}
\centering
\includegraphics[width=\linewidth]{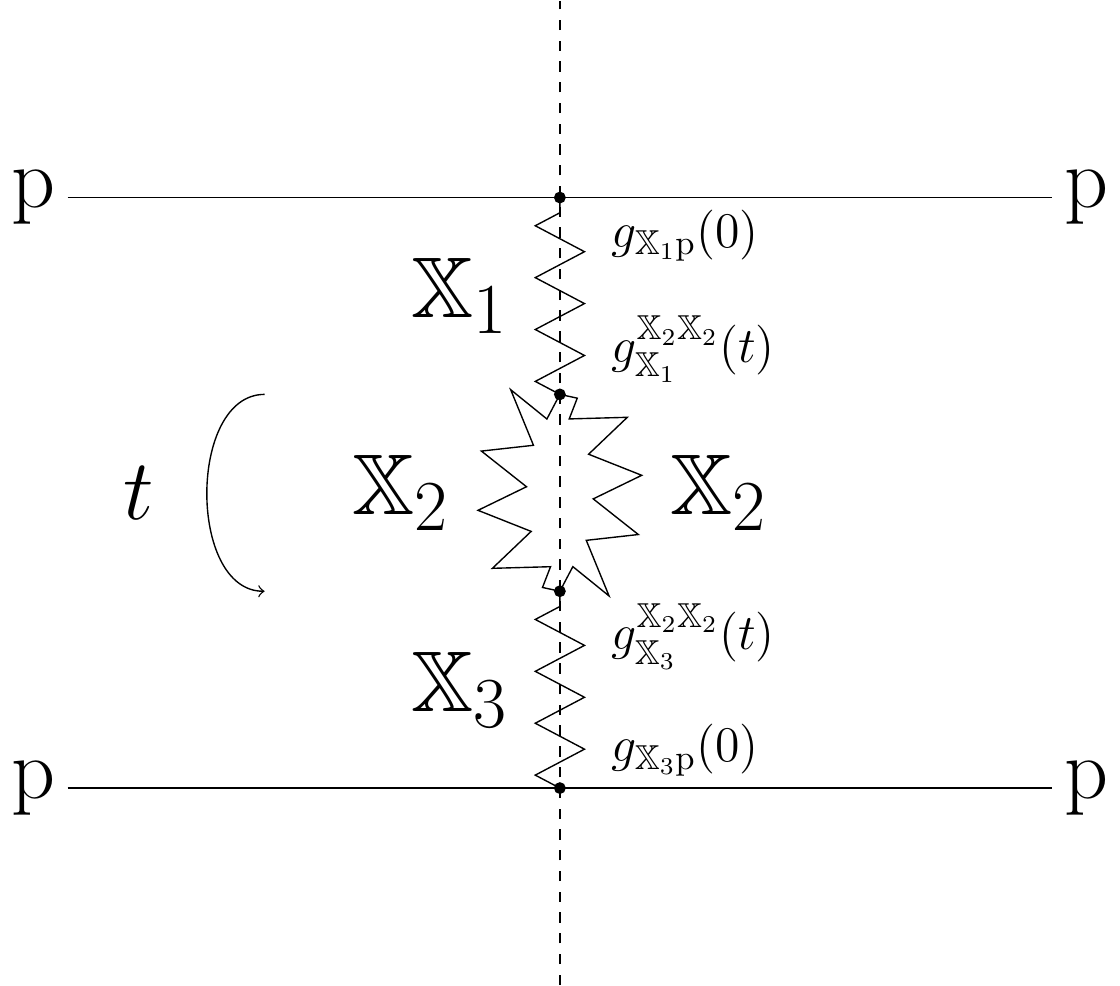}\\
(e)
\end{minipage}
\hfill
\begin{minipage}[c]{0.3\linewidth}
\centering
\includegraphics[width=\linewidth]{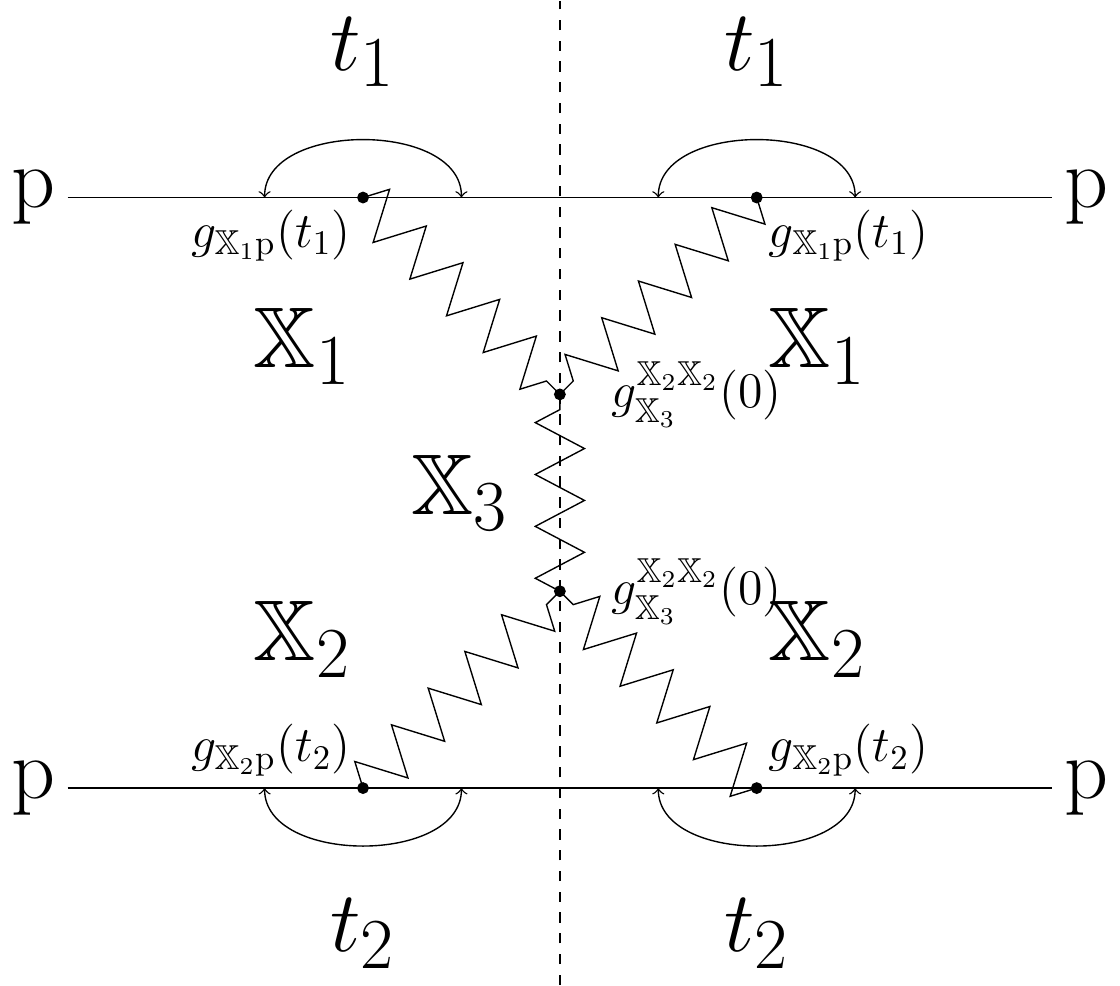}\\
(f)
\end{minipage}
\caption{\label{Fig:ALLXS}
The squared matrix element for the total (a), elastic (b), single (c,d),
double (e) and central (f) diffractive cross sections.
}
\end{figure}

The LHC has provided new information on any number of topics,
including total, (differential) elastic and (differential) diffractive
cross sections, or $\sigma_{\mrm{TED}}$ for short. The
$\sigma_{\mrm{TED}}$ kind of quantities cannot be predicted from the
QCD Lagrangian, although this is where they have their
origin. Therefore $\sigma_{\mrm{TED}}$ results are often overshadowed
by results from the perturbative domain, where comparisons with the
Standard Model, and searches for physics beyond it, are more directly
related to the underlying theory. Nevertheless, there are good reasons
to study the old and new $\sigma_{\mrm{TED}}$ data now available. One
is to assess how well different effective models can describe the
data, and implicitly or explicitly pave the way for better models and
better understanding, ultimately to form a stronger connection with
the underlying QCD theory.  Another is that diffractive events form
part of the ``underlying event'' and pileup backgrounds that have a
direct impact e.g.\ on jet energy scales and jet profiles, and thereby
on many experimental studies.  In this latter aspect they combine with
the inelastic nondiffractive events into the overall inelastic event
class, with a separation that is far from unambiguous.

Historically there are two main approaches to $\sigma_{\mrm{TED}}$ in
hadron--hadron collisions, the diagrammatical and the geometrical,
although both aspects may well be represented in a specific model
\cite{Collins:1977jy,Forshaw:1997dc,Donnachie:2002en,Barone:2002cv}.  
In the diagrammatical approach new effective particles are introduced,
specifically the Pomeron(s) $\Pom$ and Reggeon(s) $\Reg$, with
associated propagators and vertex coupling strengths. A
Feynman-diagram-like expansion may be performed into different event
classes, with higher-order corrections. A subset of these are shown in
fig.~\ref{Fig:ALLXS}, with $\mathbb{X}=\Pom,\Reg$ and each of the
couplings denoted with a $g$. In the diagrammatical approach, the
dashed line (the cut) represents the diagram at amplitude level. A cut
through a $\Pom$ or $\Reg$ thus represent particle formation at
amplitude level, while an uncut Pomeron or Reggeon represents an area
void of particle production. In a geometrical approach the
impact-parameter aspects are emphasized, where diffraction largely is
related to peripheral collisions. The analogy with wave scattering
theory here is natural, and has given the diffractive event class its
name. Diffraction can also be viewed as a consequence of the
interaction eigenstates being different from the mass ones 
\cite{Good:1960ba,Flensburg:2010kq}.

Neither of these approaches address the detailed structure of
diffractive events. In olden days, at low energies, a diffractive
system was simply viewed as an excited proton state that could decay
more-or-less isotropically, a ``fireball'' 
\cite{Hagedorn:1965st,Collins:1977jy}. This is clearly not a
valid picture for higher-mass diffractive states, where the same kind
of longitudinal structure is observed as for nondiffractive ones. The
simplest partonic approach would then be for a $\Pom/\Reg$ to kick out
a single quark or gluon from a proton, giving rise to one or two
fragmenting colour strings.  The Ingelman--Schlein picture 
\cite{Ingelman:1984ns} takes it
one step further and introduces an internal structure for the $\Pom$,
such that a $\Pom\p$ collision may be viewed as an inelastic
nondiffractive $\p\p$ (or better $\pi^0\p$) one in miniature. Thereby
also hard jet activity and multiparton interactions (MPIs) become
possible within a diffractive system, as supported by data.

A key aspect of MPI modelling is the relation to colour reconnection
(CR), whereby partons in the final state may be related in colour so as
to reduce the total string length relative to naive expectations. This
opens for another view on diffraction, where CR can generate rapidity 
gaps dynamically \cite{Edin:1995gi,Buchmuller:1995qa}.
Then the diffractive and inelastic nondiffractive event classes have a
common partonic origin, and only differ by the event-by-event
fluctuations in colour topologies. Even in models that do not go quite
as far, the dividing line between the two kinds of events may be
fuzzy. This is even more so since the experimental classification in
terms of a rapidity gap allows for misidentification in both
directions, relative to the classification in a specific
model. High-mass diffraction need not give a gap in the central
detector, while nondiffractive events by chance (CR or not) can have a
large rapidity gap.

What should now be clear is that description of the
$\sigma_{\mrm{TED}}$ physics, and especially the diffractive part, is
too multifaceted to be based purely on analytical calculations. The
implementation into Monte Carlo Event Generators is crucial to test
different approaches.  One of the most commonly used generators is
\textsc{Pythia} \cite{Sjostrand:2006za,Sjostrand:2014zea}, 
which by default is based on a rather old diagrammatical ``tune'' 
for the $\sigma_{\mrm{TED}}$ issues \cite{Schuler:1993wr}, combined
with an Ingelman--Schlein-style approach to the diffractive event
structure \cite{Navin:2010kk}. In particular the first part does not 
agree well with LHC data, and so needs an overhaul.

For the total and elastic cross sections we have chosen to implement
two different parametrizations, the parametrization from the COMPAS
group as found in the Review of Particle Physics 2016 
\cite{Patrignani:2016xqp} and a model
developed by Appleby and collaborators (ABMST) \cite{Appleby:2016ask}. In
addition to a better fit to the integrated cross sections, these also
include a more detailed description of the differential elastic cross
sections.

The ABMST model also addresses single diffraction. It is in an
ambitious diagrammatical approach, supplemented with a careful
description of the resonance shape in the low-mass region, based on
comparisons with low-energy data. Unfortunately, as is common in such
ans\"atze, the diffractive cross section asymptotically grows faster
with energy than the total one, making it marginally acceptable
already at LHC energies and definitely unacceptable for FCC ones. We
therefore study possible modifications that would give a more
reasonable energy behaviour. Further, while ABMST does not address
double or central diffraction, we use the framework of the model to
extend it also to these event classes, and in the process need
to make further adjustments. Results for the ABMST-based modelling 
implemented in \textsc{Pythia} are compared with the already existing 
default framework of Schuler-Sj\"ostrand (SaS) and Donnachie-Landshoff
(DL) \cite{Donnachie:1992ny,Schuler:1993wr}, and confronted with LHC
data.

Furthermore we study the sensitivity to CR by comparing with the
Christiansen--Skands QCD-based CR model (CSCR) \cite{Christiansen:2015yqa}.
This model has no protection against ``accidental'' rapidity gaps in 
nondiffractive events, unlike the default CR framework. But it is also 
not intended to describe (the bulk of) diffraction, and therefore it 
requires a retuning to provide a sensible combined description. It 
therefore offers an interesting case study for a tuning task that is 
likely to become more common in the future.

The plan of the article is as follows: In section 2 we begin by
summarising the current status of \textsc{Pythia}~8, the default cross
section parametrizations along with the hadronic event properties of
diffractive events. In section 3 we describe the new models for total
and elastic (differential) cross sections. In section 4, 5, 6 we
extend these to single-, double- and central diffractive
(differential) cross sections, respectively. In section 7 we provide
some comparisons to LHC data and provide new tunes of the default
\textsc{Pythia}~8 model. We end with section 8, where we summarise and
provide an outlook to further studies.

\section{The current status of \textsc{Pythia}~8}

\textsc{Pythia}~8 is a multi-purpose event generator aimed at the
generation of high-energy events. This includes collisions both of
a perturbative and a non-perturbative character, each of which gives 
contributions to the total collision cross section. In perturbative 
collisions, the description begins with the matrix element of the 
hard scattering process in combination with parton distribution
functions. This core is dressed up with several other elements such 
as multiparton interactions, parton showers and hadronisation. In 
non-perturbative scattering collisions, on the other hand, no standard 
formulation exists for the core process, and phenomenological models 
are needed. After the model-dependent choices of the key kinematical 
variables have been made, the event generation may be continued in a 
similar manner as for perturbative events, where relevant.  

In this paper we focus on the non-perturbative scattering processes, and
the generation of these. To set the stage for further improvements, 
the purpose of this section is to describe the current status of the
event generator. This we have split into two parts, beginning with the
description of the default cross section models, the SaS/DL one, and 
then go on to describe the event property aspects that are the same 
regardless of the choice of model. 

\subsection{Differential cross sections}

In the current version of \textsc{Pythia}~8, the predictions for the 
total, elastic and diffractive cross sections do not agree so well with 
measurements performed at the LHC. The current implementation is the  
parametrization of DL \cite{Donnachie:1992ny} for the total 
cross section, 
\begin{align}\label{eq:DLtot}
\sigma_{\mrm{tot}}(s) =& 
  X^{AB}s^{\epsilon} + Y^{AB}s^{-\eta},
\end{align}
with $s=E_{\mrm{CM}}^2$, $\epsilon=0.0808$, $\eta=0.4525$. $A$ and
$B$ denote the initial-state particles, and $X^{AB},\,Y^{AB}$ are 
specific to each such state. The elastic and diffractive cross 
sections are described using the parametrization of SaS 
\cite{Schuler:1993wr},
\begin{align}
\frac{\d\sigma_{\mrm{el}}}{\d t} =&
  (1+\rho^2) \, \frac{\sigma_{\mrm{tot}}^2(s)}{16\pi} \,
  \exp(B_{\mrm{el}}(s) \, t) \label{eq:sasEL} ~,\\
\frac{\d\sigma_{XB}(s)}{\d t \, \d M_X^2} =&
  \frac{g_{3\Pom}}{16\pi} \, \frac{\beta_{A\Pom}(s) \, \beta_{B\Pom}^2(s)}{M_X^2} 
  \, \exp(B_{XB}(s) \, t) \, F_{\mrm{SD}}(M_X^2,s) ~, \label{eq:sasXB} \\
\frac{\d\sigma_{AX}(s)}{\d t \, \d M_X^2} =&
  \frac{g_{3\Pom}}{16\pi} \, \frac{\beta_{A\Pom}^2(s) \, \beta_{B\Pom}(s)}{M_X^2}
  \, \exp(B_{AX}(s) \, t) \, F_{\mrm{SD}}(M_X^2,s) ~,\label{eq:sasAX} \\
\frac{\d\sigma_{XY}(s)}{\d t \, \d M_X^2 \, \d M_Y^2} =&
  \frac{g_{3\Pom}^2}{16\pi} \, \frac{\beta_{A\Pom}(s) \, \beta_{B\Pom}(s)}
  {M_X^2 \, M_Y^2} \, \exp(B_{XY}(s) \, t) \, F_{\mrm{DD}}(M_X^2,M_Y^2,s)
  \label{eq:sasXY} ~,
\end{align}
where indices $X$ and $Y$ here represent diffractive systems
(not to be confused with the coefficients of eq.~(\ref{eq:DLtot})), 
$\rho$ is the ratio of real to imaginary parts of the elastic scattering 
amplitude at $t=0$ , $\beta_{A\Pom}$ and  $\beta_{B\Pom}$ are hadron 
couplings strengths to the Pomeron, and $g_{3\Pom}$ the triple-Pomeron
vertex strength. The slope parameters are defined as
\begin{align}
B_{\mrm{el}}(s) =& 2b_{A} + 2b_B + 4s^{\epsilon} - 4.2,\nonu\\
B_{XB}(s) =& 2b_B +
  2\alpha_{\Pom}'\ln\left(\frac{s}{M_X^2}\right)\nonu\\
B_{AX}(s) =& 2b_A +
  2\alpha_{\Pom}'\ln\left(\frac{s}{M_X^2}\right)\nonu\\
B_{XY}(s) =& 
  2\alpha_{\Pom}'\ln\left(e^4+\frac{s \, s_0}{M_X^2 \, M_Y^2}\right),
\end{align}
where $b_i=2.3$ for $i=\p,\pbar$, $\alpha_{\Pom}'=0.25$ GeV$^{-2}$, 
$s_0=1/\alpha_{\Pom}'$, and the term $e^4$ is added by hand in order to 
avoid $B_{\mrm{DD}}(s)$ to break down for large values of $M_X^2 \, M_Y^2$.
Special care was taken to avoid unphysical high-energy behaviours;
e.g.\ a logarithmic $s$ dependence of $B_{\mrm{el}}$ would
have lead to $\sigma_{\mrm{el}}(s) > \sigma_{\mrm{tot}}(s)$ for large $s$.

Fudge factors are introduced to dampen large (overlapping) mass
systems as well as increasing the low-mass ``resonance'' region, without 
describing the resonances individually,
\begin{align}
F_{\mrm{SD}}(M_X^2,s) =& \left(1-\frac{M_X^2}{s}\right) 
  \left(1 + \frac{c_{\mrm{res}} \, M_{\mrm{res}}^2}
  {M_{\mrm{res}}^2 + M_X^2}\right)\nonu\\
F_{\mrm{DD}}(M_X^2, M_Y^2, s) =& \left(1-\frac{(M_X+M_Y)^2}{s}\right) 
  \left(\frac{s\,m_{\p}^2}{s\,m_{\p}^2 + M_X^2M_Y^2}\right)\cdot\nonu\\
  &\left(1 + \frac{c_{\mrm{res}} \, M_{\mrm{res}}^2}
  {M_{\mrm{res}}^2 + M_X^2}\right)
  \left(1 + \frac{c_{\mrm{res}} \, M_{\mrm{res}}^2}
  {M_{\mrm{res}}^2 + M_Y^2}\right),
\end{align}
where $c_{\mrm{res}}=2$ and $M_{\mrm{res}}=2$~GeV for $\p\p$ and $\p\pbar$.

Central diffraction has been added to \textsc{Pythia}~8, but is not
widely used in the experimental communities, hence have not been
maintained properly after its inclusion. It is off by default, and is 
not included in any of the tunes performed by the \textsc{Pythia}~8 
collaboration or the experimental communities. Thus the results obtained 
with it included should not be trusted too far. The cross section is 
\begin{align}
\sigma_{\mrm{CD}}(s) =& \sigma_{\mrm{CD}}^{\mrm{ref}}
  \frac{\ln^{1.5}\left(\frac{0.06s}{s_{\mrm{min}}}\right)}
  {\ln^{1.5}\left(\frac{0.06s_{\mrm{ref}}}{s_{\mrm{min}}}\right)} ~, 
\label{eq:sasAXB}
\end{align}
with $\sigma_{\mrm{CD}}^{\mrm{ref}}=1.5$ mb, $s_{\mrm{ref}}=4$ TeV$^2$ and 
$s_{\mrm{min}}=1$ GeV$^2$. The diffractive mass is chosen from a 
$(1 - \xi_1) (\d\xi_1/\xi_1) (1 - \xi_2) (\d\xi_2 / \xi_2)$
distribution, with $\xi_{1,2}$ being the momentum fraction taken from the 
respective incoming hadron, such that $M_X^2 = \xi_1 \xi_2 s$.  
The two $t$ values are selected according to exponentials with slope
$2b_A + \alpha_{\Pom}'\ln(1/\xi_1)$ and $2b_B + \alpha_{\Pom}'\ln(1/\xi_2)$,
respectively.  

The expressions in eqs.\ (\ref{eq:sasXB}) -- (\ref{eq:sasXY}) can be 
integrated to give the total elastic and diffractive cross sections. 
This worked reasonably well up to Tevatron energies, but it overshot
diffractive cross sections observed at the LHC \cite{ATLAS:2010kza}. 
Simple overall modification factors were therefore introduced 
\cite{Corke:2010yf} to dampen the growth of the diffractive cross 
sections (including the CD one in eq.~(\ref{eq:sasAXB})),
\begin{align}\label{Eq:SaSmod}
\sigma_i^{\mrm{mod}}(s) =& \frac{\sigma_i^{\mrm{old}}(s) \, \sigma_i^{\mrm{max}}}%
{\sigma_i^{\mrm{old}}(s) + \sigma_i^{\mrm{max}}} ~, 
\end{align}
where the $\sigma_i^{\mrm{max}}$ are free parameters. The ansatz 
allows phenomenology at lower energies to be preserved while giving some 
reasonable freedom for LHC tunes. It gives asymptotically constant
diffractive cross sections, but typically with asymptotia so far away
that it is not an issue for current studies. 

The kinematical limits for $t$ are determined by all the masses in the
system. We define the scaled variables
$\mu_1=m_A^2/s,\,\mu_2=m_B^2/s,\,\mu_3=M_X^2/s,\,\mu_4=M_Y^2/s$ where
$M_X=m_A$ if $A$ scatters elastically and $M_Y=m_B$ if $B$ scatters
elastically. Thus the combinations
\begin{align}
C_1 =& 1 - (\mu_1 + \mu_2 + \mu_3 + \mu_4) +
  (\mu_1-\mu_2)(\mu_3-\mu_4)\nonu\\
C_2 =& \sqrt{(1-\mu_1-\mu_2)^2-4\mu_1\mu_2}
  \sqrt{(1-\mu_3-\mu_4)^2-4\mu_3\mu_4}\nonu\\
C_3 =& (\mu_3-\mu_1)(\mu_4-\mu_2) +
  (\mu_1+\mu_4-\mu_2-\mu_3)(\mu_1\mu_4-\mu_2\mu_3),\nonu
\end{align}
will lead to the kinematical limits $t_{\mrm{min}}<t<t_{\mrm{max}}$.
\begin{align}\label{Eq:tRange}
t_{\mrm{min}}=&-\frac{s}{2}(C_1+C_2)\nonu\\
t_{\mrm{max}}=&\frac{s^2C_3}{t_{\mrm{min}}}.
\end{align}
These expressions are directly applicable for elastic scattering and 
for single and double diffraction. For central diffraction $AB \to AXB$
they can be applied twice, with $\mu_4 = M_{XB}^2 / s$ for $t_1$ and 
$\mu_3 = M_{AX}^2 / s$ for $t_2$. 

An electromagnetic Coulomb term can be added to describe low-$|t|$ 
elastic scattering. The implementation is here based on the formalism 
as outlined e.g.\ in \cite{Bernard:1987vq,Antchev:2016vpy}. 
Introducing an electromagnetic low-$|t|$ form factor as 
\begin{align}
G(t) \approx& \frac{\lambda^2}{(\lambda - t)^2}~, ~~~~ \lambda \approx 
0.71~\mathrm{GeV}^2~,   
\end{align}
and a Coulomb term phase factor approximation \cite{West:1968du,Cahn:1982nr}
\begin{align}
\phi(t) \approx& \pm \aem \left(- \gamma_{\mathrm{E}} 
- \log\left( - \frac{B_{\mrm{el}}(s) \, t}{2} \right) \right)
~, ~~~~\gamma_{\mathrm{E}} \approx 0.577~,
\end{align} 
with $+$ for $\p\p$ and $-$ for $\p\pbar$, Coulomb and interference 
terms are added to the hadronic $\d\sigma_{\mrm{el}}/\d t$ above
\begin{align}
\frac{\d\sigma_{\mrm{el}}^{\mrm{C+int}}}{\d t} =&
\frac{4\pi\aem^2\,G^4(t)}{t^2} \pm \frac{\aem\,G^2(t)}{t}\,
\left( \rho \, \cos\phi(t) + \sin\phi(t) \right) \, 
\sigma_{\mrm{tot}}(s) \, \exp\left(\frac{B_{\mrm{el}}(s) \, t}{2}\right) ~.
\end{align} 
The same expression can also be added to the Minimum Bias Rockefeller
(MBR) model \cite{Ciesielski:2012mc} (and a flexible ``set your 
own'' one), while the ABMST and RPP formalisms each introduce the 
Coulomb corrections as one extra amplitude term, 
with the full phase expressions of \cite{Cahn:1982nr}. Numerically 
the three implementations give very similar results.

\subsection{Hadronic event properties}

To model a diffractive system, it is convenient to view its internal 
structure as a consequence of the interaction between two hadronlike 
objects, e.g.\ as a $\Pom B$ subcollision for the $AB \to AX$ process,
in the same spirit as a high-energy nondiffractive $\p\p$ event, where 
perturbative processes largely shape its structure. Such an approach 
is not viable for low-mass diffractive systems, however. Therefore the 
diffractive event generation is split into two regimes, a high-mass
and a low-mass one, with a smooth transition between the two. The 
probability for applying the high-mass description is given by 
\cite{Navin:2010kk}
\begin{align}
P_{\mrm{pert}}=& 1 - \exp\left( -
\frac{\max(0, M_X - m_{\mrm{min}})}{m_{\mrm{width}}}\right),\nonu
\end{align}
with $m_{\mrm{min}}$ and $m_{\mrm{width}}$ free parameters, both by
default 10~GeV. Note how $P_{\mrm{pert}}$ vanishes when below 
$m_{\mrm{min}}$. 

For very low masses, $M_X \leq m_B + 1~$GeV for a $\Pom B$ subcollision, 
the diffractive system is allowed to decay isotropically into a 
two-hadron state. Above this limit, but still in the nonperturbative
regime, the collision process is viewed as the $\Pom$ kicking out  
either a valence quark or a gluon from the incoming hadron $B$.
The relative rate of the two is is mass-dependent,
\begin{align}
\frac{P(q)}{P(g)}=&\frac{N}{M_X^p},\nonu
\end{align}
with $N$ and $p$ as free parameters, and $M_X$ in GeV. 
In the former case a single string will be stretched between the 
kicked-out quark and the left-behind diquark, whereas the latter   
gives a ``hairpin'' string topology, going from one remnant valence 
quark via the struck gluon and back to the remnant diquark. These strings 
are then allowed to fragment using the Lund fragmentation model 
\cite{Andersson:1983ia}. The default values $N = 5$ and $p = 1$
ensures that the double-string topology wins out at higher masses,
consistent with what the exchange of a single gluon (a.k.a.\ a cut 
Pomeron) is expected to give in $\p\p$ collisions.

In the high-mass regime it is assumed that the diffractive cross 
section factorises into a Pomeron flux, a Pomeron--proton cross 
section, and a proton form factor. Together these determine the mass 
$M_X$ of the diffractive system and the squared momentum transfer $t$
in the process. Neither the $\Pom$ flux nor the $\Pom\p$ cross section
are known from first principles; therefore seven similar but somewhat
different $\Pom$ flux options are available in \textsc{Pythia}~8. 

The internal structure of the $\Pom\p$ system is then considered in an 
Ingelman--Schlein-inspired picture. Thus perturbative processes are
allowed, and $\Pom$ parton distribution functions (PDFs) are introduced 
like for a hadron. Standard factorization can be assumed, i.e.\ 
cross sections are given by hard-scattering matrix elements convoluted 
with the PDFs of two incoming partons. Furthermore, the full interleaved 
shower machinery of \textsc{Pythia}~8 is enabled, giving rise both to 
initial- and final-state showers and to multiparton interactions in 
the $\Pom\p$ system. This results in a more complex colour string 
structure than in the low-mass regime, which can also be subjected to 
additional colour reconnection, owing to overlap and crosstalk between 
the multiple subsystems. 

The activity in the $\Pom\p$ system, as represented e.g.\ by the average 
charged multiplicity, can be tuned to roughly reproduce that of a 
non-diffractive $\p\p$ collision of the same mass. This activity
is closely related to the average number of MPIs per event, 
the calculation of which differs between the two systems by a $\Pom$ 
vs.\ a $\p$ PDF in the numerator, and by $\sigma_{\Pom\p}^{\mrm{eff}}$ 
vs.\ $\sigma_{\p\p}^{\mrm{nondiffractive}}$ in the denominator. 
Given a $\Pom$ PDF, and assuming the same MPI-framework parameters 
as in $\p\p$, the $\sigma_{\Pom\p}^{\mrm{eff}}$ thus becomes the
main (mass-dependent) tuning parameter. In reality the two systems can 
be different, however, so experimental information on diffractive mass 
and multiplicity distributions can be used to refine the tune.
Be aware that a different choice of PDFs is likely to require a different 
$\sigma_{\Pom\p}^{\mrm{eff}}$ value. Ten different $\Pom$ PDF sets are 
implemented
\cite{Aktas:2006hy,Aktas:2007bv,Alvero:1998ta,Goharipour:2018yov}, 
plus a few toy ones for special purposes. Many of these have been fixed
by some convention for the $\Pom$ flux normalization, that in 
\textsc{Pythia} could be set differently. Hence all $\Pom$ PDFs are 
implemented with the option to be rescaled, e.g.\ in order to 
approximately impose the momentum sum rule.

In the MPI framework \cite{Sjostrand:2017cdm} the joint probability 
distribution for extracting several partons from a Pomeron needs 
to be defined. This is done in the same spirit as for protons 
\cite{Sjostrand:2004pf}. MPIs are ordered in a sequence of decreasing 
$\pT$ scales, and for the hardest interaction the normal PDFs are used. 
For subsequent ones the $x$ value is interpreted as a fraction of the 
then remaining $\Pom$ momentum, thereby ensuring that the momentum sum 
is not violated. If a quark is kicked out, flavour conservation ensures 
that a companion antiquark must also be present, and vice versa. Such a 
companion is introduced as an extra component of the $\Pom$ PDF, 
with normalization to unity. Overall momentum is preserved by scaling 
down the gluon and ordinary sea quark distributions to compensate. 
If the companion is selected for a subsequent MPI, then that component 
is removed, and gluon and sea are scaled up. 

Also initial-state radiation (ISR) requires special attention in the MPI 
framework. ISR is generated starting from the hard interaction and then
evolving backwards, to lower scales and larger $x$ values 
\cite{Sjostrand:1985xi}. Such ISR branchings are combined with the 
MPI generation into one interleaved sequence of falling $\pT$ scales.  
As above special consideration has to be given to branchings that
change the flavour of the incoming parton, and that can either induce 
or remove a companion (anti)quark. 

Similar to a proton \cite{Sjostrand:2004pf}, the Pomeron will leave 
behind a remnant after 
the MPIs and showers have removed momentum and removed or added partonic 
content. To begin, assume that only one gluon is kicked out of the 
incoming $\Pom$. The remnant will then be in a net colour octet state, 
which means that two colour strings eventually are stretched to the 
outgoing partons of the hard collision (or to the other beam remnant). 
The remnant could only consist of gluons and sea $\q\qbar$ pairs, 
since the $\Pom$ has no valence flavour content, so the simplest 
representation is as a single gluon or a single $\q\qbar$ pair.
From a physical point of view the two options would give very closely 
the same end result, since the hairpin string via a gluon remnant
eventually would break by the production of $\q\qbar$ pairs. 
For convenience, the choice is therefore made to represent the 
remnant as an octet $\u\ubar$ or $\d\dbar$ pair with equal probability.
In the general case, further unmatched companion quarks are added to
represent the full flavour content needed in the remnant. Most MPI 
initiators are gluons, however, which carry colour that should be
compensated in the remnant. This is addressed by attaching the gluon 
colour lines to the already defined remnants, which implicitly 
introduces colour correlations between the initiator partons. 
Such initial-state correlations can be further enhanced by colour 
reconnections in the final state. The final colour topology decides
how strings connect the outgoing partons after the collision, and
thereby sets the stage for the hadron production by string 
fragmentation.

\subsection{Hard diffraction}

Recently a framework for truly hard diffractive processes have been
implemented into \textsc{Pythia} \cite{Rasmussen:2015qgr}. This 
allows for diffractive subprocesses to generate eg.\ hard jets, 
electroweak particles and other internal \textsc{Pythia} processes, 
unlike the soft-to-medium QCD-only processes that were allowed in the 
framework described above.
This framework decides on whether or not a process is diffractive by
evaluating the diffractive part of the proton PDF,
\begin{align}\label{Eq:TentativeProb}
f_{i/\p}^{\mrm{D}}(x, Q^2) =&  \int_0^1 \d x_{\Pom} \, 
  f_{\Pom/\p}(x_{\Pom})\,  \int_0^1 \d x' \, 
  f_{i/\Pom}(x', Q^2) \, \delta(x - x_{\Pom} x') \nonu \\
  =&  \int_x^1  \frac{\d x_{\Pom}}{x_{\Pom}} \,  
  f_{\Pom/\p}(x_{\Pom}) \, f_{i/\Pom} \left( 
  \frac{x}{x_{\Pom}}, Q^2 \right),
\end{align}
where $f_{\Pom/\p}(x_{\Pom}) = \int f_{\Pom/\p}(x_{\Pom},t) \, \d t$,
as $t$ for the most part is not needed. The ratio
$f_{i/\p}^{\mrm{D}}/f_{i/\p}$ defines the tentative probability for
diffraction. A full evolution of the $\p\p$ system is then performed
and only the fraction of events passing the evolution without any
additional MPIs is kept as diffractive. Additional MPIs between the
two hadrons gives rise to hadronic activity, which could destroy
the rapidity gap between the elastically scattered hadron and the
interaction subsystem, which is one of the clear experimental
signatures of a diffractive event. If the event survives the no-MPI
criterion and is classified as diffractive, the partonic sub-collision
is assumed to have happened in a $\Pom\p$ sub-system. The $\Pom\p$
system is set up and a full evolution is performed in this
subsystem, similar to the method described above. 

The no-MPI requirement introduces a gap survival probability 
determined on an event-by-event basis, unlike other methods used 
in the literature. As MPIs only occur in hadron-hadron collisions, the
framework provides a simple explanation of the differences between the
diffractive event rates obtained at HERA and Tevatron. Diffractive 
fractions and survival probabilities obtained with the new framework 
show good agreement with experiments, while some distributions show 
less-than-perfect agreement, see \cite{Rasmussen:2015qgr} for a
discussion. The model is currently only available for single
diffraction; future work would be to extend this to both double and
central diffraction. 

\section{Total and elastic cross sections}\label{Sec:TotalAndElasticXS}

The parametrizations of the total and elastic cross sections are related 
through the optical theorem. The elastic cross section has
historically been well described in the framework of Regge theory,
with varying complexity based on the number of exchanges included in the
model. Up until the LHC era the simple ansatz of DL 
\cite{Donnachie:1992ny} using only a Pomeron and an 
effective Reggeon has described the total cross section surprisingly 
well. With a simple exponential $t$ spectrum, the SaS parametrization 
\cite{Schuler:1993wr} extended this to the elastic  cross section,
and here at least the low-$t$ 
data was well described. But with the higher energies probed at the
LHC it has become obvious that these simple parametrizations fail.
More complex trajectories have to be introduced in order to describe
both the rise of the total cross section and the $t$ spectrum
of the elastic cross section. 

We have chosen to implement two additional models in \textsc{Pythia}~8. 
One, the model from the COMPAS group as presented in the Review of
Particle Physics 2016 \cite{Patrignani:2016xqp}, is of great complexity, 
using six different single exchanges as well as some combinations of 
double exchanges, along with the exchange of three gluons, the latter 
becoming important at high $|t|$. The other, the newly developed ABMST 
model \cite{Appleby:2016ask}, is somewhat 
simpler, extending the original DL model to four single trajectories 
and all possible combinations of double exchanges between these, along 
with the triple-gluon exchange for high $|t|$ values. 

Recent TOTEM collaboration data on elastic scattering hint that none of 
the traditional models describe all aspects of their data. Specifically, 
TOTEM  obtains a decreasing $\rho$ parameter \cite{Antchev:2017yns}, and
observes no structure in the high-$|t|$ region (unpublished, but
see eg. \cite{TOTEMunpublished}). There is an ongoing discussion 
in both the theoretical and experimental community on how to describe
all data simultaneously. None of the models implemented here do that,
specifically they do not predict a decreasing $\rho$ value. 
Further, the ABMST model does not show any sign of structure at
high $|t|$, while the COMPAS one does. Models could be extended to 
include a  maximal odderon, similar to the work of Avila et al.\ 
\cite{Avila:2006wy,Martynov:2007kn} (AGN) and Martynov et al.\ 
\cite{Martynov:2017zjz} (FMO), which would be able to 
describe the decrease in $\rho$. At the time of writing the former has 
not been fitted to the new TOTEM data and the latter has not been extended 
to $t \neq 0$. Thus, for now, we have chosen not to implement either 
in \textsc{Pythia}̣~8, but we show the FMO model in the relevant figures 
for completeness. Below we will give short descriptions of each of the 
fully implemented models. 

\subsection{The COMPAS model}

For the Review of Particle Physics 2016 the COMPAS group 
\cite{Patrignani:2016xqp} has fitted a parametrization of the elastic 
differential cross section to all available $\p\p$ (upper signs) 
and $\p\pbar$ (lower signs) data, using a set of 37 free parameters. 
The cross sections are functions of the nuclear and amplitude, 
$T_{\pm}$, as well as the Coulomb amplitude, $T_{\pm}^c$,
\begin{align}
\sigma_{\mrm{tot}}(\sqrt{s}) =&
  \frac{\mrm{Im}\left[T_{\pm}(s,0)\right]}{\sqrt{s(s-4m_p^2)}}\nonu\\
\frac{\d\sigma_{\mrm{el}}}{\d t}(\sqrt{s},t) =&
  \frac{| T_{\pm}(s,t) + T_{\pm}^c |^2}{16\pi
  (\hbar c)^2\,s(s-4m_p^2)}\nonu\\
\sigma_{\mrm{el}}(\sqrt{s}) =& 
  \frac{1}{16\pi(\hbar c)^2\,s(s-4m_p^2)}
  \int_{t_{\mrm{min}}}^{t_{\mrm{max}}}\d t  \, | T_{\pm}(s,t) |^2.
\end{align}

The Coulomb term, $T_{\pm}^c$, and the nuclear term, $T_{\pm}$, are 
given as 
\begin{align}
T_{\pm}^c(s,t) =& \pm 8\pi\,\alpha_{\mrm{em}}\,\exp
  \left(\mp i\,\alpha_{\mrm{em}}\,\phi_{\pm}^{\mrm{NC}}(s,t)\right)
  \frac{s}{t} \left( 1 - \frac{t}{\Lambda^2} \right)^{-4} \nonu\\
T_{\pm}(s,t) =& F_+(\hat{s},t) \pm F_-(\hat{s},t)\nonu\\
F_+(\hat{s},t) =& F_+^H(\hat{s},t) + F_+^P(\hat{s},t) + 
  F_+^{PP}(\hat{s},t) + F_+^R(\hat{s},t) + F_+^{RP}(\hat{s},t) + 
  N_+(\hat{s},t)\nonu\\
F_-(\hat{s},t) =& F_-^{MO}(\hat{s},t) + F_-^O(\hat{s},t) + 
  F_-^{OP}(\hat{s},t) + F_-^R(\hat{s},t) + F_-^{RP}(\hat{s},t) + 
  N_-(\hat{s},t).
\end{align}
with the exact definitions of the different terms given as stated in
\cite{Patrignani:2016xqp}. It should be noted that earlier versions of
the PDG contains misprints in the definitions above as well as in the
crossing of even and odd functions, and the current still contains sign 
errors for the Coulomb term, so these should be used with care.

\subsection{The ABMST model}

A somewhat simpler scattering model was proposed by  
Appleby et al.\ describing $\p\p$ and $\p\pbar$ data from ISR 
to Tevatron energies \cite{Appleby:2016ask}. The model is based on 
work by Donnachie and Landshoff 
\cite{Donnachie:2013xia, Donnachie:2011aa} 
describing both elastic scattering and single diffractive scattering, 
but includes new and more sophisticated fits compared to the ones
from Donnachie and Landshoff. In this section the details 
on the elastic scattering will be given, while the single diffractive 
scatterings are presented in Sec. \ref{Sec:SDXS}.

The ABMST model includes both the Coulomb and nuclear
amplitudes, as well as the interference between the two. The cross
sections are given as 
\begin{align}
\frac{\d \sigma_{\mrm{el}}}{\d t} =& \pi \, | f_c(s,t)e^{i\alpha\phi(t)} +
f_n(s,t) |^2\nonu\\
\sigma_{\mrm{el}}(s) =& \pi
\int_{t_{\mrm{min}}}^{t_{\mrm{max}}} \d t \, |f_n(s,t)|^2\nonu\\
\sigma_{\mrm{tot}}(s) =& \mrm{Im}\left[f_n(s,0)\right], 
\end{align}
where the triple-gluon amplitude is left out of the nuclear
amplitude \cite{Donnachie:2013xia} when evaluating the total cross
section. The Coulomb amplitude from \cite{Amos:1985wx} is used
and the nuclear amplitude consists 
of five terms: A hard Pomeron ($\Pom_h$), a soft Pomeron ($\Pom_s$), 
the $f_2,a_2$ Regge trajectory ($\Reg_1$), the $\rho,\omega$ Regge 
trajectory ($\Reg_2$) and a triple-gluon exchange amplitude,
\begin{align}\label{Eq:A_El_ABMST}
f_n(s,t) =& A_{ggg}(t) + \sum_{i=\Pom_h, \Pom_s, \Reg_1, \Reg_2} A_i(s,t).
\end{align}
Also included is a double exchange term, where eg. two
Pomerons are exchanged. Exact definitions of the various terms are
found in \cite{Appleby:2016ask,MolsonThesis}. 
It should be noted that that the cross sections are only valid down to 
$\sqrt{s}=10$ GeV, and that the fits have only been performed up to
UA1 energies. We thus expect good agreement in this energy range,
whereas the fit might disagree with data outside of it. 

\subsection{The FMO model} 

The FMO model \cite{Martynov:2017zjz} includes the maximal odderon, 
excluded by hand in the COMPAS model. The odderon has been a 
controversial subject ever since its introduction, and so far no 
signs of it has been observed. The main feature of its introduction 
is that the difference between $\p\p$ and $\pbar\p$ total cross 
sections is not vanishing at high energies. Similarly the $\rho$ 
values will deviate at high energies. The FMO model only includes 
the $t=0$ contribution and can be written as
\begin{align}
\sigma_{\mrm{tot}}(s) =& \frac{\mrm{Im}T_{\pm}(s,0)}{\sqrt{s(s - 4
m_{\p}^2}}\nonu\\
T_{\pm} =& F_+^H \pm F_-^{MO} + F_+^R \pm F_-^R,
\end{align}
with the exact definitions of the crossing-odd and -even amplitudes
found in \cite{Martynov:2017zjz}. 

\subsection{Comparisons with data}

\begin{figure}[ht!]
\begin{minipage}[c]{0.475\linewidth}
\centering
\includegraphics[width=\linewidth]{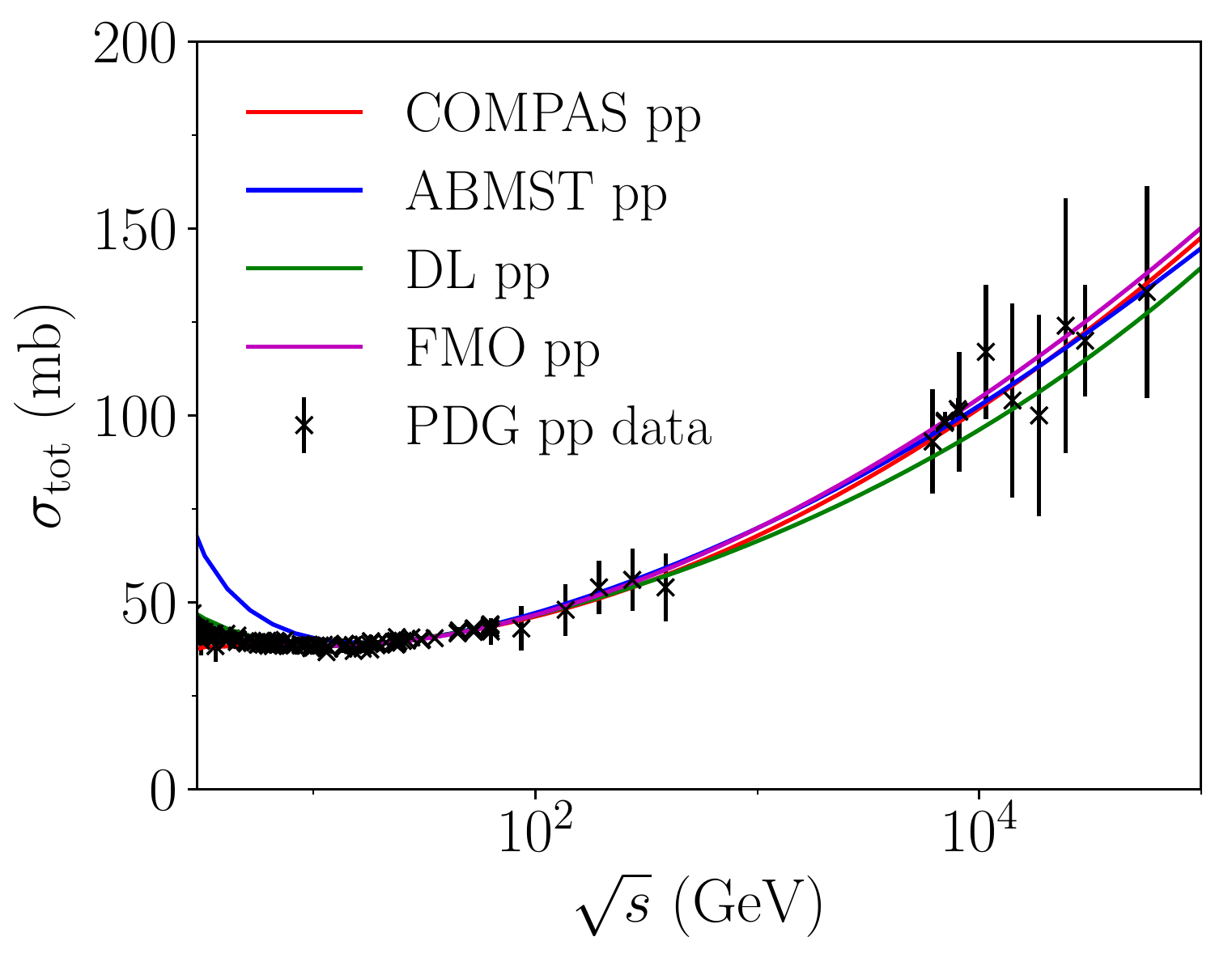}\\
(a)
\end{minipage}
\hfill
\begin{minipage}[c]{0.475\linewidth}
\centering
\includegraphics[width=\linewidth]{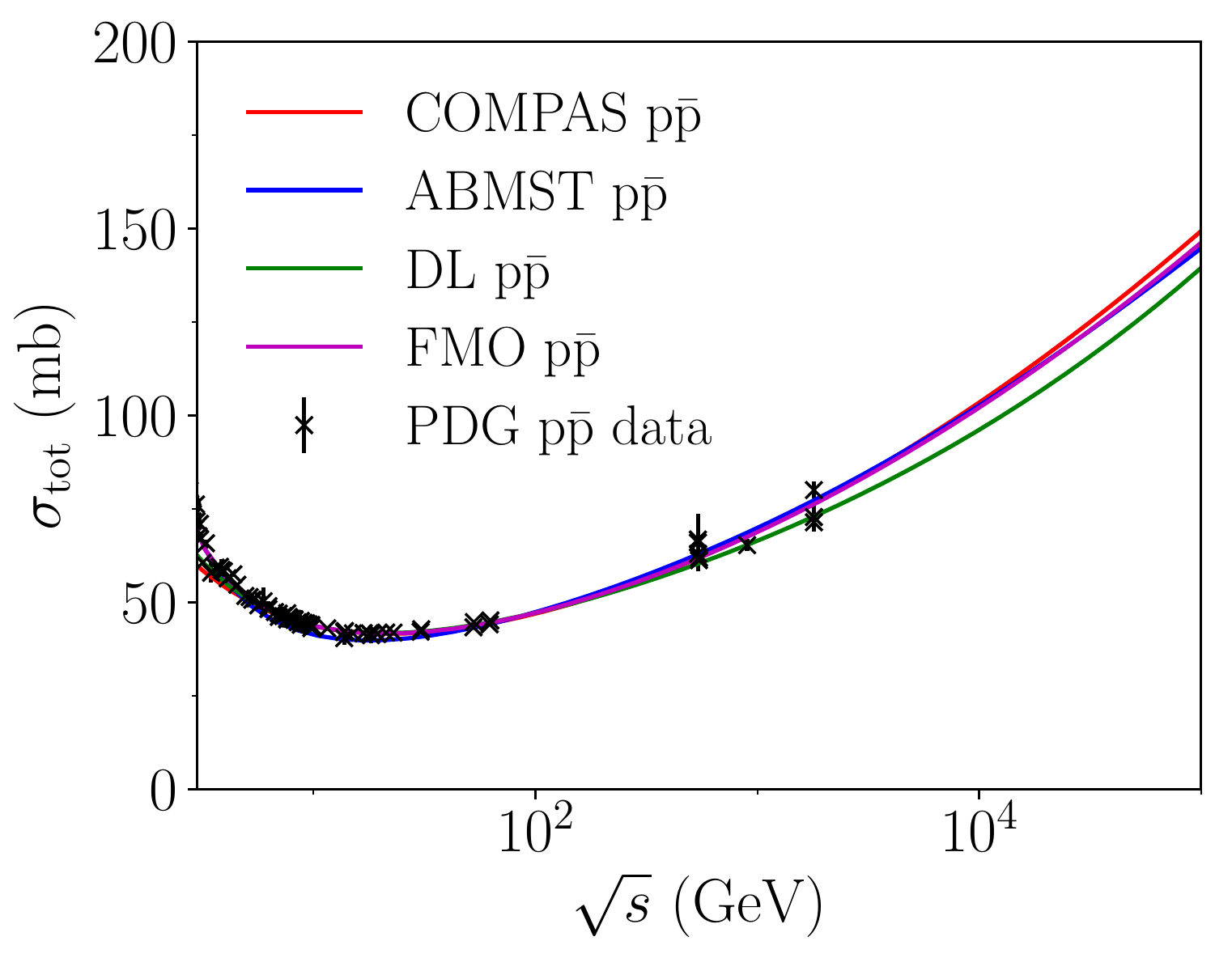}\\
(b)
\end{minipage}
\begin{minipage}[c]{0.475\linewidth}
\centering
\includegraphics[width=\linewidth]{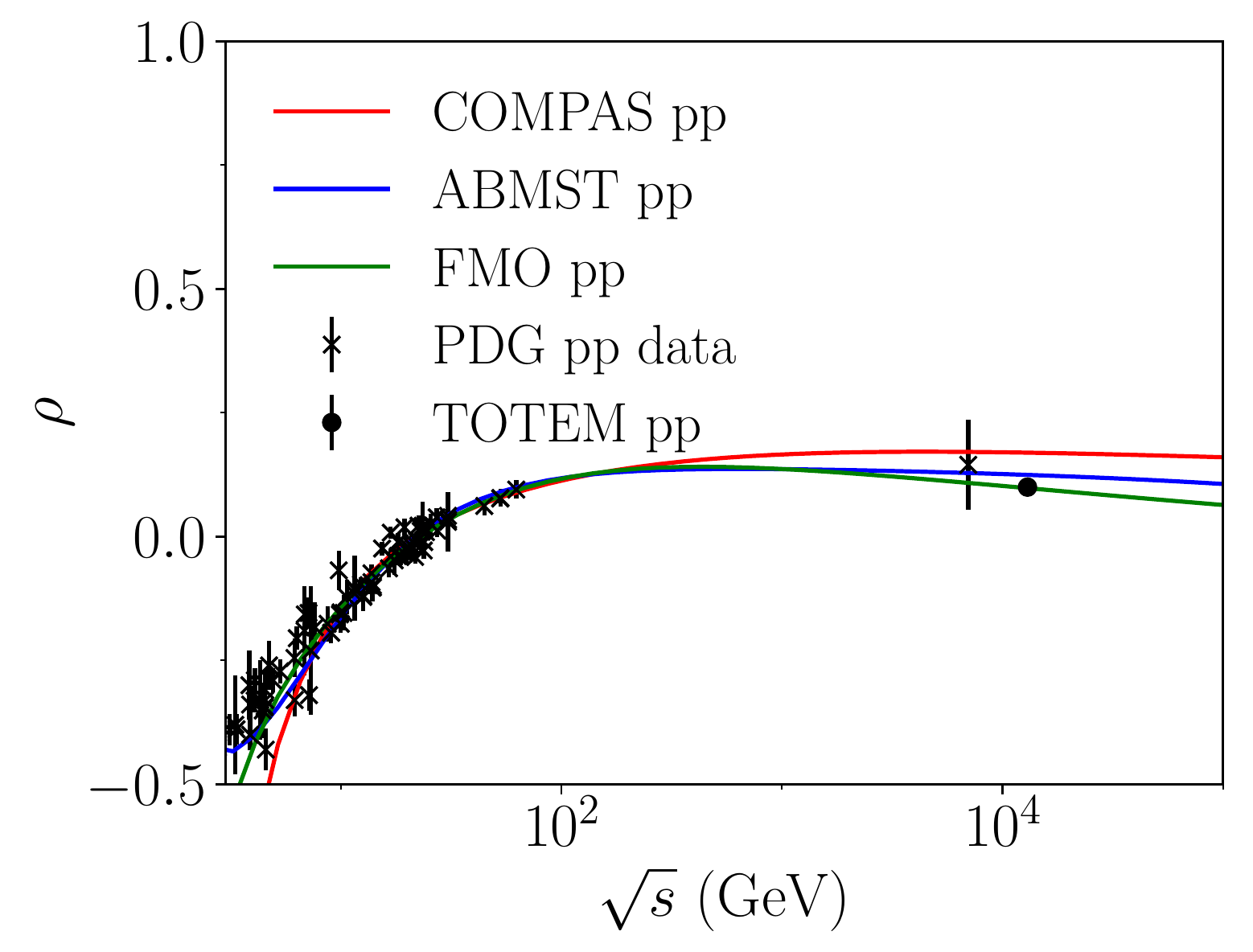}\\
(c)
\end{minipage}
\hfill
\begin{minipage}[c]{0.475\linewidth}
\centering
\includegraphics[width=\linewidth]{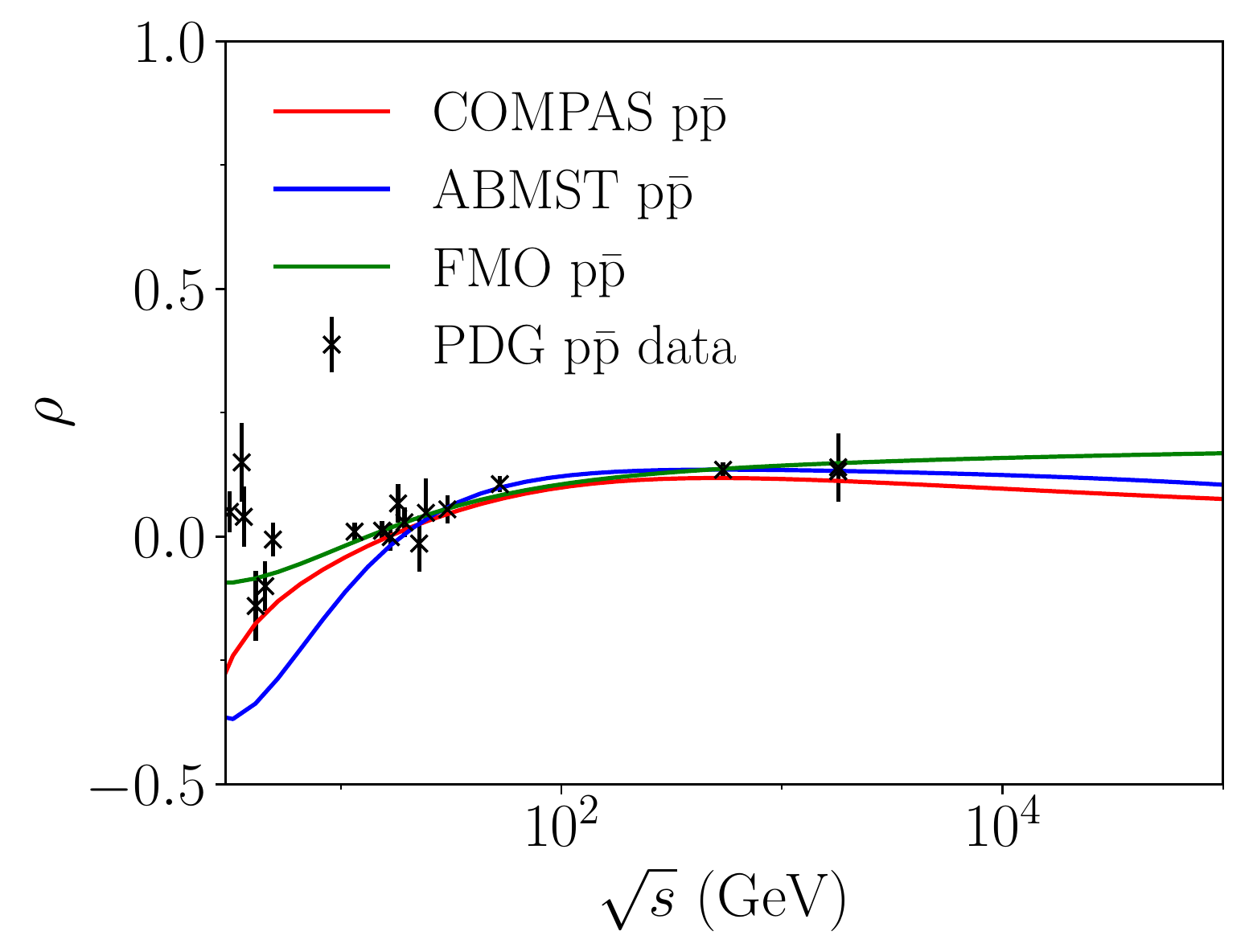}\\
(d)
\end{minipage}
\caption{\label{Fig:SigTotSigEl}
The total cross section parametrizations in (a) $\p\p$ and (b)
$\p\pbar$ processes. The ratio of real to imaginary parts of the
elastic amplitude at $t=0$ for $\p\p$ (c) and $\p\pbar$ (d). 
Note that the SaS model has been left out in (c) and (d), as $\rho$ is
a constant here, that can be set freely by the user. Data from PDG
\cite{Patrignani:2016xqp}.}
\end{figure}

In fig.~\ref{Fig:SigTotSigEl}a,b we show the above parametrizations 
of the total cross section and in fig.~\ref{Fig:SigTotSigEl}c,d the 
$\rho$ parameter, for $\p\p$ and $\p\pbar$ processes respectively. 
Note how the ABMST $\sigma_{\mrm{tot}}^{\p\p}$ parametrization 
rises at $\sqrt{s}<10$ GeV, a consequence of it not being fitted to this 
range. We do not aim to describe so low energies in \textsc{Pythia}~8, 
so this is not an issue. Both the ABMST and COMPAS parametrizations
well describe the LHC data points in $\p\p$, and seem to favour the higher 
of the Tevatron data points in $\p\pbar$ processes, unlike the original 
DL parametrization available in \textsc{Pythia}~8. In 
fig.~\ref{Fig:SigTotSigEl}c the $\rho$ is well described by all three
parametrizations, below LHC energies. But at LHC the latest TOTEM value
\cite{Antchev:2017yns} is described only by the FMO model, which 
explicitly includes the maximal odderon term in order for $\rho$ to 
decrease here. This term also gives rise to the difference in $\rho$ 
for $\p\p$ and $\p\pbar$ processes, as seen in 
figs.~\ref{Fig:SigTotSigEl}c,d, a difference not present in the other 
two models.

\begin{figure}[ht!]
\begin{minipage}[c]{0.475\linewidth}
\centering
\includegraphics[width=\linewidth]{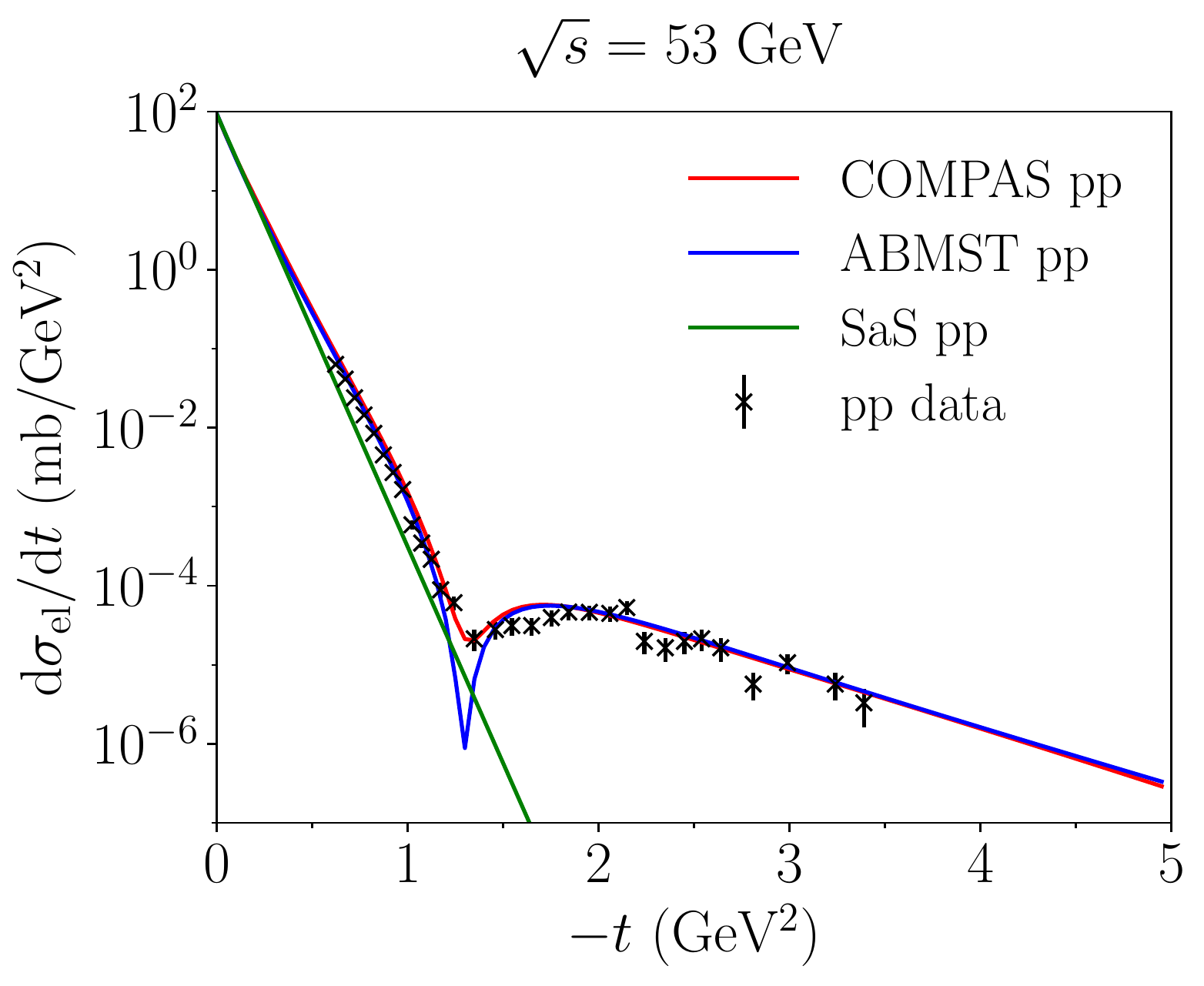}\\
(a)
\end{minipage}
\hfill
\begin{minipage}[c]{0.475\linewidth}
\centering
\includegraphics[width=\linewidth]{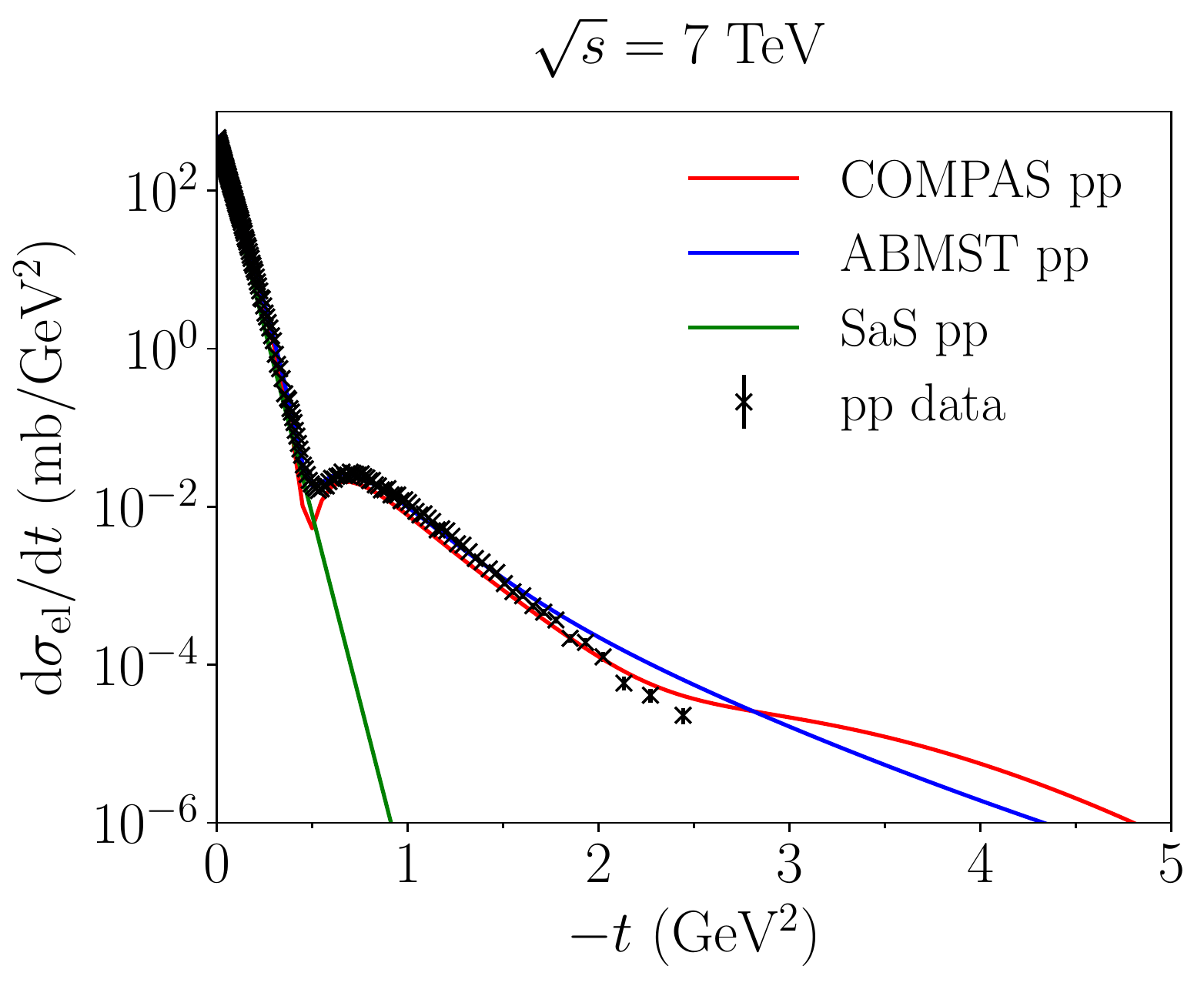}\\
(b)
\end{minipage}
\begin{minipage}[c]{0.475\linewidth}
\centering
\includegraphics[width=\linewidth]{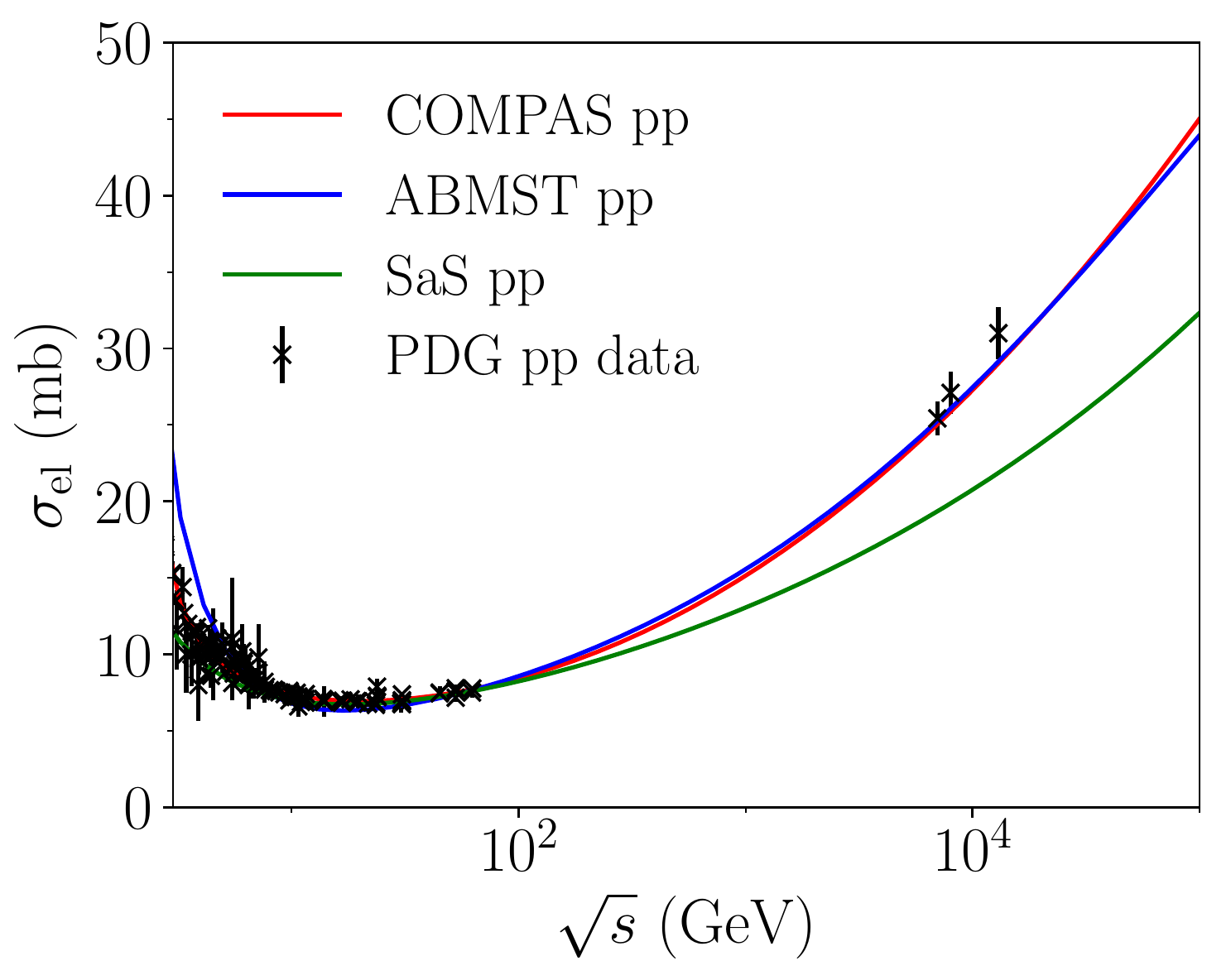}\\
(c)
\end{minipage}
\hfill
\begin{minipage}[c]{0.475\linewidth}
\centering
\includegraphics[width=\linewidth]{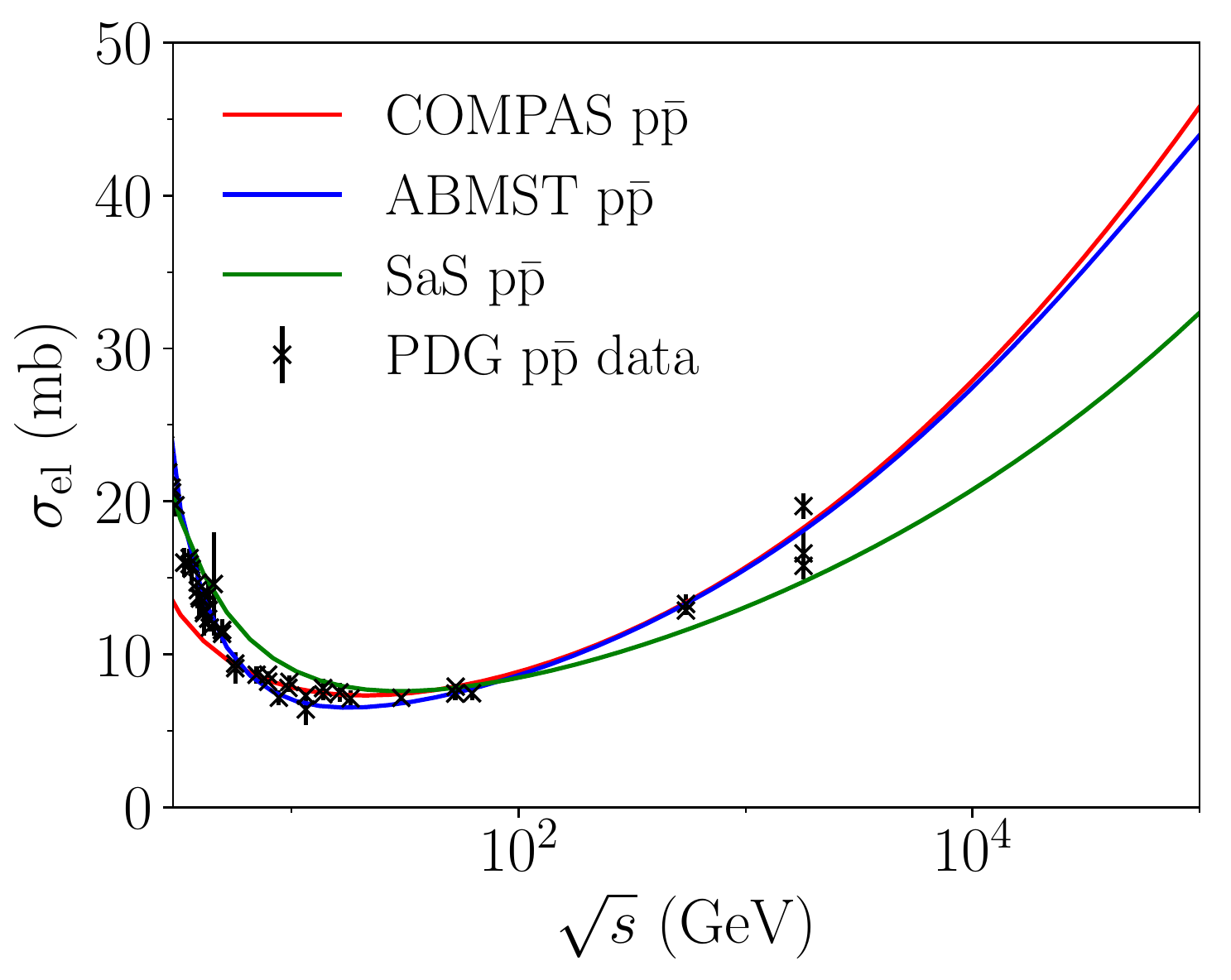}\\
(d)
\end{minipage}
\caption{\label{Fig:dSigEl_dt}
The elastic differential cross section parametrizations in 
$\p\p$ collisions at 53 GeV (a) and 7 TeV (b).
The integrated elastic cross section parametrizations in (c) 
$\p\p$ and (d) $\p\pbar$ processes. Data from PDG
\cite{Patrignani:2016xqp}.
}
\end{figure}

In fig.~\ref{Fig:dSigEl_dt} we show the available parametrizations 
of the elastic differential (a,b) and integrated  (c,d) cross
sections for $\p\p$ and $\p\pbar$ processes. Here it is 
evident that the pure exponential description used by SaS 
only makes sense for small $|t|$. Both the COMPAS and ABMST 
parametrizations have been fitted to the $\sqrt{s}=23$ GeV data, but 
not to the 7 TeV data. Here it seems that the COMPAS parametrization 
prefers a larger dip than seen in data, while it captures the high-$|t|$ 
region slightly better than the ABMST parametrization. It is also 
evident that SaS underestimates the rise of the total elastic cross 
section, whereas the other two do quite well.

\section{Single diffractive cross sections}\label{Sec:SDXS}

As we proceed to the topologies of diffraction, the 
situation is more complicated than for total and elastic cross sections. 
The experimental definition of diffraction is based on the presence of 
rapidity gaps, but such gaps are subject to random fluctuations in the 
hadronization process, and therefore cannot be mapped one-to-one to an 
underlying colour-singlet-exchange mechanism. Also the separation 
between single, double and central diffraction is not always so 
clearcut. Some single-diffractive data is available at lower energies, 
but much of it is old and of varied quality. This will of course affect 
any model trying to describe these topologies, as usually there are model
parameters that have to be fitted to data. To the best of our knowledge, 
only a few models actually try to fit data fully differentially in both 
$s$, $M_X^2$ and $t$. The normal ansatz is instead to define an 
$s$-independent $\Pom$ flux, with factorized $\xi$ and $t$ distributions, 
e.g.\ of the form $(\d\xi / \xi^{1+ \delta}) \, \exp(b \, t) \, \d t$ 
\cite{Donnachie:1984xq,Berger:1986iu,Jung:1993gf,Aktas:2006hy}
where $\delta$ is a small number. The t-integrated $\xi$ distribution is 
then directly mapped on to an $M_X^2 = \xi s$ spectrum.
 
The COMPAS group has not made any attempts to describe other topologies 
that the elastic, neither has the FMO model. Hence, in addition to the
 already implemented SaS and MBR models, we are left with the ABMST model 
as a new alternative, that gives a full description of the single 
diffractive topologies. This model has been fitted to differential data 
in the energy range $17.2 < \sqrt{s} < 546$ GeV and in the $t$ range 
$0.015 < |t| < 4.15$ GeV$^2$, and is thus expected to give a reasonable 
prediction in this range. The model, however, has some unfortunate 
features, which we will discuss in a later section. But first an 
introduction to the basics of the model itself.

\subsection{The ABMST model}

In \cite{Appleby:2016ask} the authors present a model for single diffractive
dissociation inspired by Donnachie and Landshoff. They operate in two
regimes, high and low mass diffraction, separated at 
\begin{align}\label{Eq:Mcut}
M_{\mrm{cut}}(s) =&
\begin{cases}
3 & s < 4000\,\mrm{GeV}^2\\
3 + 0.6\ln\left(\frac{s}{4000}\right)& s > 4000\,\mrm{GeV}^2\\
\end{cases}.
\end{align}
In the high mass regime, they use a triple-Regge model with two
components; An effective Pomeron and a degenerate Reggeon term. In
order for the unknown phases of the propagators to vanish, they
require that the two $t$-dependent propagators in the diagrams
contributing to the single diffractive cross section are equal. This
results in four diagrams; $\Pom\Pom\Pom$, $\Pom\Pom\Reg$, 
$\Reg\Reg\Pom$, $\Reg\Reg\Reg$. The authors also include pion
exchange in the differential cross section arriving at
\begin{align}
\frac{\d^2\sigma_{\mrm{HM}}}{\d t\d\xi}(\xi, s,t) =&
  f_{\Pom\Pom\Pom}(t)\xi^{\alpha_{\Pom}(0)-2\alpha_{\Pom}(t)}
  \left(\frac{s}{s_0}\right)^{\alpha_{\Pom}(0)-1}\nonu\\
  +&f_{\Pom\Pom\Reg}(t)\xi^{\alpha_{\Reg}(0)-2\alpha_{\Pom}(t)}
  \left(\frac{s}{s_0}\right)^{\alpha_{\Reg}(0)-1}\nonu\\
  +&f_{\Reg\Reg\Pom}(t)\xi^{\alpha_{\Pom}(0)-2\alpha_{\Reg}(t)}
  \left(\frac{s}{s_0}\right)^{\alpha_{\Pom}(0)-1}\nonu\\
  +&f_{\Reg\Reg\Reg}(t)\xi^{\alpha_{\Reg}(0)-2\alpha_{\Reg}(t)}
  \left(\frac{s}{s_0}\right)^{\alpha_{\Reg}(0)-1}\nonu\\
  +&\frac{g_{\pi\pi\p}^2}{16\pi^2}\frac{|t|}{(t-m_{\pi}^2)^2}
  F^2(t)\xi^{1-2\alpha_{\pi}(t)}\sigma_{\pi^0\p}(s\xi),
\end{align}
with trajectories and parameter choices found in \cite{Appleby:2016ask}.
Each of the effective three-Reggeon couplings are given as 
\begin{align}\label{Eq:Couplings}
f_{kki}(t) =& A_{kki}e^{B_{kki}t}+C_{kki},
\end{align}
except for the triple-Pomeron coupling, which is modified as
\begin{align}
f_{\Pom\Pom\Pom}(t) =& 
\begin{cases}
0.4 + 0.5t & -0.25 \leq t < -10^{-4}\\
(A_{\Pom\Pom\Pom}e^{B_{\Pom\Pom\Pom}t}+C_{\Pom\Pom\Pom})
\left(\frac{t}{t-0.05}\right) & -1.15 \leq t < -0.25\\
(A_{\Pom\Pom\Pom}e^{B_{\Pom\Pom\Pom}t}+C_{\Pom\Pom\Pom})
\left(\frac{t}{t-0.05}\right) \times & \\
\times (1 + 0.4597(|t|-1.15) + 5.7575(|t|-1.15)^2) & -4 \leq t < 1.15\\
\end{cases}.
\end{align}

Four resonances are modelled in the low-mass regime, along with a 
background from the high-mass regime and a contact term matching 
the two regimes smoothly. The resonances are excited states of the 
proton, each a unit of angular momentum higher than the previous one. 
The resonances are parametrized by Breit--Wigner shapes with masses
$m_i$, widths $\Gamma_i$ and couplings $c_i$,
\begin{align}
\frac{\d^2\sigma_{\mrm{res}}}{\d t\d\xi}(\xi,s,t) =& 
\frac{e^{13.5(t+0.05)}}{\xi}\sum_{i=1}^4\left[\frac{c_im_i\Gamma_i}{(\xi
s - m_i^2)^2 + (m_i\Gamma_i)^2}\right] ~,
\end{align}
with exact definitions found in the paper.
The background is assumed quadratic and vanishes
at a threshold, $\xi_{\mrm{th}} = \frac{(m_{\p} + m_{\pi})^2}{s}$,
\begin{align}\label{Eq:Bkg}
A_{\mrm{bkg}}(\xi,s,t) =& a(s,t)(\xi - \xi_{\mrm{th}})^2 +
b(s,t)(\xi-\xi_{\mrm{th}}).
\end{align}
A matching term between the high- and low-mass regions is subtracted 
from the resonances to avoid any discontinuities at
$\xi_{\mrm{cut}}$, and parametrized such that it is equal to the 
magnitude of the resonance term at the matching point.


\subsection{Comments on the ABMST model}

\begin{figure}[ht!]
\begin{minipage}[c]{0.475\linewidth}
\centering
\includegraphics[width=\linewidth]{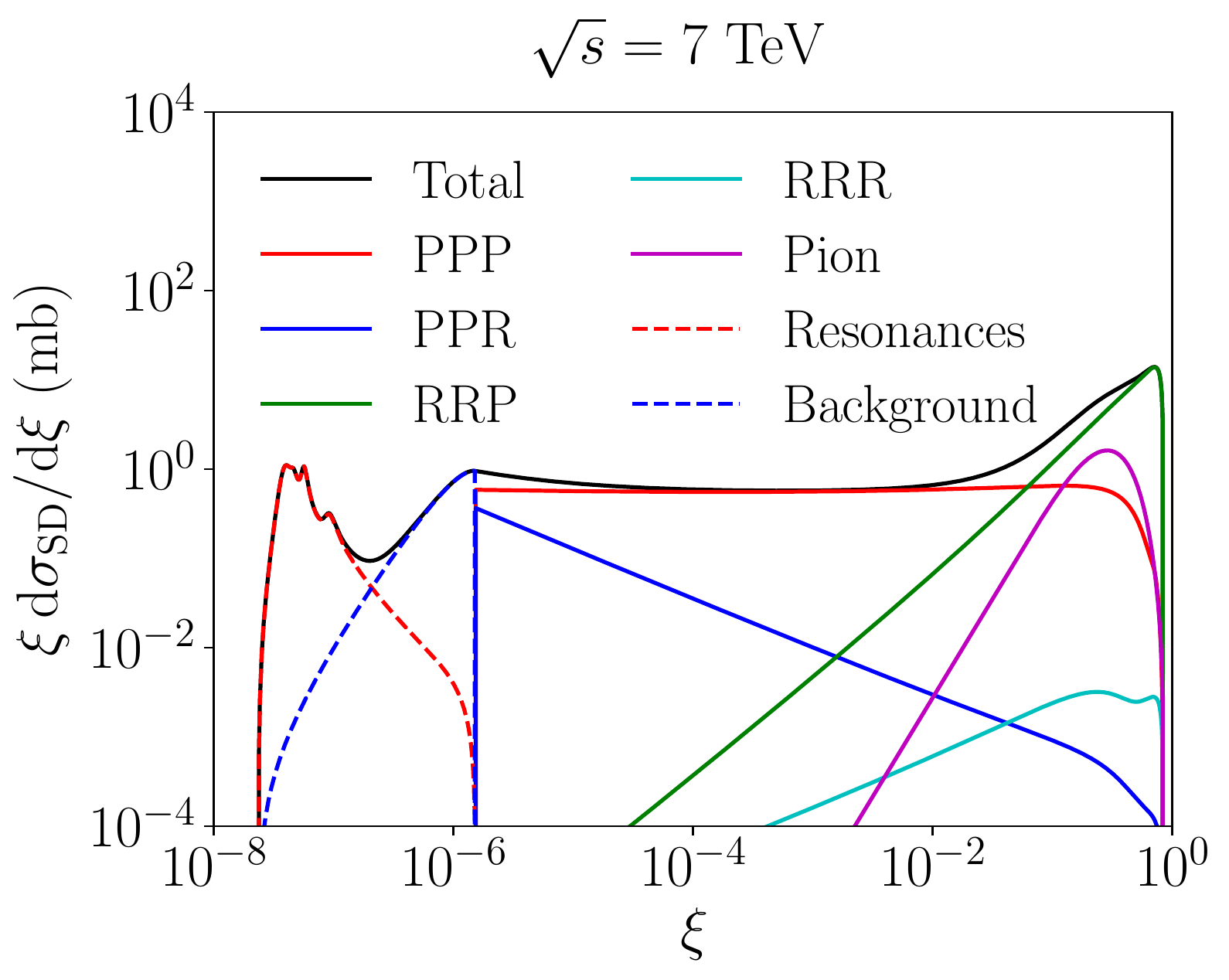}\\
(a)
\end{minipage}
\hfill
\begin{minipage}[c]{0.475\linewidth}
\centering
\includegraphics[width=\linewidth]{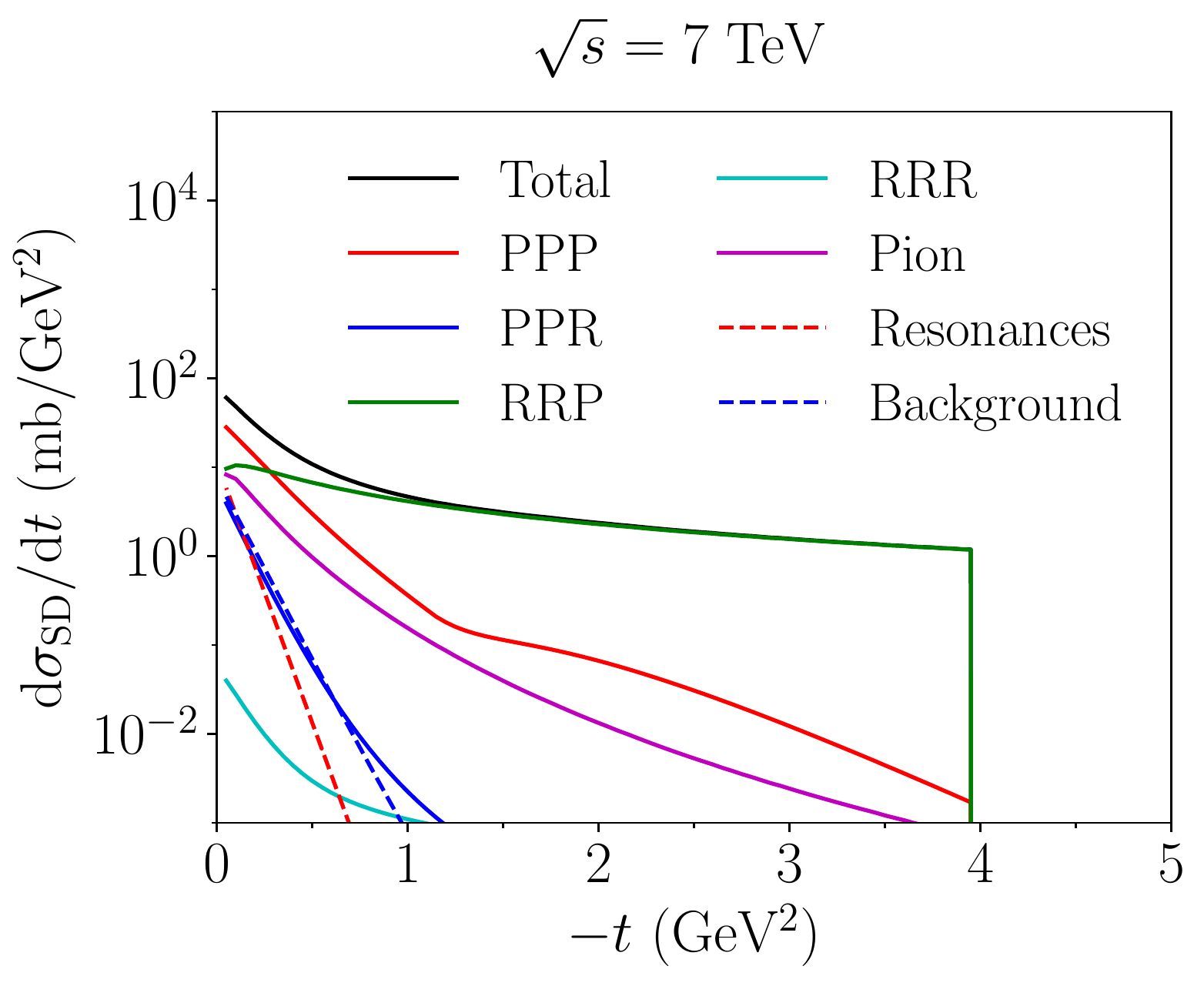}\\
(b)
\end{minipage}
\begin{minipage}[c]{0.475\linewidth}
\centering
\includegraphics[width=\linewidth]{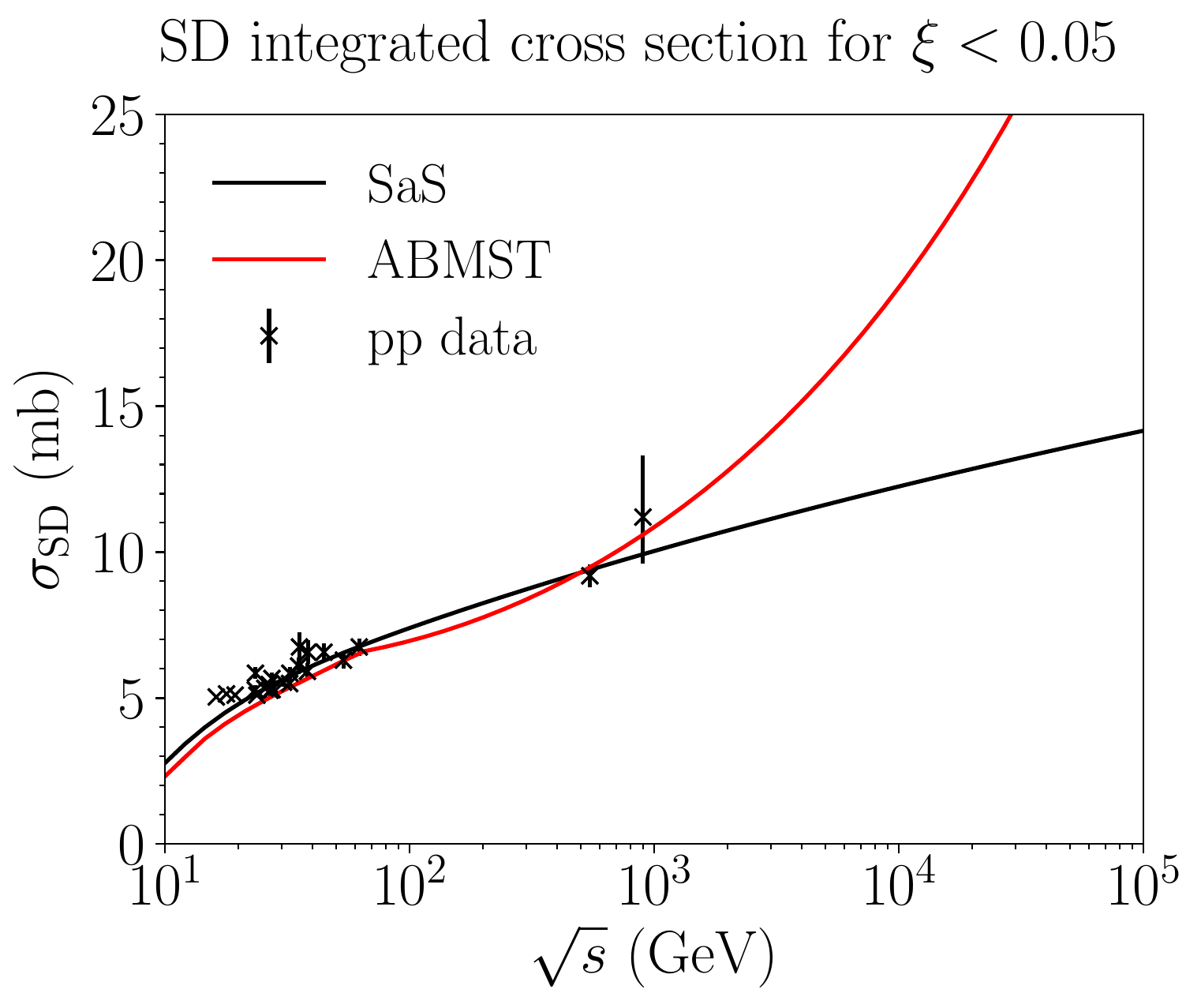}\\
(c)
\end{minipage}
\hfill
\begin{minipage}[c]{0.475\linewidth}
\centering
\includegraphics[width=\linewidth]{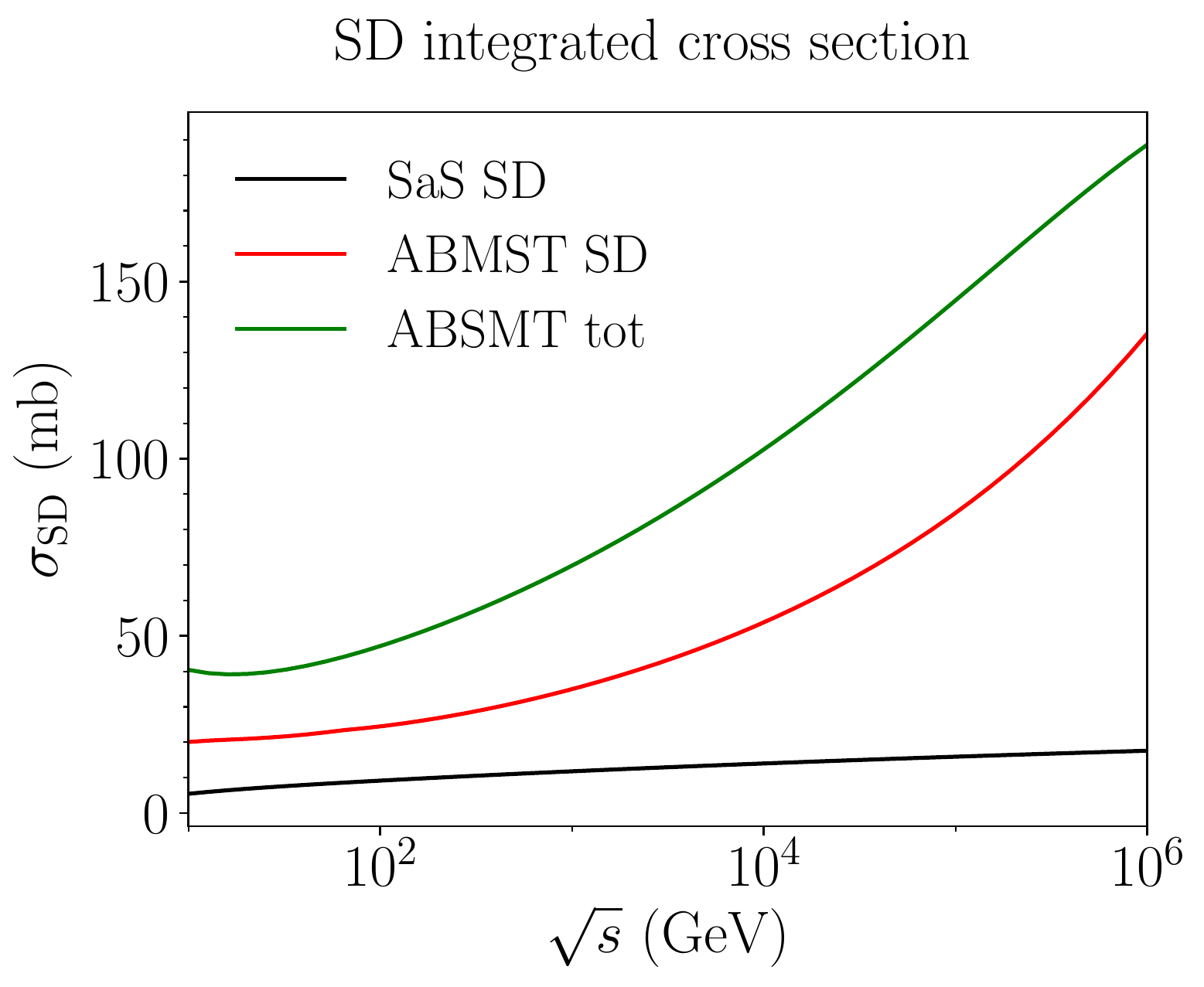}\\
(d)
\end{minipage}
\caption{\label{Fig:ABMSTbreakdown} The different components of the
ABMST model for single diffraction as a function of (a) $\xi$ and (b)
$t$ at 7 TeV. The integrated single diffractive cross section as a
function of $\sqrt{s}$ for $\xi<0.05$ (c) and in the full single 
diffractive phase space (d). Data from references in \cite{Appleby:2016ask}.
}
\end{figure}

In fig.~\ref{Fig:ABMSTbreakdown}a,b we show the different components of
the ABMST model at an energy of $\sqrt{s}=7$ TeV along with the
integrated cross sections in fig.~\ref{Fig:ABMSTbreakdown}c,d. 
We have several comments to these distributions, as they show some 
unexpected features. 

To begin, consider the differential distribution in 
fig.~\ref{Fig:ABMSTbreakdown}a. Here the cross 
section (multiplied by a factor of $\xi$ for visibility) is shown as a 
function of $\xi$, displaying both the low-mass resonances and the 
high-mass Regge terms. Note, however, the dip between these two regimes, 
a decrease of a factor of 10. This is a feature of the background 
modelling, whereas one would expect a more smooth transition between 
the two regimes. There is no physical motivation as to why the Regge 
trajectories should have a quadratic behaviour at low masses, since 
none of the terms show this behaviour at higher masses. One could 
imagine a simple continuation of the high-mass background to lower 
masses, with the resonances added on top. But this would likely cause 
too high a cross section in the low-mass region, hence requiring a 
remodelling of the background description to avoid too high a low-mass 
cross section.

Similarly unexpected is the increase of the cross section at higher 
masses ($\xi \sim 1$), induced by the triple-Reggeon and pion terms. 
The larger the mass of the system the smaller the rapidity gap between 
the diffractive system and the elastically scattered proton. The rule 
of thumb is that $\Delta y_{\mrm{gap}} \approx − \ln(\xi)$, so for large 
$\xi$ there will essentially be no gap at all. The diffractive system 
will simply look like a non-diffractive one, making it impossible to 
distinguish between the two experimentally. The rise at $\xi \sim 1$ 
also introduces a vast increase with energy in the integrated cross 
section, making the single-diffractive cross section dominate at large 
energies, which leaves little room for other processes,
see fig.~\ref{Fig:ABMSTbreakdown}d. The authors themselves have tried 
to dampen the increase of the cross section by allowing the mass cut, 
separating the low- and high-mass regimes, to vary with $s$, 
eq.~(\ref{Eq:Mcut}). Unfortunately the introduced dampening gives rise to 
a kink in the integrated cross section where the dampening kicks in, 
at $\sqrt{s}\sim 60$ GeV, and does not dampen the cross section 
sufficiently at high energies.

In fig.~\ref{Fig:ABMSTbreakdown}b we show the ABMST model differential in 
$t$. Noteworthy are
the $t$-independent terms $C_{kki}$ and the sharp cutoff at $t=-4$ GeV$^2$,
both of which are unphysical on their own. That is, if the sharp cutoff 
is disregarded, then all but the pion and triple-Pomeron terms become 
constant at large $|t|$, lacking any form factor suppression for scattering 
a proton without breaking it up.  
The choice of $t$ parametrization shape was based on the goodness-of-fit, 
and not on any physical grounds. The authors note that the 
parametrization as such gives too large a cross section at high energies, 
hence the modification of the Pomeron coupling, as this dominates at 
high energies. The $t$ ansatz may also cause problems if used in other 
diagrams, e.g.\ in the extension to double and central diffraction that
we will introduce later.

As \textsc{Pythia}~8 aims to describe current and future colliders,
the need for a more sensible high-energy behaviour of the ABMST model 
is evident. It is not realistic to have a model where single diffraction 
and elastic scattering almost saturates the total cross section at FCC
energies (at $10^5$~GeV $\sigma_{\mrm{tot}} - \sigma_{\mrm{el}} 
- \sigma_{\mrm{SD}} \approx 145 - 45 - 80 \approx 20$ mb).
At the same time we want to make use of the effort already 
put into the careful tuning to low-energy and low-diffractive-mass data.
We have thus chosen to provide a modified version of the ABMST model, 
addressing the problems discussed above, as described in the next section, 
while retaining the good aspects of the ABMST model. Both the modified 
and the original version of the ABMST model are made available in the 
latest \textsc{Pythia}~8 release.
 
\subsection{The modified ABMST model}  

To smoothen the dip between the low-mass and high-mass regions,
several background terms have been studied, such as a linear background 
becoming constant at threshold, a combination of the linear and the
quadratic background and, as an extreme, a continuation of the high-mass
background. The best results was found with the combination of the
linear and quadratic,
\begin{align}
A_{\mrm{bkg}}(s) =&
\begin{cases}
A_{\mrm{bkg}}^{\mrm{quadratic}} & M_X < M_{4}\\
A_{\mrm{bkg}}^{\mrm{linear}} & M_{4} < M_X < M_{\mrm{cut}}\\
\end{cases},
\end{align}
where $M_4$ is the mass of the fourth resonance.
 
The new parametrization of the high-mass background in the
low-mass region does smoothen the decrease between the two regions,
but in itself does increase the integrated cross section. We 
tame the integrated cross section by introducing a multiplicative 
rescaling of the high-mass region, as well as a different $M_{\mrm{cut}}$
parametrization. Again several possibilities have been tried, and
best results were obtained for a $\ln^2(s)$-dependent $M_{\mrm{cut}}$
and rescaling. That is, $M_{\mrm{cut}} = 3+c\,\ln^2(s/s_0)$ GeV and
the rescaling factor is $3 / (3+c\,\ln^2(s/s_0))$, with $c$ a free 
parameter and $s_0 = 100$ GeV$^2$, which is also where the rescaling 
begins, so as to avoid kinks in the distributions.
 
While this change reduces the cross section at intermediate $\xi$ 
values, it does not address the strong rise near $\xi = 1$. This is 
an unobservable behaviour, as already argued, and therefore we also 
introduce a dampening factor $1 / (1 + (\xi\exp(y_{\mrm{min}}))^p)$ 
for the high-mass region. Here $y_{\mrm{min}}$ is the gap size where 
the dampening factor is $1/2$ and $p$ regulates how steeply this 
factor drops around $y_{\mrm{min}}$; by default $y_{\mrm{min}} = 2$ 
and $p = 5$. 

Separately, we wish to remove the artificial cut at $t = -4$~GeV$^2$,
in favour of a shape that is valid at all $t$ scales. To this end, 
couplings are modified as
\begin{align}
f_{kki}^{\mrm{ABMST}}(t) \rightarrow& f_{kki}^{\mrm{mod}}(t)
  = (A_{kki}+C_{kki}^{\mrm{mod}})e^{B_{kki}^{\mrm{mod}}t},
\end{align}
where two new parameters $C_{kki}^{\mrm{mod}}$ and $B_{kki}^{\mrm{mod}}$
are introduced. These are fixed by the two requirements that the 
integral over $t$ and the average $t$ value should remain unchanged
relative to the original ABMST values. Note, however, that 
we do not modify the $\Pom\Pom\Pom$ part, as this already 
has the desired decreasing behaviour at high $|t|$. 
Besides these modifications, a minimum diffractive slope
$B_{SD} = 2$ is introduced, to avoid any unphysical situations where 
the slope could become negative.

\begin{figure}[ht!]
\begin{minipage}[c]{0.475\linewidth}
\centering
\includegraphics[width=\linewidth]{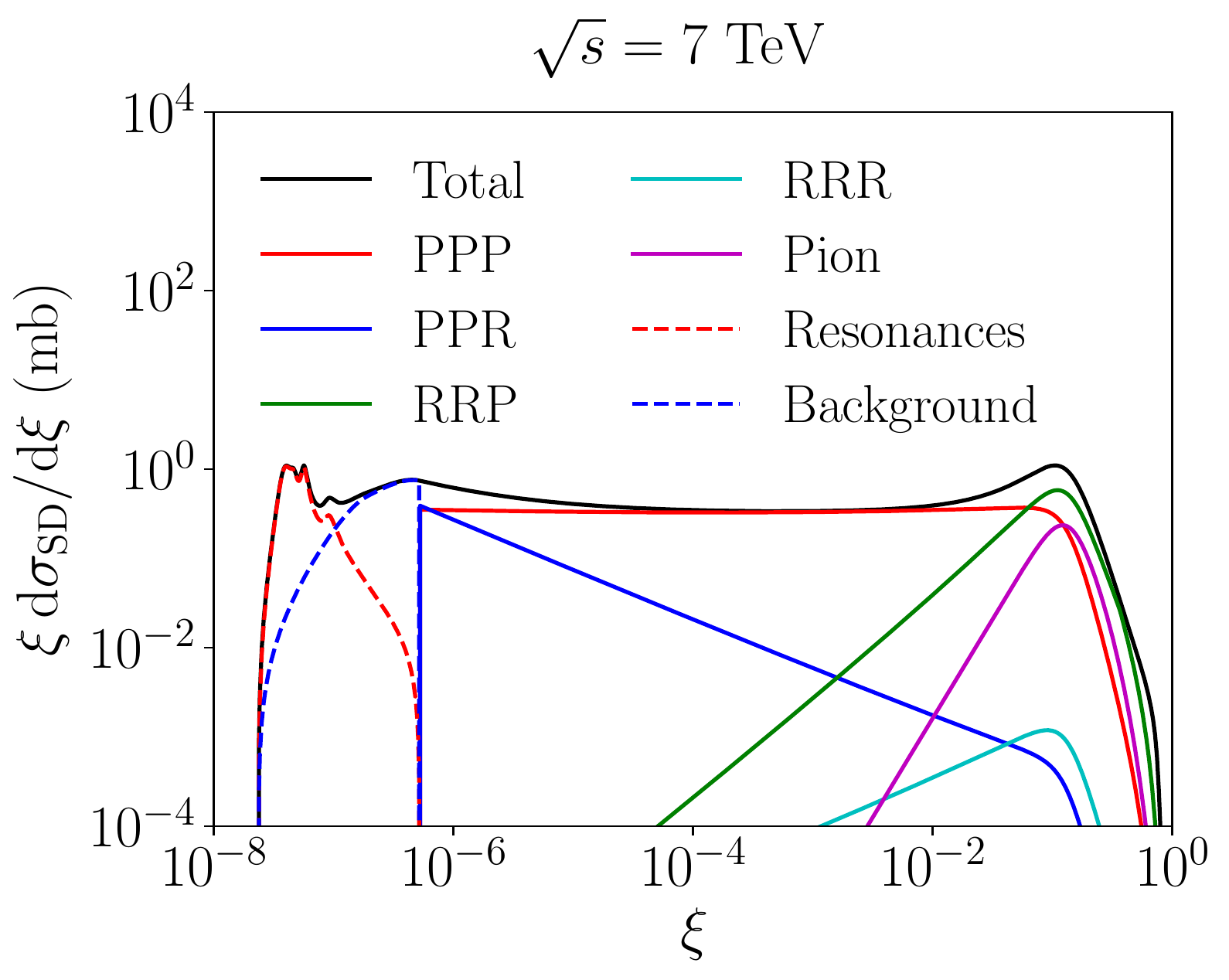}\\
(a)
\end{minipage}
\hfill
\begin{minipage}[c]{0.475\linewidth}
\centering
\includegraphics[width=\linewidth]{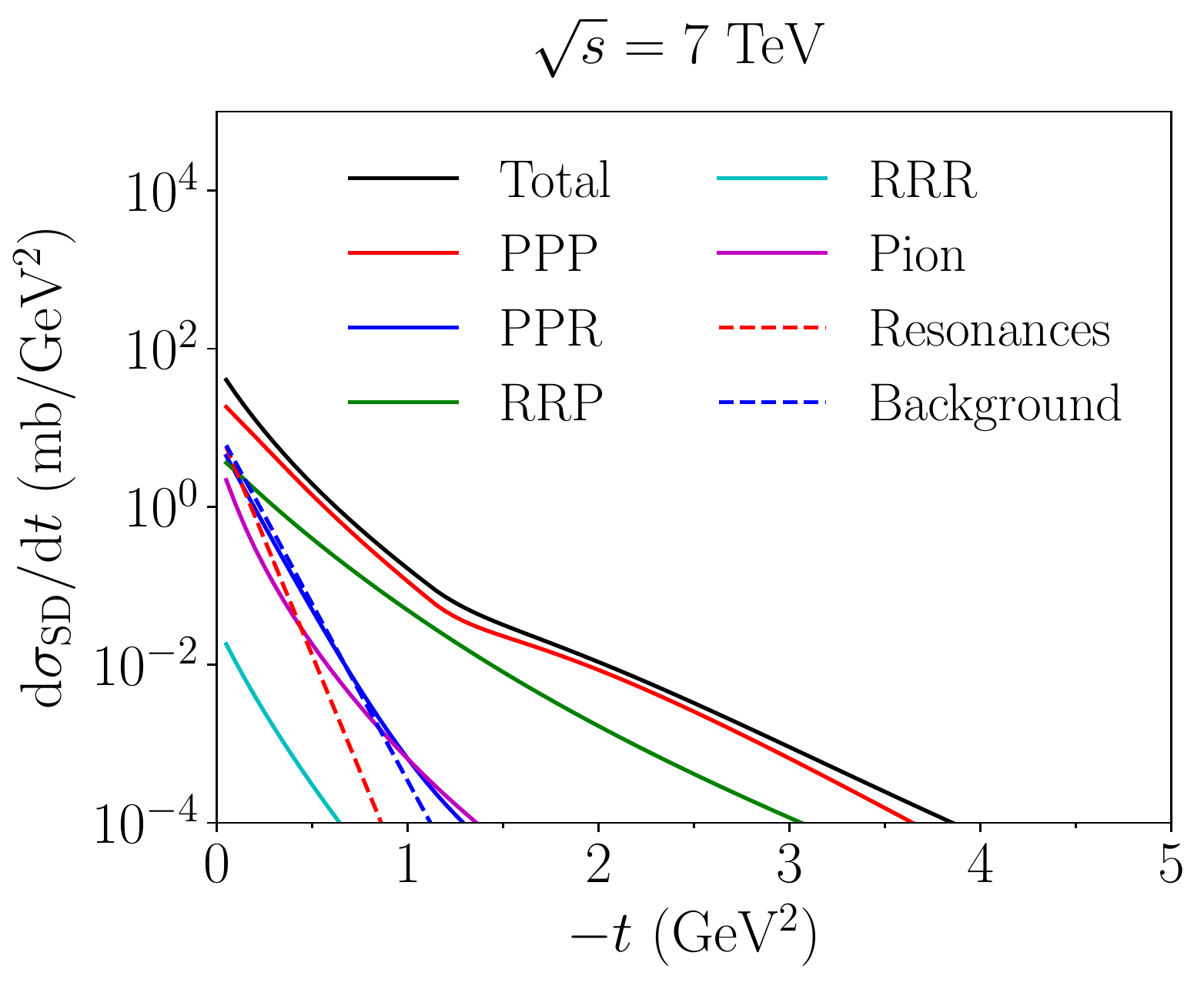}\\
(b)
\end{minipage}
\begin{minipage}[c]{0.475\linewidth}
\centering
\includegraphics[width=\linewidth]{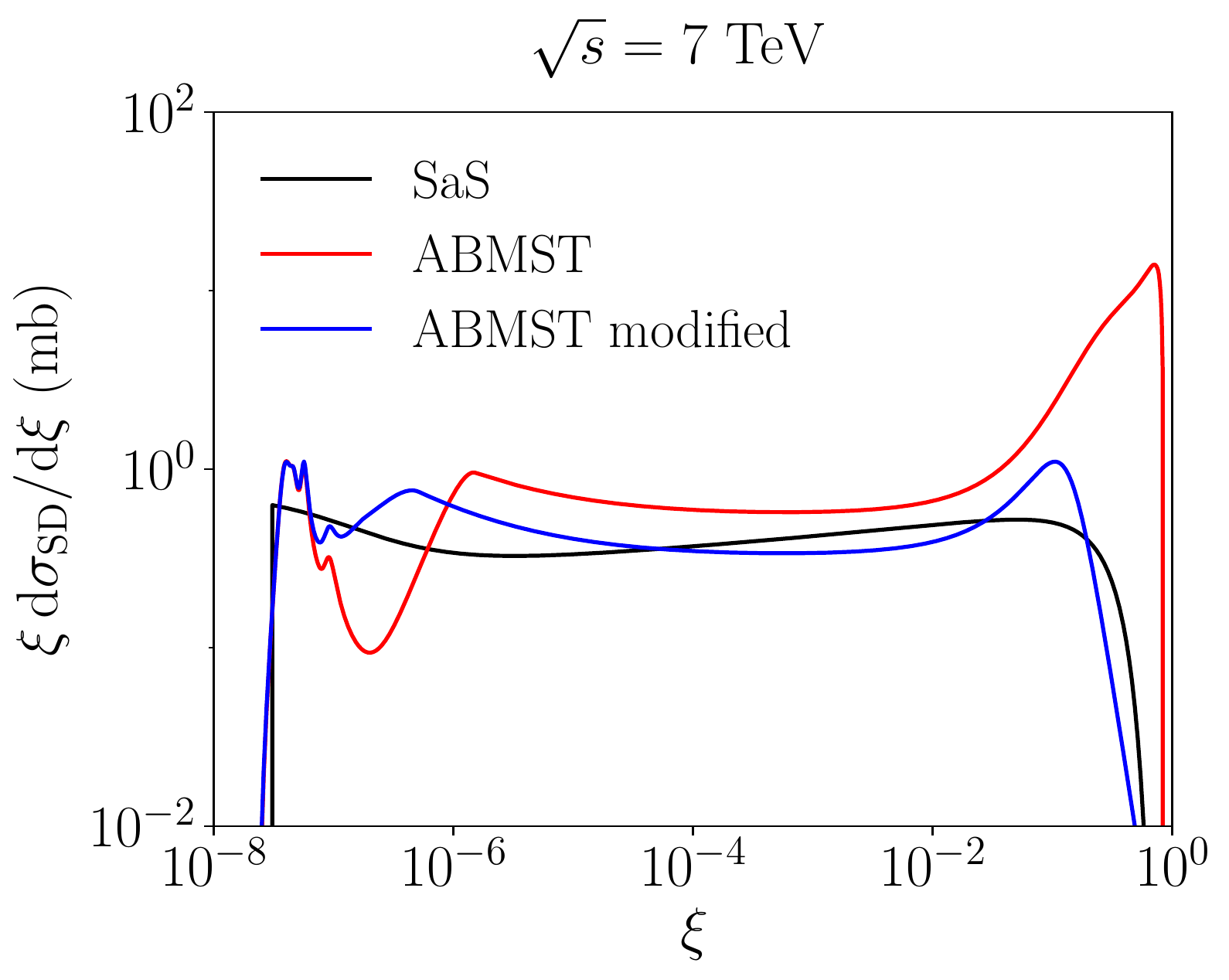}\\
(c)
\end{minipage}
\hfill
\begin{minipage}[c]{0.475\linewidth}
\centering
\includegraphics[width=\linewidth]{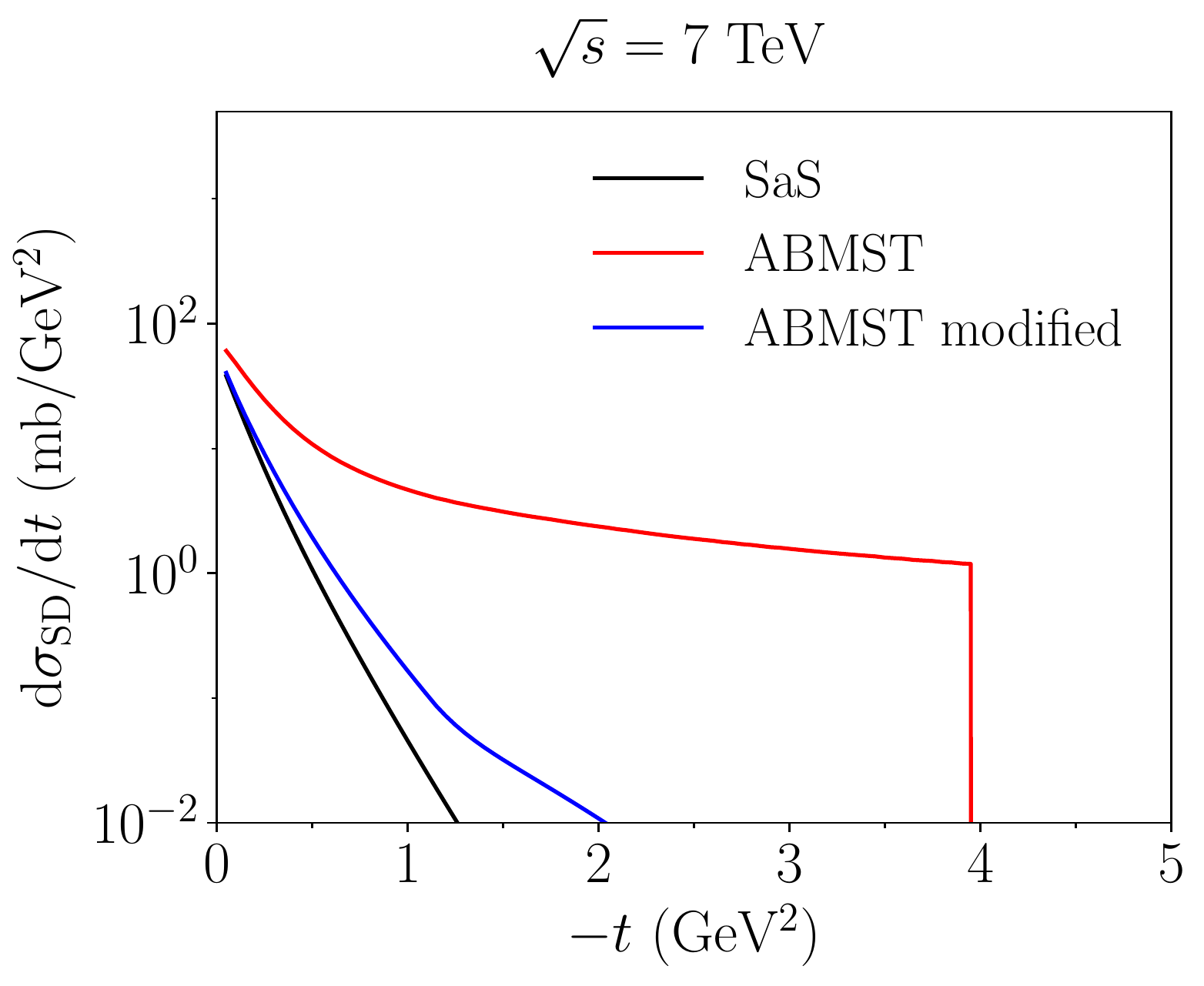}\\
(d)
\end{minipage}
\caption{\label{Fig:SDmodels} The different components of the
modified ABMST model for single diffraction as a function of (a) $\xi$ 
and (b) $t$ at 7 TeV. The same distributions are shown in (c) and (d),
where we compare the two models ABMST and ABMST modified to the SaS
model.}
\end{figure}

In fig.~\ref{Fig:SDmodels} we show the components of the modified ABMST
model as a function of $\xi$ (a) and $t$ (b). The improvements of the
modifications are clearly seen, as the dip between the low- and
high-mass description has decreased, the high-$\xi$ region has been
dampened and none of the components become constant at large $|t|$. In
figs.~\ref{Fig:SDmodels}c,d the two ABMST models are compared to the 
SaS model available in \textsc{Pythia}~8 as default. We note that the 
modified ABMST model shows better agreement with the SaS model
at intermediate $\xi$ values, where SaS is in rough agreement with data,
while retaining some features of the ABMST model, such as the detailed
resonance structure.

\begin{figure}[ht!]
\begin{minipage}[c]{0.475\linewidth}
\centering
\includegraphics[width=\linewidth]{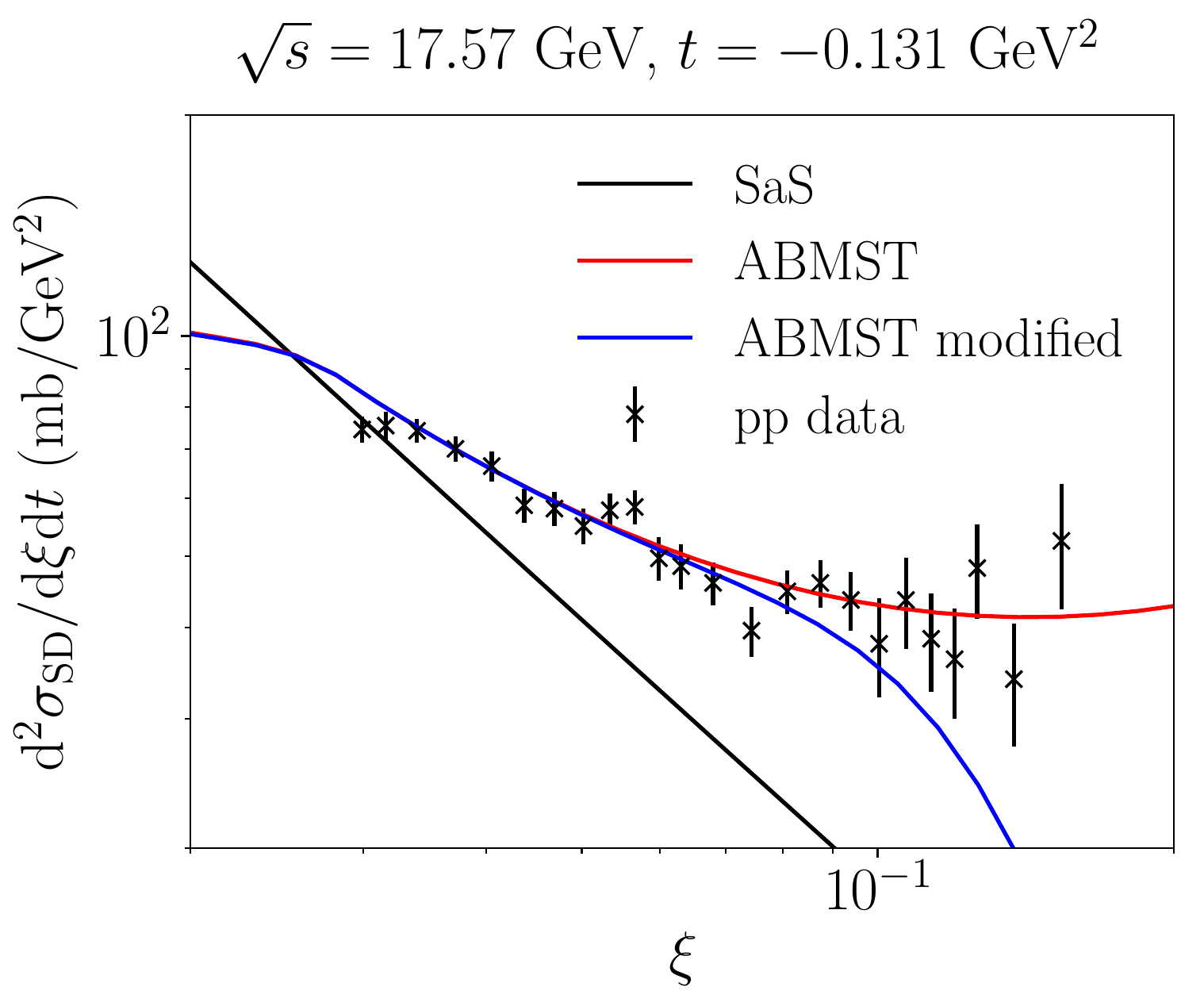}\\
(a)
\end{minipage}
\hfill
\begin{minipage}[c]{0.475\linewidth}
\centering
\includegraphics[width=\linewidth]{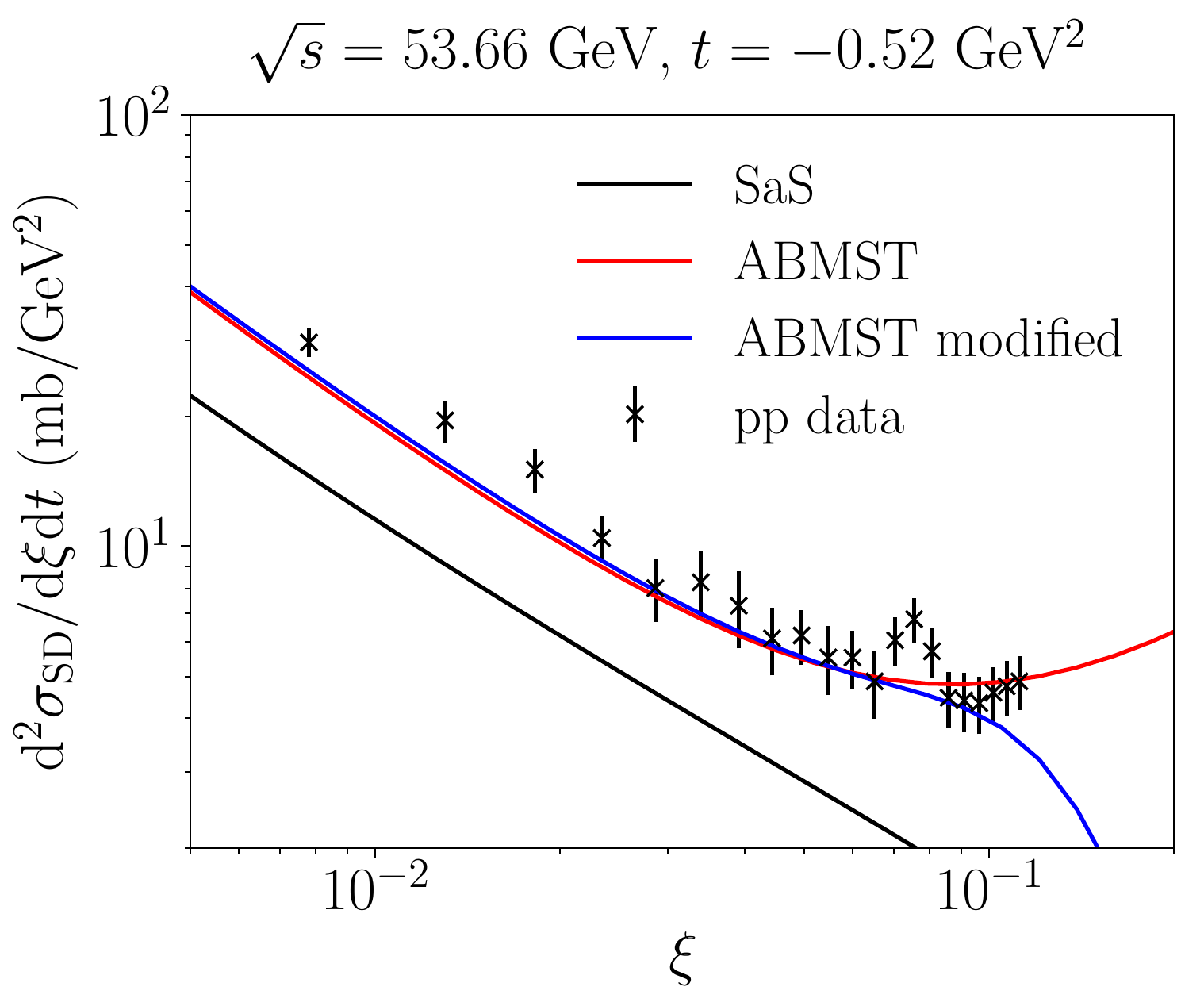}\\
(b)
\end{minipage}
\begin{minipage}[c]{0.475\linewidth}
\centering
\includegraphics[width=\linewidth]{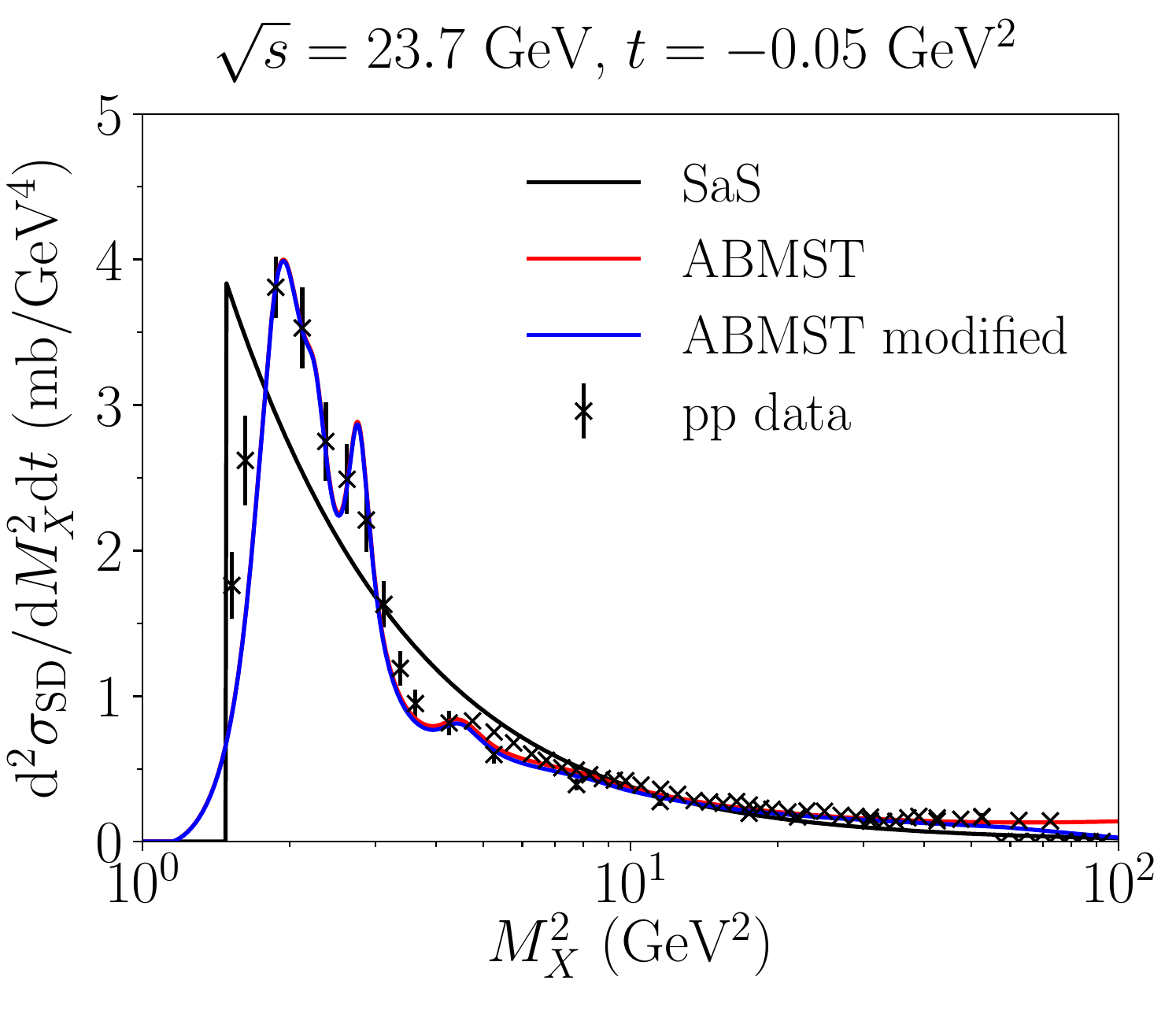}\\
(c)
\end{minipage}
\hfill
\begin{minipage}[c]{0.475\linewidth}
\centering
\includegraphics[width=\linewidth]{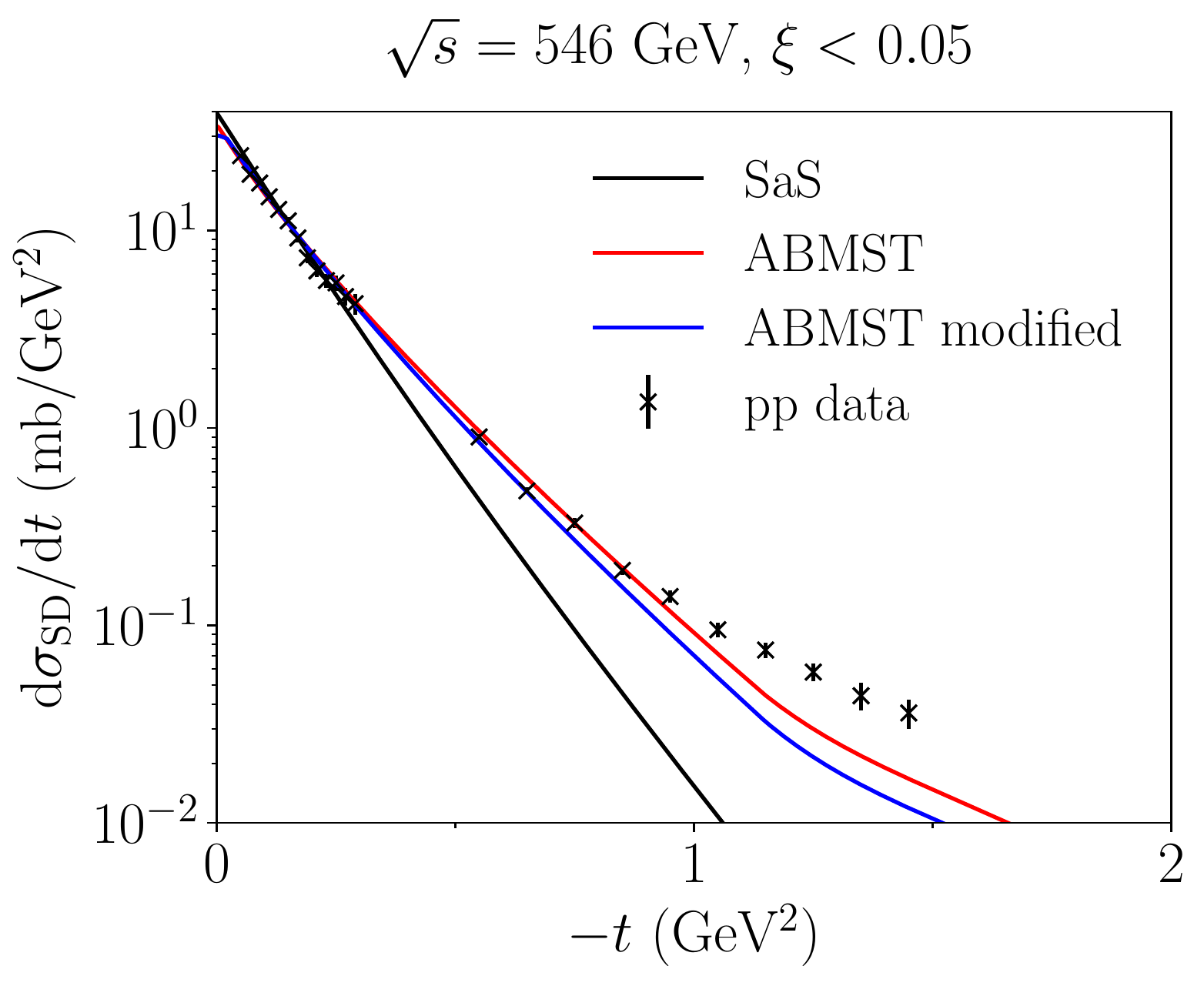}\\
(d)
\end{minipage}
\caption{\label{Fig:SD_data}
The single diffractive differential cross section parametrizations in 
$\p\p$ collisions at $\sqrt{s}$ 17.57 GeV with $t=-0.131$ GeV$^2$ (a) 
and 53.66 GeV with $t=-0.52$ GeV$^2$ (b). The mass-spectrum showing
the resonances at $\sqrt{s}=$ GeV and $t=-$ GeV$^2$ (c). The integrated
$t$ spectrum at $\sqrt{s}=$ GeV (d). Data from references in 
\cite{Appleby:2016ask}.
}
\end{figure}

\begin{figure}[ht!]
\begin{minipage}[c]{0.475\linewidth}
\centering
\includegraphics[width=\linewidth]{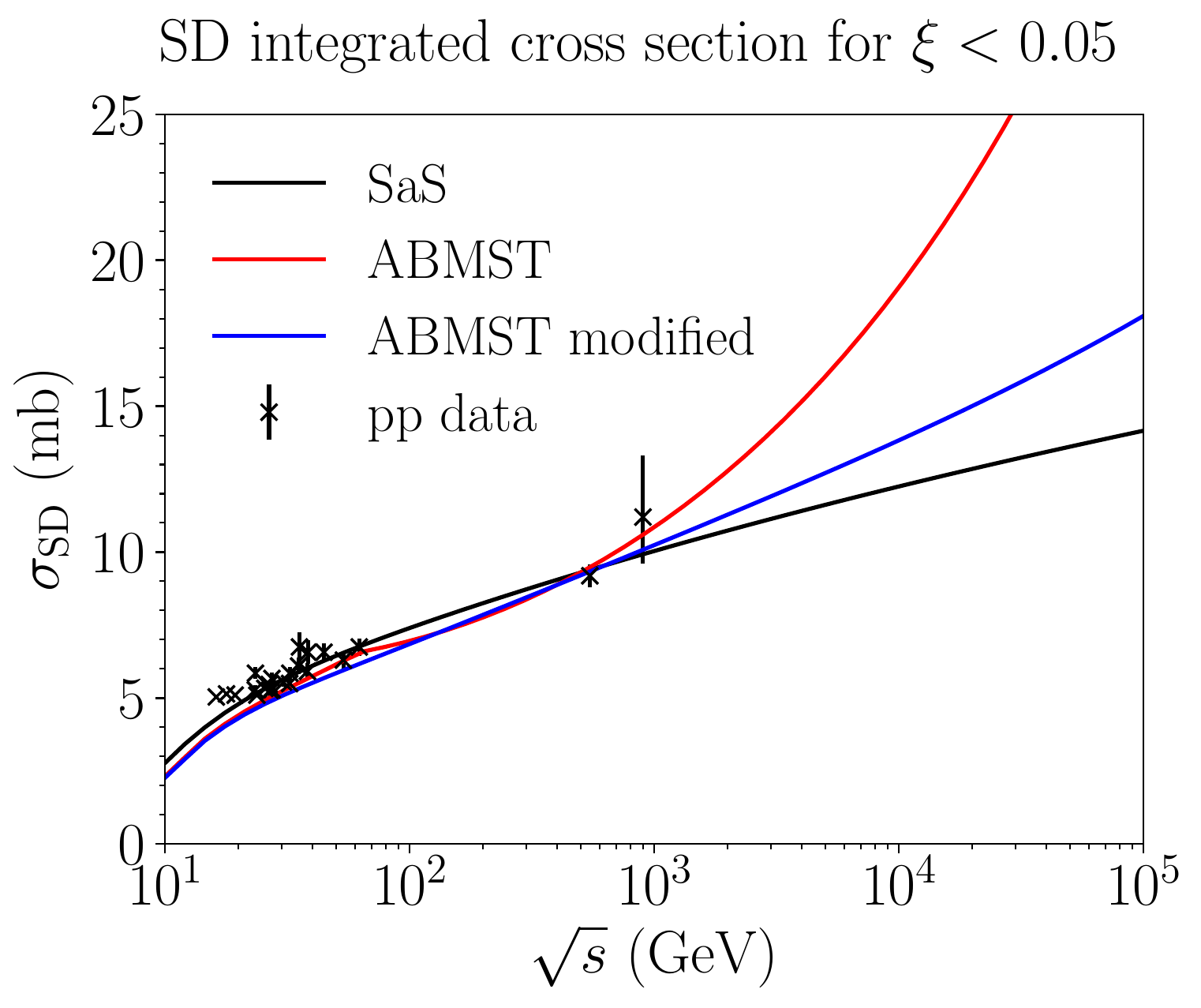}\\
(a)
\end{minipage}
\hfill
\begin{minipage}[c]{0.475\linewidth}
\centering
\includegraphics[width=\linewidth]{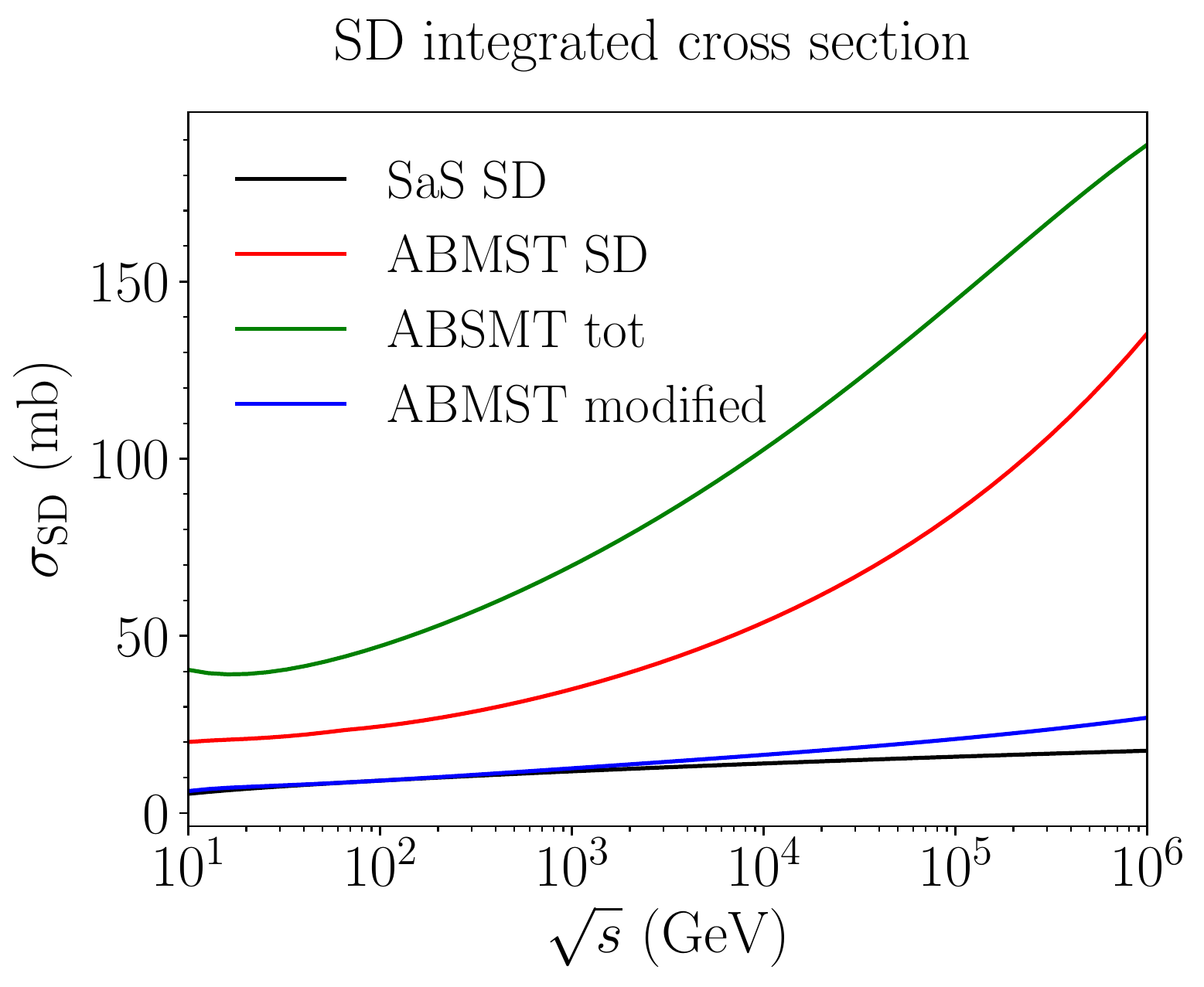}\\
(b)
\end{minipage}
\caption{\label{Fig:SDint}
The integrated single diffractive cross section at different energies
for $\xi<0.05$ (a) and in the full phase space (b). Data from 
references in \cite{Appleby:2016ask}.
}
\end{figure}

In fig.~\ref{Fig:SD_data} we show the comparison between the 
implemented models and the low-energy data used in \cite{Appleby:2016ask}. 
It is clear that the SaS model does not agree with data, while
both the original and the modified ABMST model describe data 
reasonably well. In figs.~\ref{Fig:SDint}a,b the integrated cross 
sections of all three models
are shown in the restricted (a) and full (b) phase space. The growth of
the ABMST model has been tamed by our modifications. Insofar as the SaS
model seems to be on the high side relative to data, and the modified 
ABMST is slightly higher, it may become necessary to finetune further
for LHC applications. To this end we have introduced an optional overall
scaling factor $k(s/m_p^2)^p$, with $k,p$ being tuneable parameters.

The bulk of the modifications applied to the ABMST framework are
intended to tame the high-energy behaviour of the model. One could 
have used an eikonal approach to the same end, e.g.\ in the spirit of 
\cite{Aurenche:1991pk}. This would require a different set of 
assumptions, however, such as the impact-parameter shape of the 
different diffractive topologies, and therefore not be any less 
arbitrary. For now we therefore stay with the current framework
and instead proceed to address other shortcomings of the ABMST model,
namely the lack of double and central diffraction.

\section{Double diffractive cross sections}\label{Sec:DDXS}

The ABMST model only provides a description of the single diffractive
differential cross section. We can extend this to double diffractive
systems, by extracting the vertices and propagators from the single
diffractive framework and using them in double diffractive diagrams. 
Fig. \ref{Fig:ALLXS}e shows a double diffractive diagram, where
$\mathbb{X}$ is one of the Reggeons used in the single diffractive 
framework. Thus several diagrams are obtained with Reggeons $i,j,k$ 
(where $i,j$ are connected to the proton and $k$ are in the loop).
Similar as for single diffraction, in order for the unknown phases 
in the propagators to vanish, the requirement of equal Reggeons is
enforced in the loop. The fact that there are two different mass 
regimes (low and high) for the two diffractive systems $X$ and 
$Y$ gives four different combinations.

If both systems have high mass, $M_{X,Y} > M_{\mrm{cut}}$, 
the diagram of fig.~\ref{Fig:ALLXS}e implies a cross section
\begin{align}
16\pi M_X^2 M_Y^2
\frac{\mrm{d}^3\sigma}{\mrm{d}t\mrm{d}M_X^2\mrm{d}M_Y^2} 
=&\sum_{ijk}g_{i\p}(0)g_i^{kk}(t)g_{j\p}(0)g_j^{kk}(t) \nonu\\
  & \left(\frac{M_X^2}{s_0}\right)^{\alpha_i(0)-1}
    \left(\frac{ss_0}{M_X^2M_Y^2}\right)^{2\alpha_k(t)-2}
    \left(\frac{M_Y^2}{s_0}\right)^{\alpha_j(0)-1}.
\end{align}
Changing variables to $\xi=M^2/s$ and collecting the terms one obtains
\begin{align}
16\pi\frac{\mrm{d}^3\sigma}{\mrm{d}t\mrm{d}\xi_X\mrm{d}\xi_Y} =&
\sum_{ijk}
  \left[g_{i\p}(0)g_i^{kk}(t)\xi_X^{\alpha_i(0)-2\alpha_k(t)}\left(\frac{s}{s_0}\right)^{\alpha_i(0)-1}\right]\nonu\\
  &\left[g_{j\p}(0)g_j^{kk}(t)\xi_Y^{\alpha_j(0)-2\alpha_k(t)}\left(\frac{s}{s_0}\right)^{\alpha_j(0)-1}\right]
    \left(\frac{s}{s_0}\right)^{2 - 2\alpha_k(t)}.
\end{align}

From the single diffractive framework one has that
\begin{align}
16\pi\frac{\mrm{d}^2\sigma_{\mrm{HM}}}{\mrm{d}t\mrm{d}\xi} =&
\sum_{ik}
  g_{k\p}^2(t)g_{i\p}(0)g_i^{kk}(t)\xi^{\alpha_i(0)-2\alpha_k(t)}\left(\frac{s}{s_0}\right)^{\alpha_i(0)-1},
\end{align}
where we can recognise a part of the single diffractive cross
section in the double diffractive cross section,
\begin{align}
16\pi\frac{\mrm{d}^3\sigma}{\mrm{d}t\mrm{d}\xi_X\mrm{d}\xi_Y} =&
\sum_{k}
  \left[\frac{16\pi}{g_{k\p}^2(t)}\frac{\mrm{d}\sigma_{\mrm{HM}}}{\mrm{d}t\mrm{d}\xi_X}\right]
  \left[\frac{16\pi}{g_{k\p}^2(t)}\frac{\mrm{d}\sigma_{\mrm{HM}}}{\mrm{d}t\mrm{d}\xi_Y}\right]
  \left(\frac{s}{s_0}\right)^{2 - 2\alpha_k(t)}\nonu\\
=&\frac{\mrm{d}\sigma_{\mrm{HM}}}{\mrm{d}t\mrm{d}\xi_X}
  \frac{\mrm{d}\sigma_{\mrm{HM}}}{\mrm{d}t\mrm{d}\xi_Y}
  \sum_{k} \frac{(16\pi)^2}{g_{k\p}^4(t)}\left(\frac{s}{s_0}\right)^{2 - 2\alpha_k(t)}.
\end{align}
A similar diagrammatic method can be used for the low-mass region,
so all four $(M_X, M_Y)$ regions can generically be described as
\begin{align}\label{Eq:DDXSsum}
\frac{\mrm{d}^3\sigma}{\mrm{d}t\mrm{d}\xi_X\mrm{d}\xi_Y} 
=&\frac{\mrm{d}\sigma_{\mrm{SD}}}{\mrm{d}t\mrm{d}\xi_X}
  \frac{\mrm{d}\sigma_{\mrm{SD}}}{\mrm{d}t\mrm{d}\xi_Y}
  \sum_{k=\Pom,\Reg}
  \frac{16\pi}{g_{k\p}^4(t)}\left(\frac{s}{s_0}\right)^{2 -
  2\alpha_k(t)}\nonu\\
\rightarrow&\frac{\mrm{d}\sigma_{\mrm{SD}}}{\mrm{d}t\mrm{d}\xi_X}
  \frac{\mrm{d}\sigma_{\mrm{SD}}}{\mrm{d}t\mrm{d}\xi_Y}
  \frac{16\pi}{g_{\Pom\p}^4(t)}\left(\frac{s}{s_0}\right)^{2 - 2\alpha_{\Pom}(t)}.
\end{align}
In the last step we have taken the high-energy limit, where the
Pomeron term dominates. The last term can then be recognised as the 
inverse of the elastic cross section in the same limit, and hence 
\cite{Barone:2002cv}
\begin{align}\label{Eq:DDXS}
\frac{\mrm{d}^3\sigma_{\mrm{DD}}}{\mrm{d}t\mrm{d}\xi_X\mrm{d}\xi_Y} \approx&
  \frac{\mrm{d}^2\sigma_{\mrm{SD}}}{\mrm{d}t\mrm{d}\xi_X}
  \frac{\mrm{d}^2\sigma_{\mrm{SD}}}{\mrm{d}t\mrm{d}\xi_Y} \, / \, 
  \frac{\mrm{d}\sigma_{\mrm{el}}^{\Pom}}{\mrm{d}t}.
\end{align}
In principle this formulation holds only at high energies, and only when
using the Pomeron as exchanged particle in all parts of the diagram in
fig.~\ref{Fig:ALLXS}e. Nevertheless it offers the best way to introduce 
double diffraction as a natural extension of the ABMST single diffractive
machinery, and is the one we will choose.
 
\begin{figure}[ht!]
\begin{minipage}[c]{0.475\linewidth}
\centering
\includegraphics[width=\linewidth]{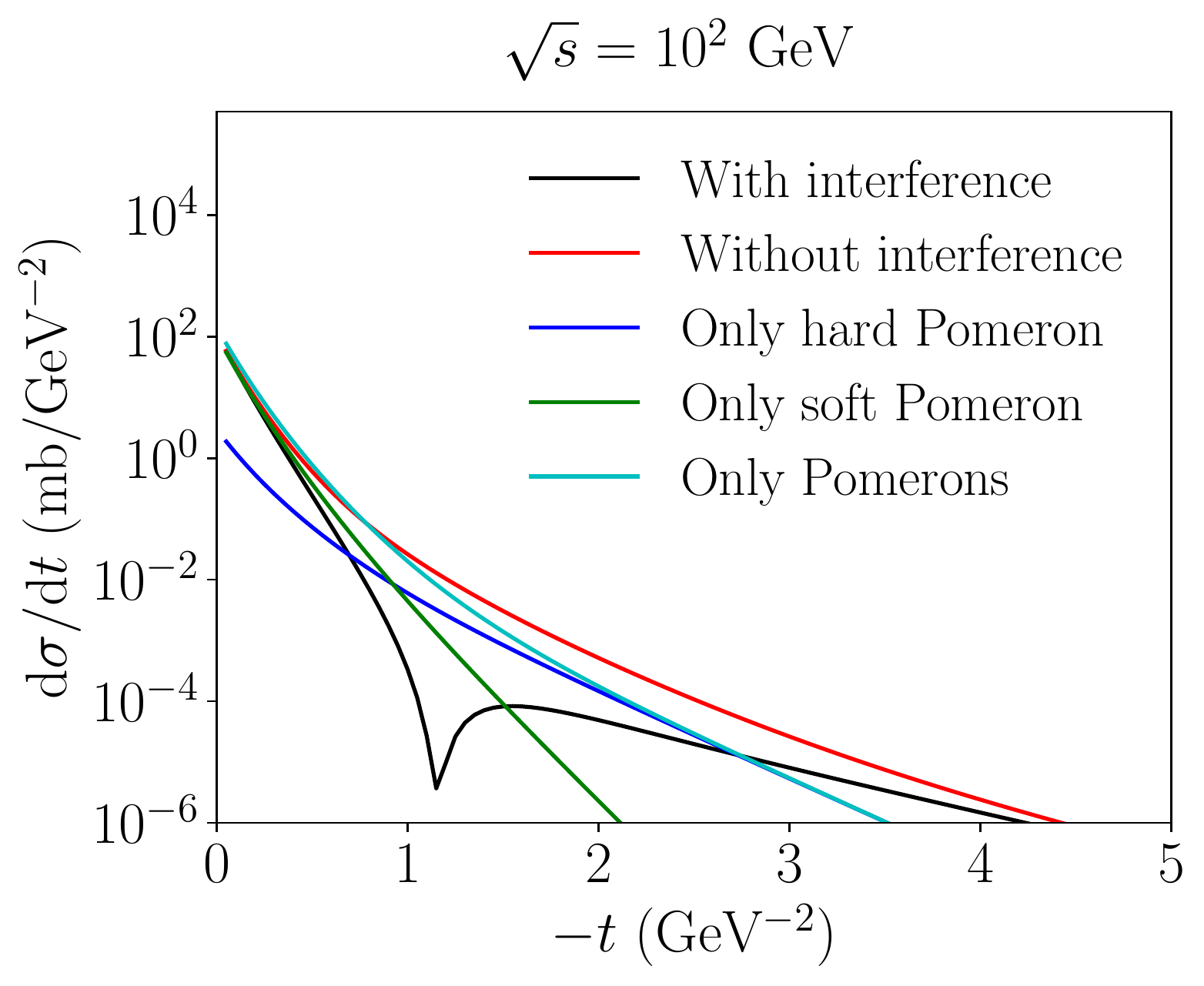}\\
(a)
\end{minipage}
\hfill
\begin{minipage}[c]{0.475\linewidth}
\centering
\includegraphics[width=\linewidth]{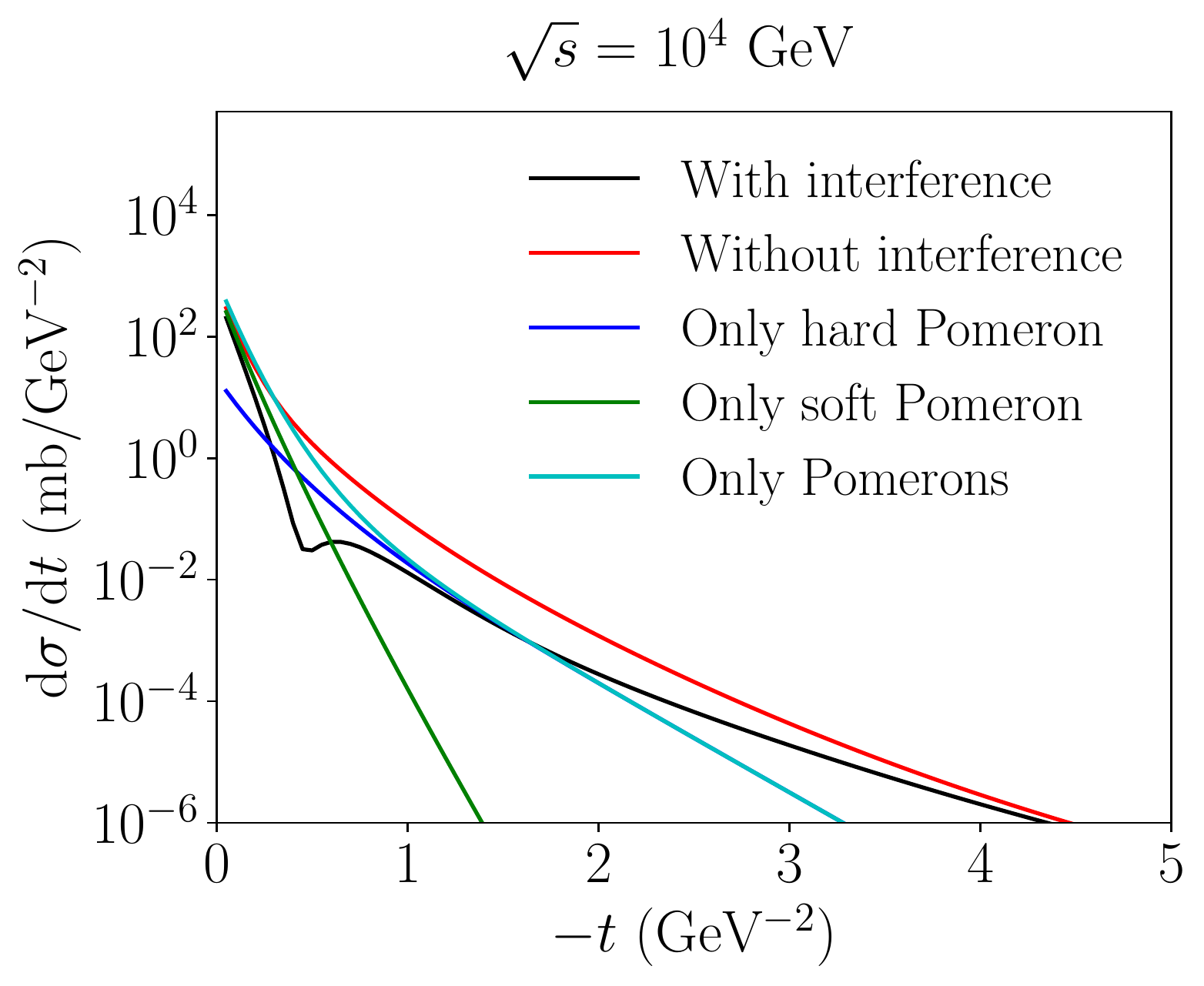}\\
(b)
\end{minipage}
\begin{minipage}[c]{0.475\linewidth}
\centering
\includegraphics[width=\linewidth]{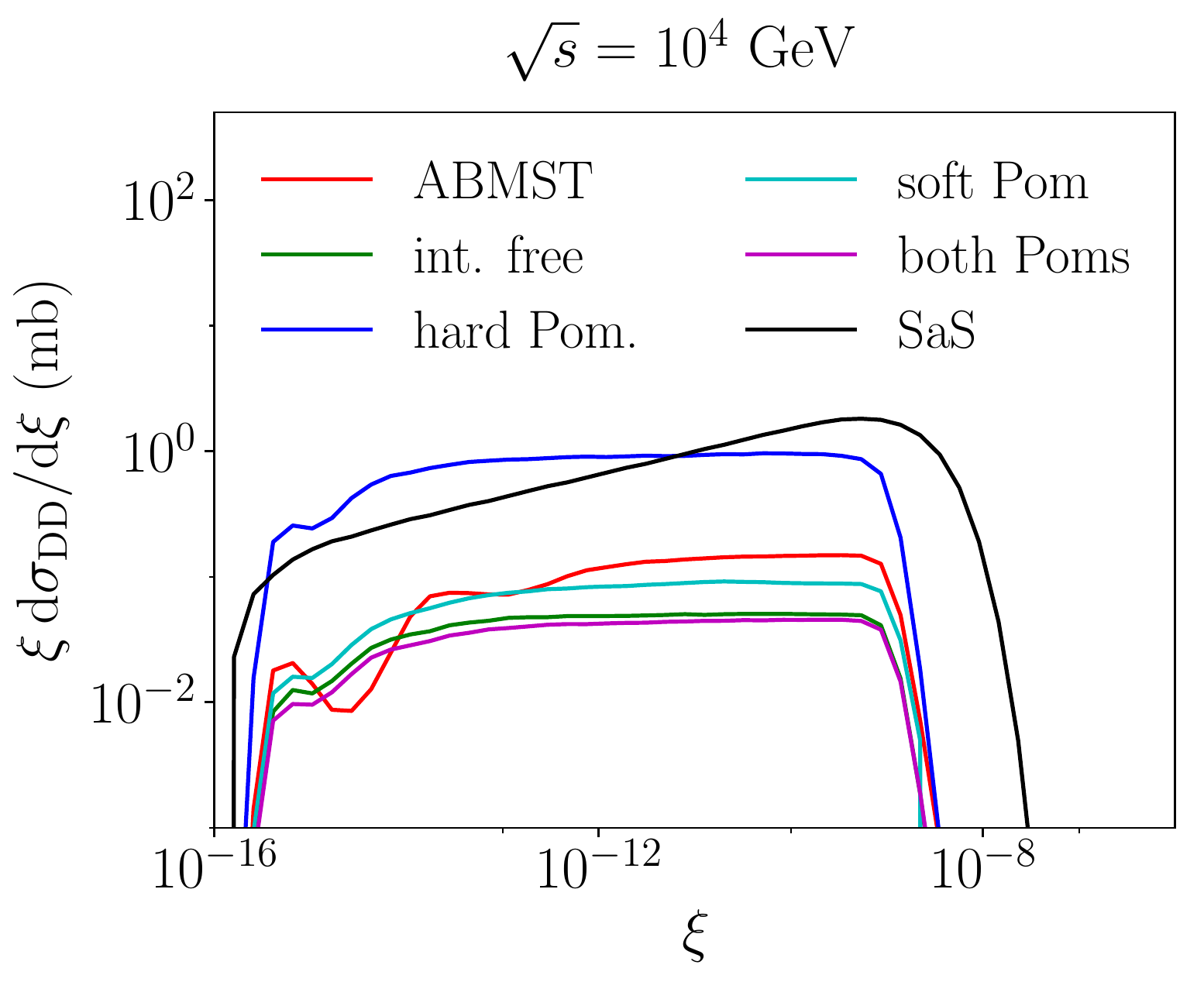}\\
(c)
\end{minipage}
\hfill
\begin{minipage}[c]{0.475\linewidth}
\centering
\includegraphics[width=\linewidth]{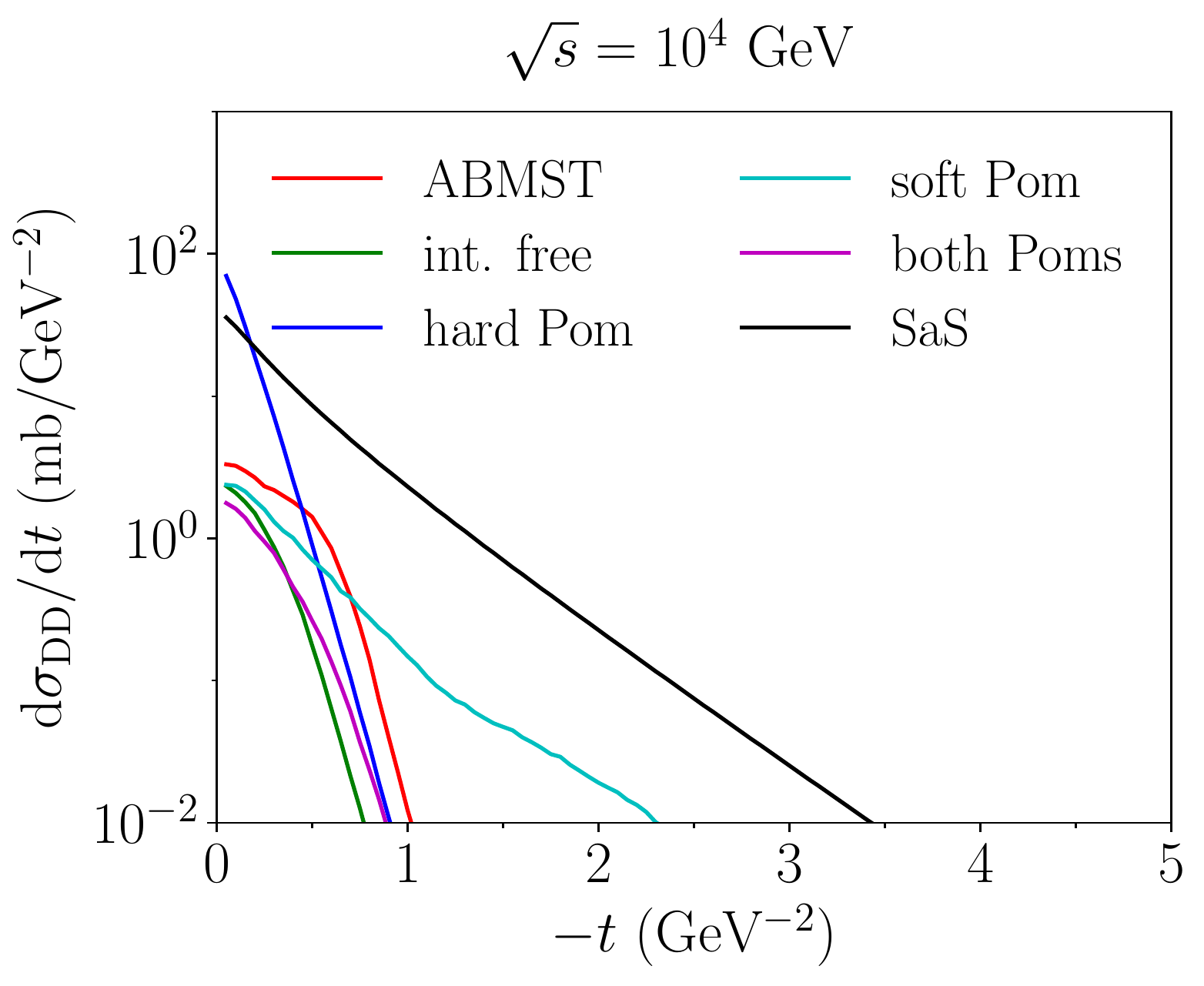}\\
(d)
\end{minipage}
\hfill
\caption{\label{Fig:ElPoms}
The effect of using only a subset of the available Pomerons in the
elastic parametrisation, as used in the expression for the double 
diffractive cross section, eq.~(\ref{Eq:DDXS}).
In (a) and (b) the elastic differential cross section is shown as a 
function of $t$ at two energies. In (c) and (d) the effect of these on 
the double diffractive distributions are shown as a
function of $\xi=\xi_1\xi_2$ and $t$, respectively. Note that the ``Pure
ABMST'' has a minimal slope of $B_{DD}=2$ such as to avoid the dip
structure of the elastic description.}
\end{figure}

One of the drawbacks of this approach is that accidental dips in the
elastic cross section denominator can come to blow up the double
diffractive cross section beyond what reasonably should be expected.
Therefore a slightly modified elastic cross section is called for in
this context. In fig.~\ref{Fig:ElPoms}a,b the different Pomeron 
contributions to the elastic differential cross section are shown at two
energies, along with the full description and an interference-free
description of the form 
\begin{align}\label{Eq:sigElapprox}
\frac{\d\sigma_{\mrm{el}}}{\d t}=&
  \frac{|\sum_iA_i(s,t)|^2}{16\pi}
  \simeq\frac{\sum_i|A_i(s,t)|^2}{16\pi}
\end{align}
where $i$ runs over all four terms. Notice that the hard and soft Pomeron
contributions dominate in two different regions. Hence a reasonable
approximation would be to use the soft Pomeron term in the low-$|t|$ range
and the hard Pomeron term in the high-$|t|$ range. In practise we use the
combination of the hard and soft Pomerons, so as to avoid splitting 
eq.~(\ref{Eq:DDXS}) into two different $t$ ranges. 

Figs. \ref{Fig:ElPoms}c,d show the effect of the various
elastic parametrizations on the double diffractive distributions. Note
the normalisation difference between the hard-Pomeron-only description
and the others, a difference that arises since the hard Pomeron term is
not the dominant one in the low-$|t|$ region, where most of the cross
section is. On the other hand, the soft-Pomeron-only $t$ spectrum
is much wider than the other distributions shown, since the soft Pomeron 
contribution to elastic scattering falls off much steeper with $t$. 
The ``Pure ABMST'', interference-free ABMST and ``ABMST both Poms'' appear 
to have the same shapes in figs.~\ref{Fig:ElPoms}c,d. They differ 
somewhat in normalisation, as is expected given that the two latter
correspond to somewhat larger elastic cross sections.

To correct for the possible suppressions arising from the chosen 
approximation of the elastic cross section, and from the underestimation
implied by the step taken in eq.~(\ref{Eq:DDXSsum}), we introduce a scaling
factor similar to the one introduced in the single diffractive
framework. A minimal double diffractive slope can also be enforced,
such as to avoid any unphysical situations. As a final modification, 
an option to reduce topologies without a rapidity gap is applied in
the region where both of the systems are of very large masses.
Again, this is to be able to distinguish the double-diffractive system
from the non-diffractive ones. 

As two different parametrizations are available in the ABMST framework
for single diffraction, several choices for the double diffractive 
framework exists. Presented here are results with three choices:
\begin{itemize}
\item Pure ABMST: the original ABMST single diffractive model together 
with the elastic cross sections using only Pomerons, with the minimal 
double diffractive slope and with reduced vanishing-gap topologies.
\item Model 1: The modified ABMST model for the single diffractive cross
section, with the only-Pomerons elastic cross section. A minimal slope
is used and the vanishing-gap topologies are also reduced here.
\item Model 2: Model 1 scaled with the tuneable factor $k(s/m_p^2)^p$,
where by default $k=2$ and $p=0.1$.
 \end{itemize}

\begin{figure}[ht!]
\begin{minipage}[c]{0.475\linewidth}
\centering
\includegraphics[width=\linewidth]{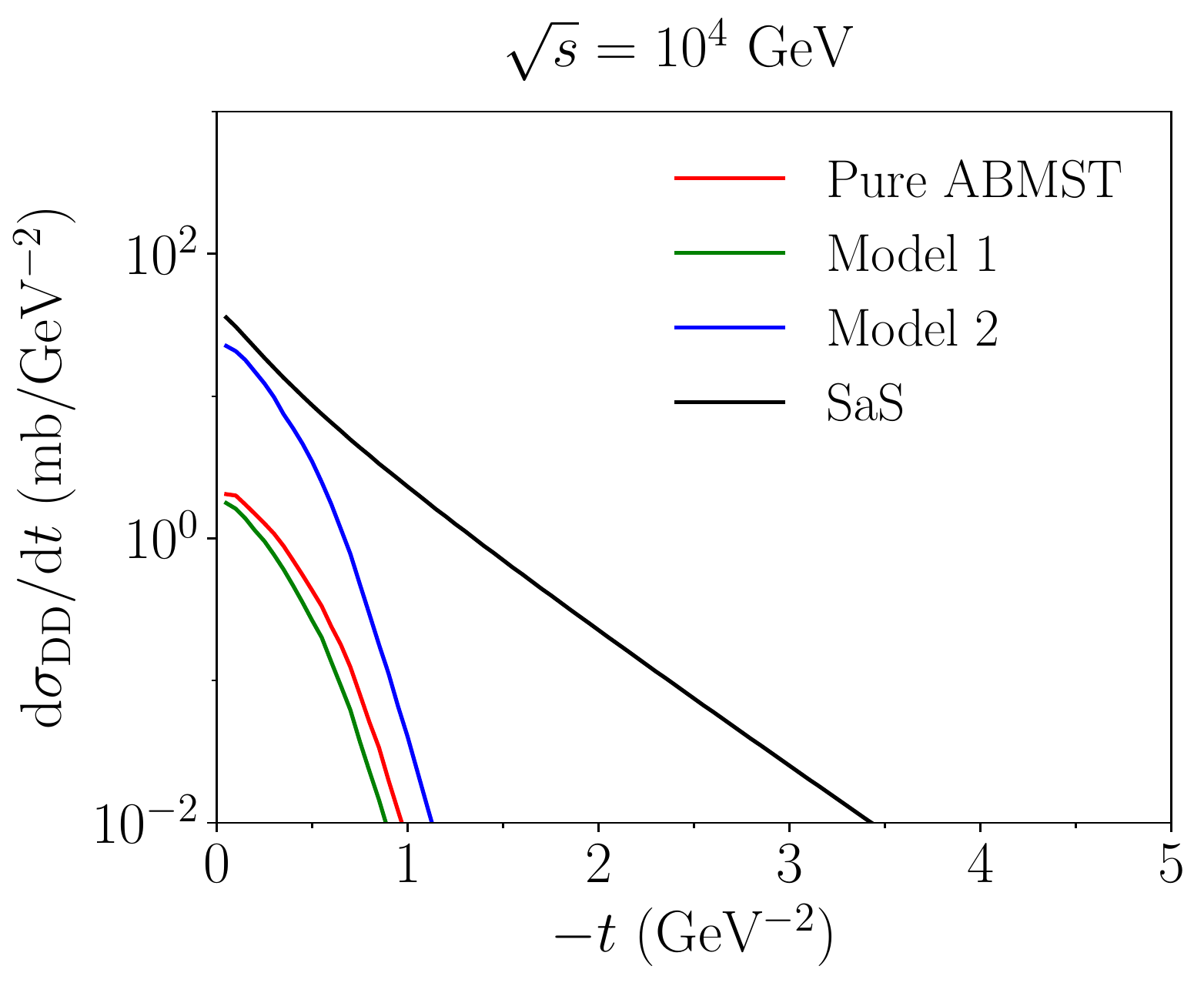}\\
(a)
\end{minipage}
\hfill
\begin{minipage}[c]{0.475\linewidth}
\centering
\includegraphics[width=\linewidth]{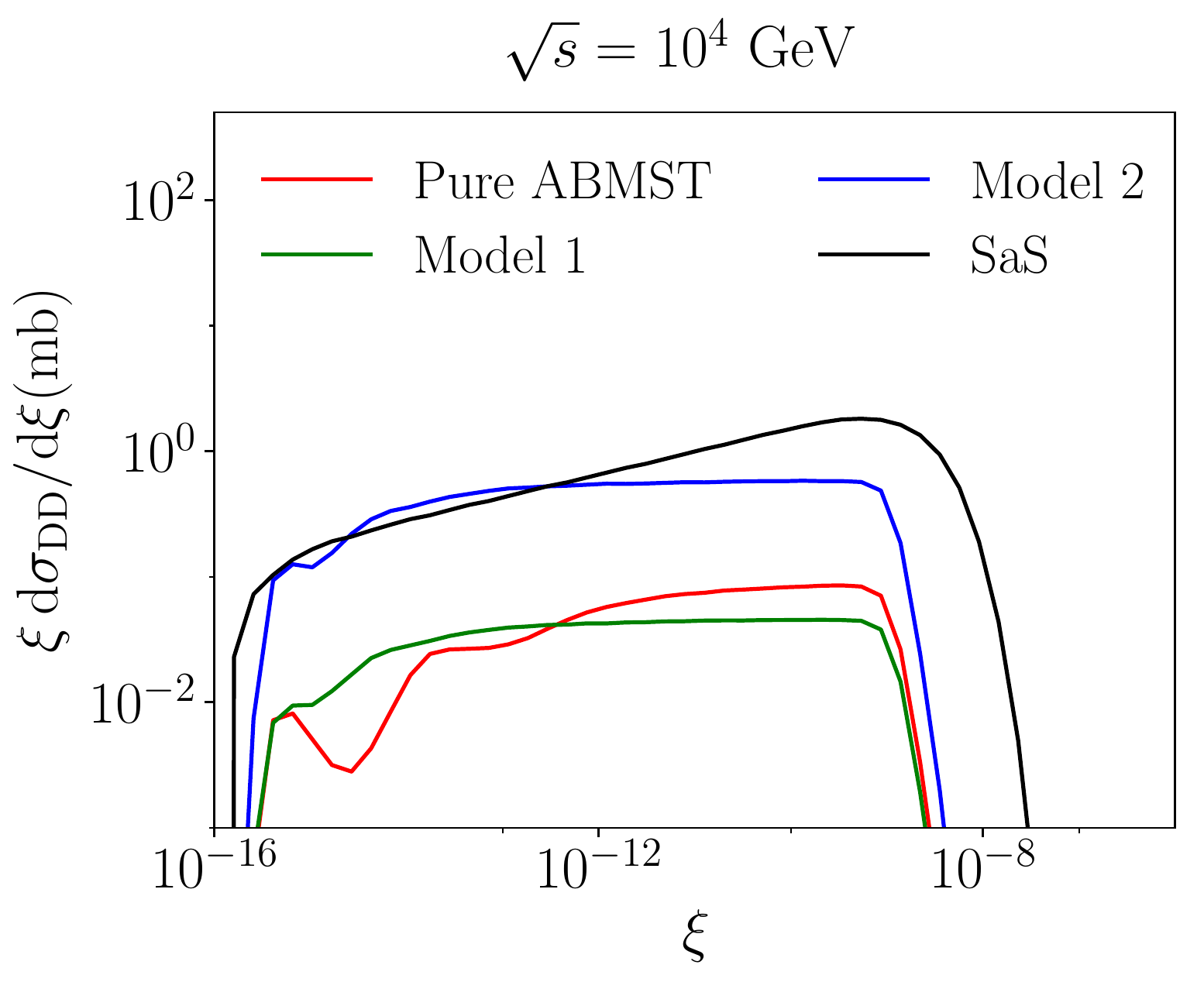}\\
(b)
\end{minipage}
\caption{\label{Fig:DD}
Some of the DD models available in \textsc{Pythia}~8. In (a) and (b) we
show the differential cross section as a function of $t$ and
$\xi=\xi_1\xi_2$, respectively. }
\end{figure}

\begin{figure}[ht!]
\centering
\includegraphics[width=0.475\linewidth]{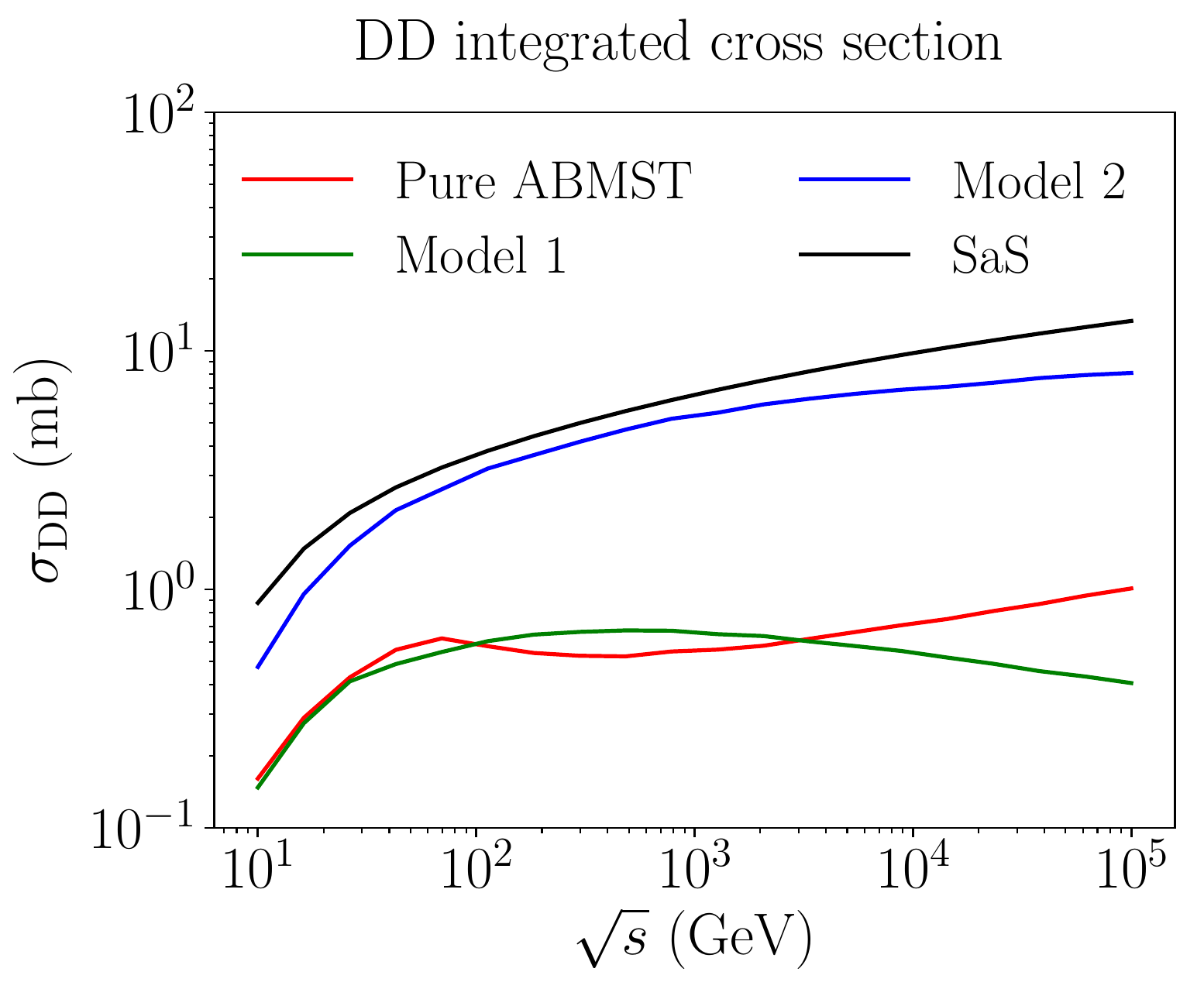}\\
\caption{\label{Fig:DDintegrated}
The integrated double diffractive cross section as a function of energy 
of some of the models available in \textsc{Pythia}~8. }
\end{figure}

Fig.~\ref{Fig:DD}a shows the $t$ spectrum of the different models
compared to the SaS model. It is evident that three models vanish 
faster than the SaS model. This is a result of the modest falloff of
the elastic $t$-spectrum in ABMST, as this affects the double diffractive 
slope less than a sharply falling elastic $t$-spectrum in SaS,
through the relation $B_{XY}=B_{AY} + B_{XB} - B_{\mrm{el}}$. 
Fig.~\ref{Fig:DD}b shows the differential 
cross section as a function of $\xi=\xi_1\xi_2$. Here, the ABMST 
models show an approximate $1/\xi$-behaviour, while the SaS model 
indicates a $1/\xi^{(1+p)}$ behaviour with $p>0$, favouring high-mass 
diffractive systems. The results of these effects are visible in 
the integrated cross section, fig.~\ref{Fig:DDintegrated}, where 
both ``Pure ABMST'' and ``Model 1'' are significantly suppressed compared
to the SaS model. The scaled version, ``Model 2'', gives more reasonable 
estimates of the cross sections, around 10 mb at LHC energies, 
but because of the choice of the power, $p=0.1$, in the scaling, 
it does not rise as steeply as the SaS prediction. Similarly to the
single-diffractive case, the SaS model predicts slightly larger cross
sections than measured, so one might expect that the scaling chosen in
Model 2 could be more in agreement with measurements.
 
\section{Central diffractive cross sections}\label{Sec:CDXS}

The central diffractive framework has long been neglected in
general-purpose event generators. Dedicated event-generators exist for
exclusive central diffractive processes, such as SuperChic 
\cite{Harland-Lang:2015cta} and ExHume \cite{Monk:2005ji}, but these
only work with a limited set of final states. \textsc{Pythia}~8
provides a description for inclusive high-mass central diffraction, 
but does not provide any such description for the exclusive processes.
As stated earlier, we stress that the framework has not been
tuned and thus is not to be trusted too far.

In this work we wish to extend the present description of
central diffraction to include the high-mass description of the ABMST 
model. We have not made any attempt to include any low-mass resonances 
of central diffraction, as some of these are still not well established.
The low-mass resonances used in ABMST are baryonic resonances, hence
they cannot be extended to the central diffractive framework, as one
expects scalar mesons, possibly scalar glueballs, to be produced in the 
collision of two Reggeons. Future work would be to extend the model to
such low-mass resonances, eg. by including a low-mass resonance 
description similar to what has been developed in \cite{EBolsThesis}. 
There the central exclusive production of a pion pair is considered and
data is used to fit a model of the scalar resonances using complex
Breit--Wigner shapes. Lacking a model for all such exclusive states,
and since some of the resonances and their decays still are not 
experimentally under control, we have decided not to include any of
the low-mass states in this framework. 

The new central diffractive cross section presented here is again 
mainly based on the ABMST single-diffractive model. By examining the 
rapidity of the different components 
in the central diffractive system, one obtains the following
relations,
\begin{align}
\Delta y_{\mrm{tot}} = \ln\frac{s}{s_0},\quad
\Delta y_X = \ln\frac{\xi_1\xi_2\,s}{s_0},\quad
\Delta y_1 = \ln\frac{1}{\xi_1},\quad
\Delta y_2 = \ln\frac{1}{\xi_2}, 
\end{align}
where $M_X^2=\xi_1\xi_2s$, $\Delta y_X$ is the rapidity span of the
diffractive system $X$, and $\Delta y_{1,2}$ are the sizes of the two
rapidity gaps. Thus, after some algebra, we obtain a central 
diffractive cross section of the form 
\begin{align}
256\pi^3\frac{\d^4\sigma_{\mrm{CD}}}{\d \xi_1\d \xi_2\d t_1\d t_2} 
  =&\sum_{ijk}
  \left[g_{i\p}^2(t_1)g^{ii}_k(t_1)
  (\xi_1)^{\alpha_k(0)- 2\alpha_i(t_1)}\right]\nonu\\
  &\left[g_{j\p}^2(t_2)g^{jj}_k(t_2)
  (\xi_2)^{\alpha_k(0)- 2\alpha_j(t_2)}\right]
  \left(\frac{s}{s_0}\right)^{\alpha_k(0) - 1}
\end{align}
Again we recognise the high-mass single diffractive cross sections and
write \cite{Donnachie:2002en}
\begin{align}\label{Eq:CDXS}
\pi\frac{\d^4\sigma_{\mrm{CD}}}{\d \xi_1\d \xi_2\d t_1\d t_2} =& 
  \frac{\d^2\sigma_{\mrm{HM}}}{\d t_1\d\xi_1}
  \frac{\d^2\sigma_{\mrm{HM}}}{\d t_2\d\xi_2}
  \sum_{k}\frac{1}{g_{k\p}^2(0)}
  \left(\frac{s}{s_0}\right)^{1 - \alpha_k(0)}\nonu\\
\rightarrow&
  \frac{\d^2\sigma_{\mrm{HM}}}{\d t_1\d\xi_1}
  \frac{\d^2\sigma_{\mrm{HM}}}{\d t_2\d\xi_2}
  \frac{1}{g_{\Pom\p}^2(0)}\left(\frac{s}{s_0}\right)^{1 -
  \alpha_{\Pom}(0)}\nonu,\\
=& \frac{\d^2\sigma_{\mrm{HM}}}{\d t_1\d\xi_1}
  \frac{\d^2\sigma_{\mrm{HM}}}{\d t_2\d\xi_2} \, / \, \sigma_{\mrm{tot}},
\end{align}
where the high-energy limit is taken in the second step and
recognised as the total cross section. 
Similar arguments on the validity of eq.~(\ref{Eq:CDXS}) applies as for the
validity of eq.~(\ref{Eq:DDXS}), i.e.\ eq.~(\ref{Eq:CDXS}) is only valid 
in the high-energy limit, where the Pomeron term dominates. In practise,
however, the expression is used over the entire energy range, using the
sum of both the soft and the hard Pomeron term from the total cross
section. A scaling factor similar to the scaling for single and
double diffraction can be applied, to compensate for the approximations, 
and the same non-vanishing gap suppression can be applied as in the 
single-diffractive framework. Finally, a minimal central diffractive 
slope can also be applied.

\begin{figure}[ht!]
\begin{minipage}[c]{0.475\linewidth}
\centering
\includegraphics[width=\linewidth]{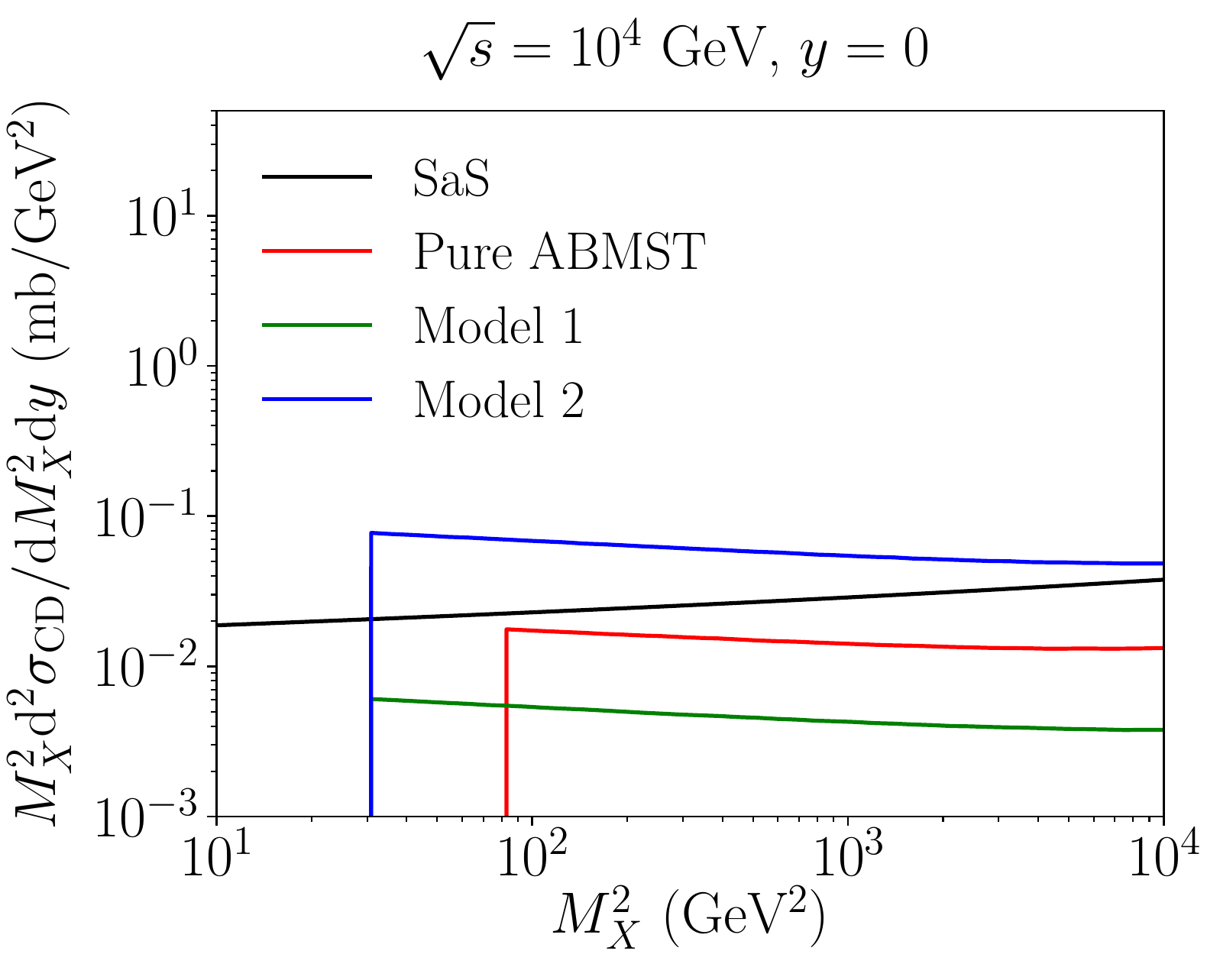}\\
(a)
\end{minipage}
\hfill
\begin{minipage}[c]{0.475\linewidth}
\centering
\includegraphics[width=\linewidth]{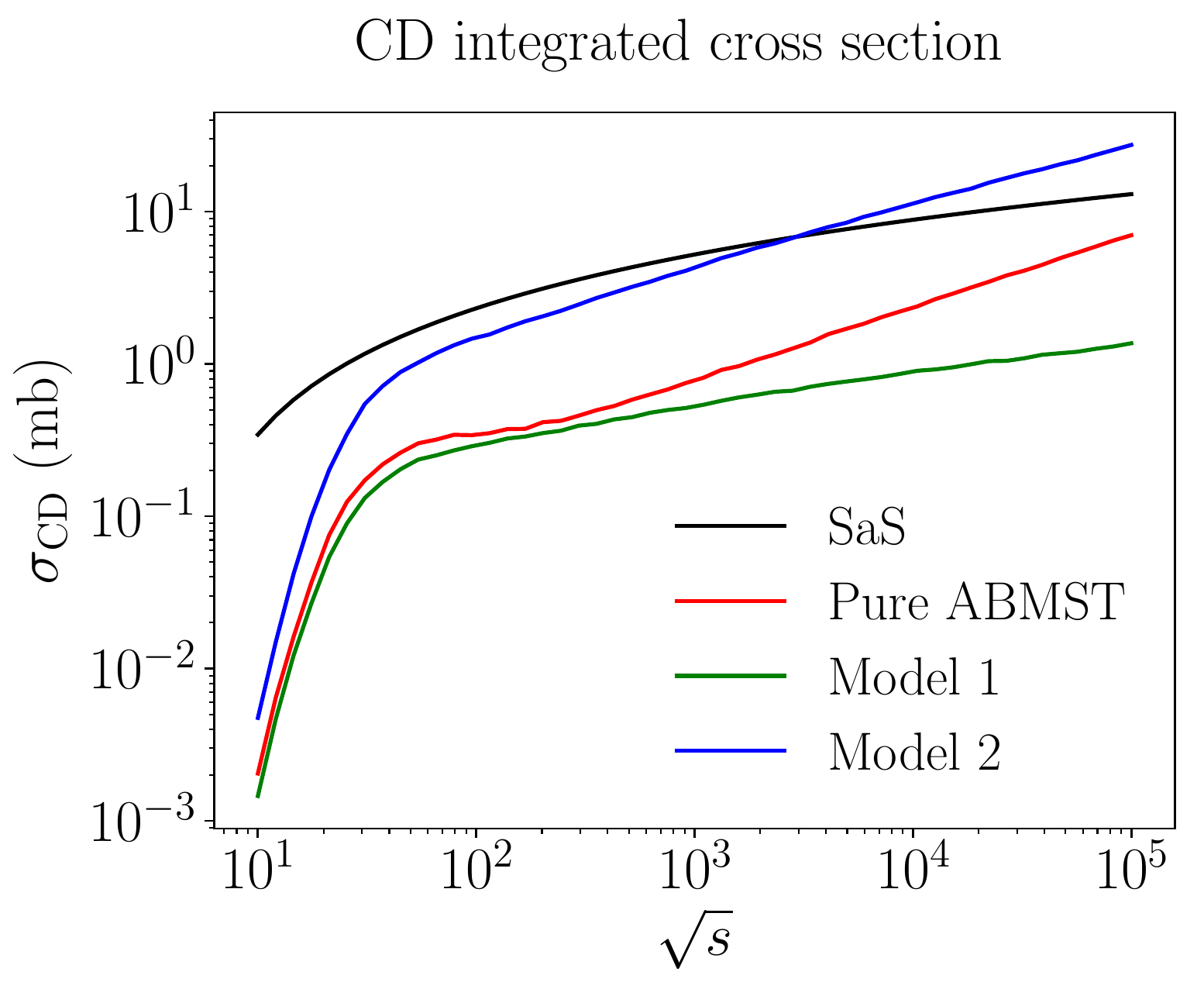}\\
(b)
\end{minipage}
\caption{\label{Fig:CDdist}
Some of the CD models available in \textsc{Pythia}~8. In (a) we show
the mass of the diffractive system produced at central rapidity and in
(b) the integrated cross section as a function of energy.}
\end{figure}

Similar to the double diffractive framework, the central diffractive
framework will depend on the choice of single diffractive framework,
thus several options exist. Fig.~\ref{Fig:CDdist} shows three choices
of models with the same name conventions as used in the double
diffractive framework. Note, however, that the $t$ spectrum is not
shown, as this is exactly that of the single diffractive model. The
mass of the diffractive system is shown in fig.~\ref{Fig:CDdist}a,
where the sharp cut at $M_X=M_{\mrm{cut}}$ is present for all
ABMST variants. The SaS model has a similar sharp cutoff, but at
$M_X=1$ GeV. Lacking both model and data in the low-mass region, the
cut allows for a clear distinction between what is included and not,
albeit being unphysical.  

Fig.~\ref{Fig:CDdist}b shows the integrated cross section as
a function of energy. Here all ABMST models lie below the SaS
prediction, although ``Model 2'' exceeds it at around LHC energies. The
lack of a low-mass model is evident at low energies ($\sqrt{s}<30$
GeV), where all three models decrease rapidly. In this energy-range 
the low-mass states make up a large part of the cross section, 
hence should not be neglected.

\section{Results}

In this section the models are confronted with more recent LHC 
data. Several experiments have performed measurements on integrated 
cross sections and diffractive fractions, but not many provide 
results on differential distributions. We focus on the analyses 
available in Rivet \cite{Buckley:2010ar}, where only two analyses 
provide differential results. First we provide a discussion of the 
available data and the tuning prospects, and end with results
obtained with the SaS model, the CSCR model and the ABMST models. 

\subsection{The 7 TeV LHC data and tuning prospects}

In 2012 and 2015 ATLAS \cite{Aad:2012pw} and CMS 
\cite{Khachatryan:2015gka} presented results on 7 TeV 
events with rapidity gaps. Both experiments measure all particles with 
transverse momenta larger than 200 MeV in pseudo-rapidity ranges of 
$|\eta| < 4.9 \, (4.7)$ for ATLAS (CMS), and define the measured gap
$\Delta\eta_F$ as the largest distance between either detector edge and 
the particle nearest to it. The two experiments, however, obtain different 
results for the shape of the distribution. 

\begin{figure}[ht!]
\begin{minipage}[c]{0.475\linewidth}
\centering
\includegraphics[width=\linewidth]{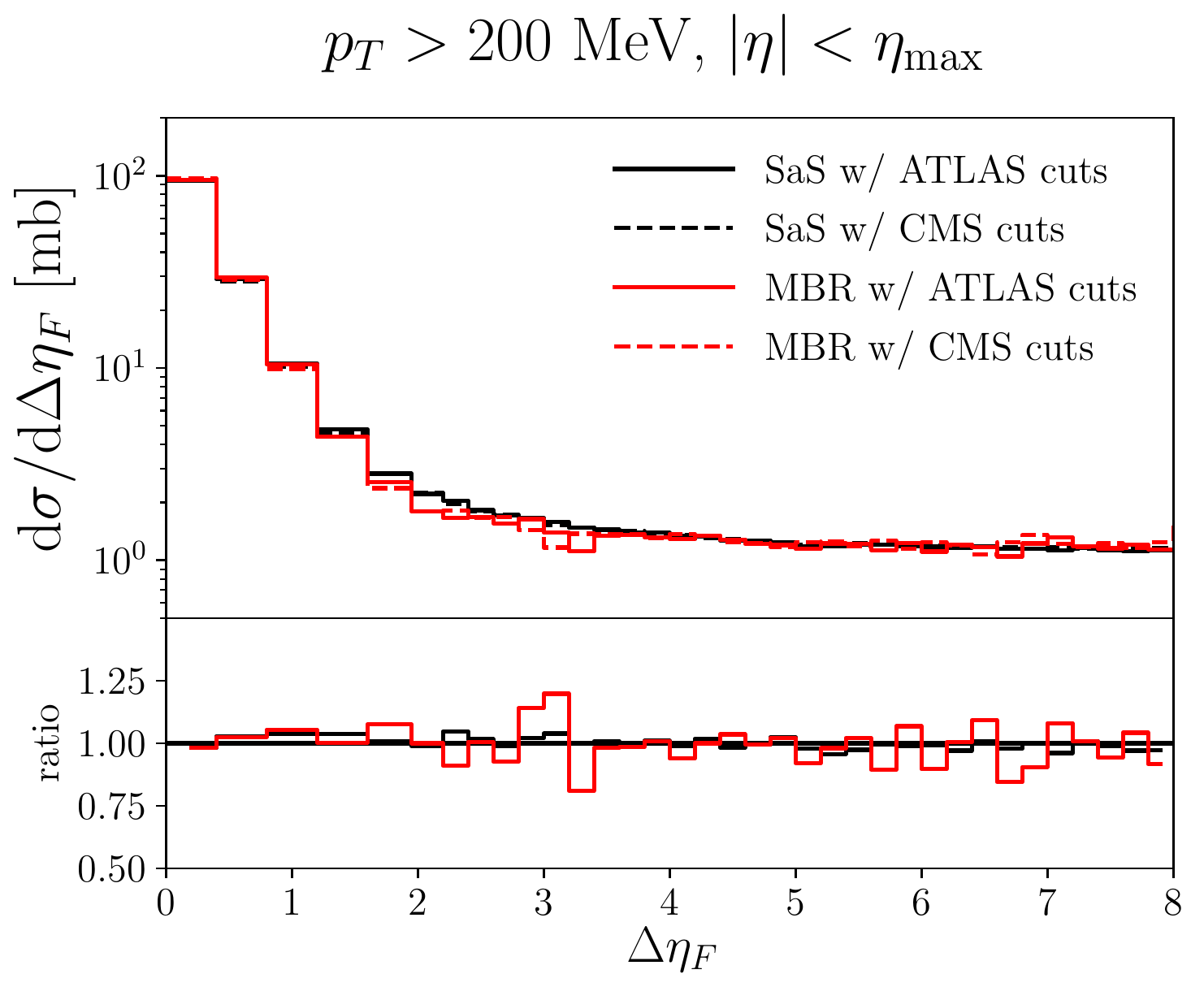}\\
(a)
\end{minipage}
\hfill
\begin{minipage}[c]{0.475\linewidth}
\centering
\includegraphics[width=\linewidth]{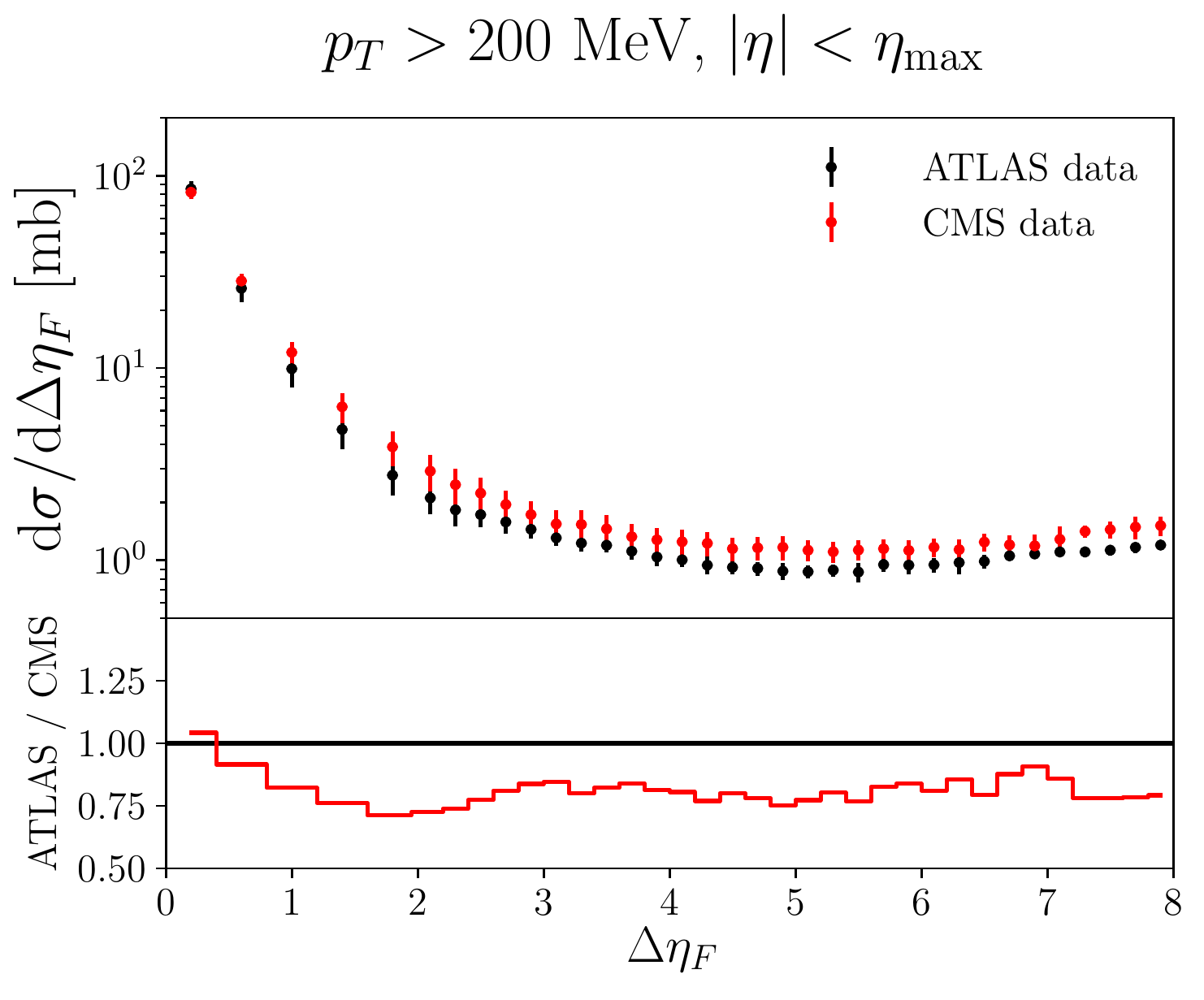}\\
(b)
\end{minipage}
\caption{\label{Fig:ATLAS_CMS} (a) The SaS and MBR models using either
ATLAS or CMS cuts along with the ratio of ATLAS to CMS cuts for both
models. (b) The ATLAS \cite{Aad:2012pw} and CMS 
\cite{Khachatryan:2015gka} data along with the ratio of ATLAS to
CMS data, showing significant differences in the entire range.}
\end{figure}

In fig.~\ref{Fig:ATLAS_CMS}a, we show the results obtained with default
\textsc{Pythia}~8 using the SaS model and the MBR model when comparing 
to either the ATLAS or CMS Rivet analyses. Both models are shown, since
ATLAS uses the SaS model for unfolding, while CMS uses the MBR one, but model
agreement is sufficiently close that unfolding differences should not be 
an issue. Further, from fig.~\ref{Fig:ATLAS_CMS}a it is evident that 
the different experimental $\eta$ cuts gives at most a 5\% effect on 
either model. This does not account for the approximately 25\% difference 
seen in data, see fig.~\ref{Fig:ATLAS_CMS}b. A tune to both 
datasets will not be able to describe either perfectly, as they so
clearly disagree. Experiment-specific tunes would likely improve the
description of that particular dataset, hence worsening the
description of the other. As we cannot decide which of the two is the 
preferred one, we instead aim for the middle ground. 

Besides the above mentioned datasets measurements of the inelastic and
diffractive cross sections have been performed by both ATLAS, CMS and
ALICE. We include the following measurements: the inelastic cross
section from ATLAS 2011 \cite{Aad:2011eu}, the inelastic cross section 
from CMS 2012 \cite{Chatrchyan:2012nj} and the inelastic and diffractive 
cross sections from ALICE 2012 \cite{Abelev:2012sea}. 

\begin{figure}[ht!]
\begin{minipage}[c]{0.475\linewidth}
\centering
\includegraphics[width=\linewidth]{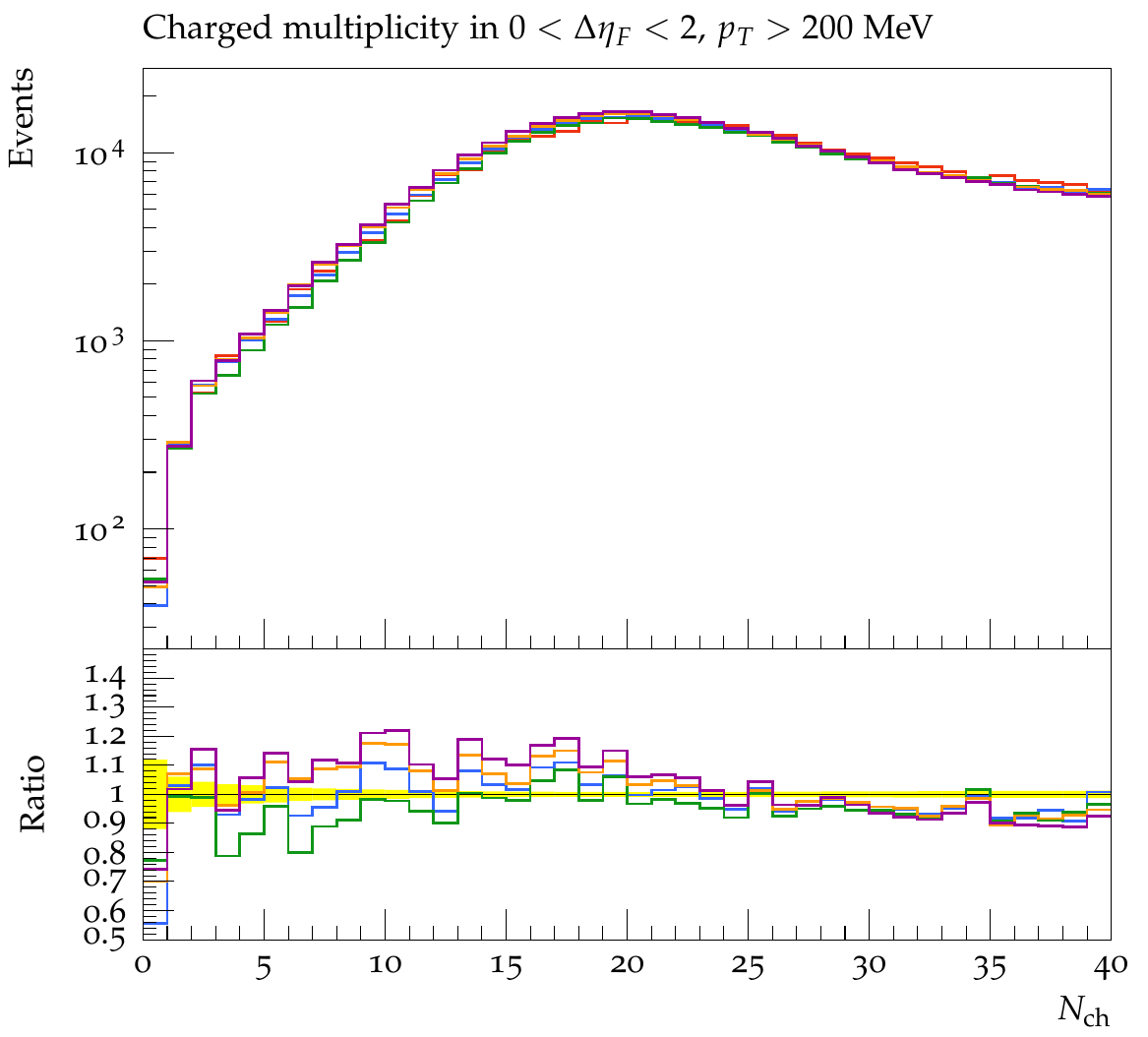}\\
(a)
\end{minipage}
\hfill
\begin{minipage}[c]{0.475\linewidth}
\centering
\includegraphics[width=\linewidth]{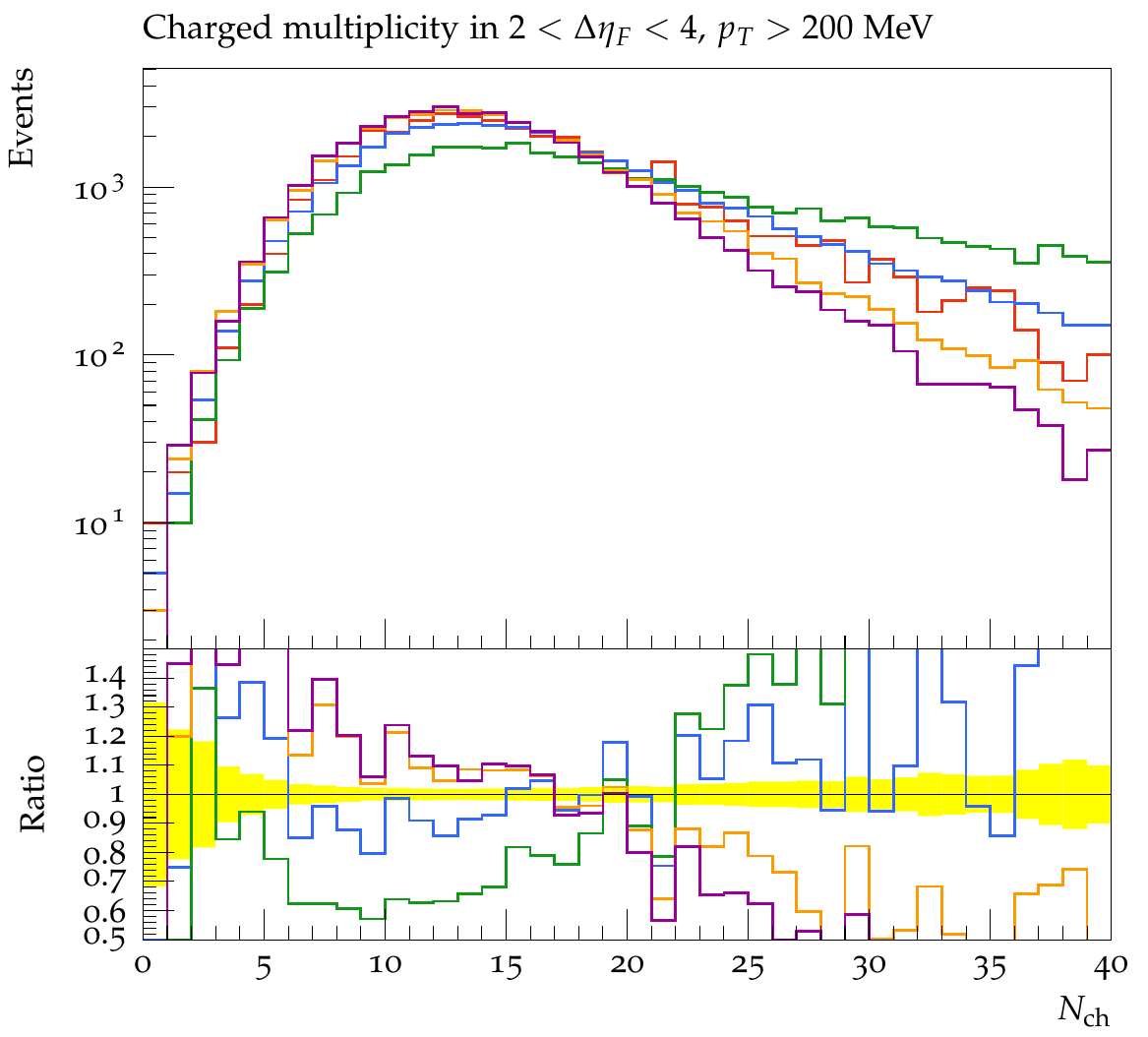}\\
(b)
\end{minipage}
\begin{minipage}[c]{0.475\linewidth}
\centering
\includegraphics[width=\linewidth]{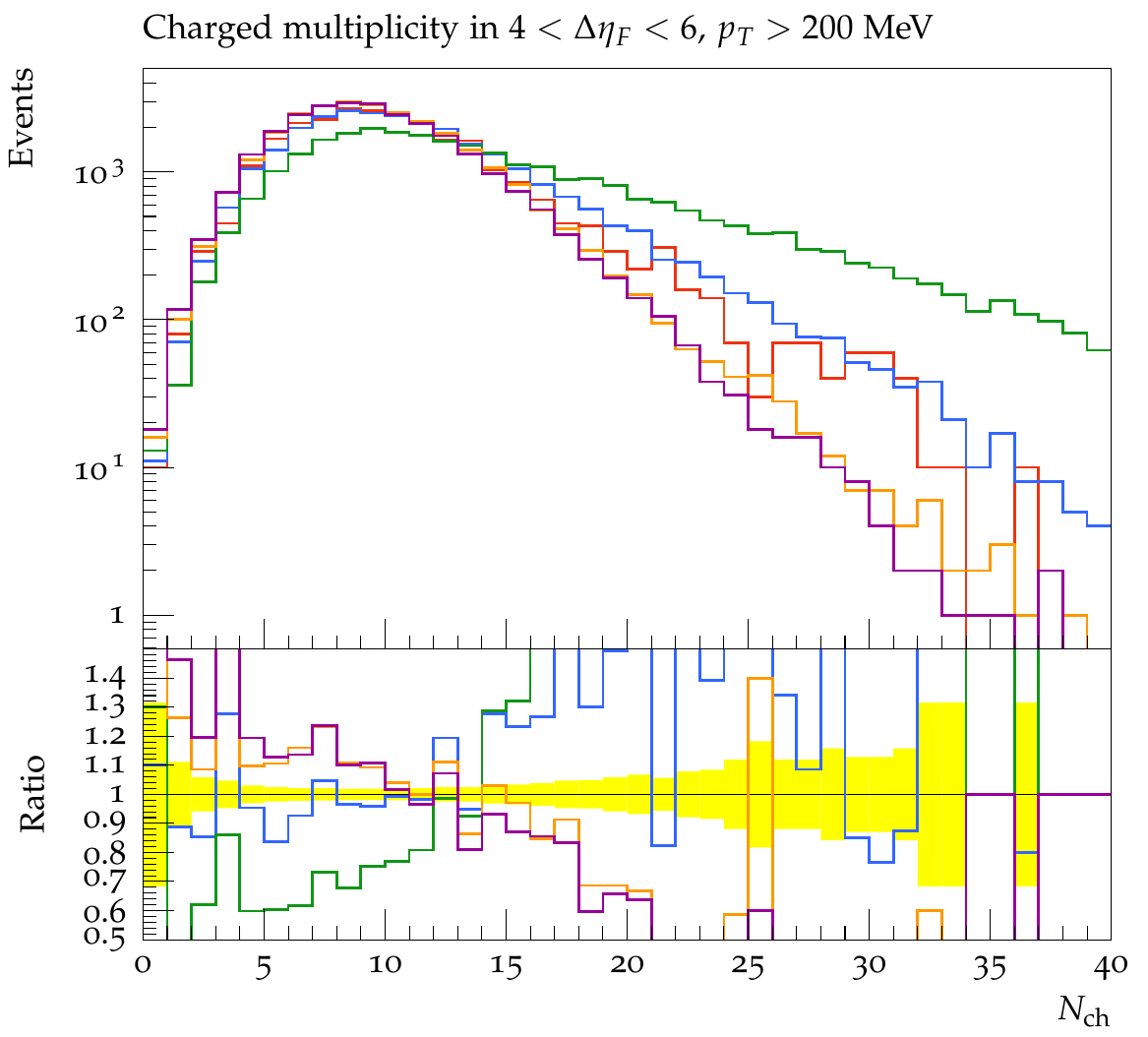}\\
(c)
\end{minipage}
\hfill
\begin{minipage}[c]{0.475\linewidth}
\centering
\includegraphics[width=\linewidth]{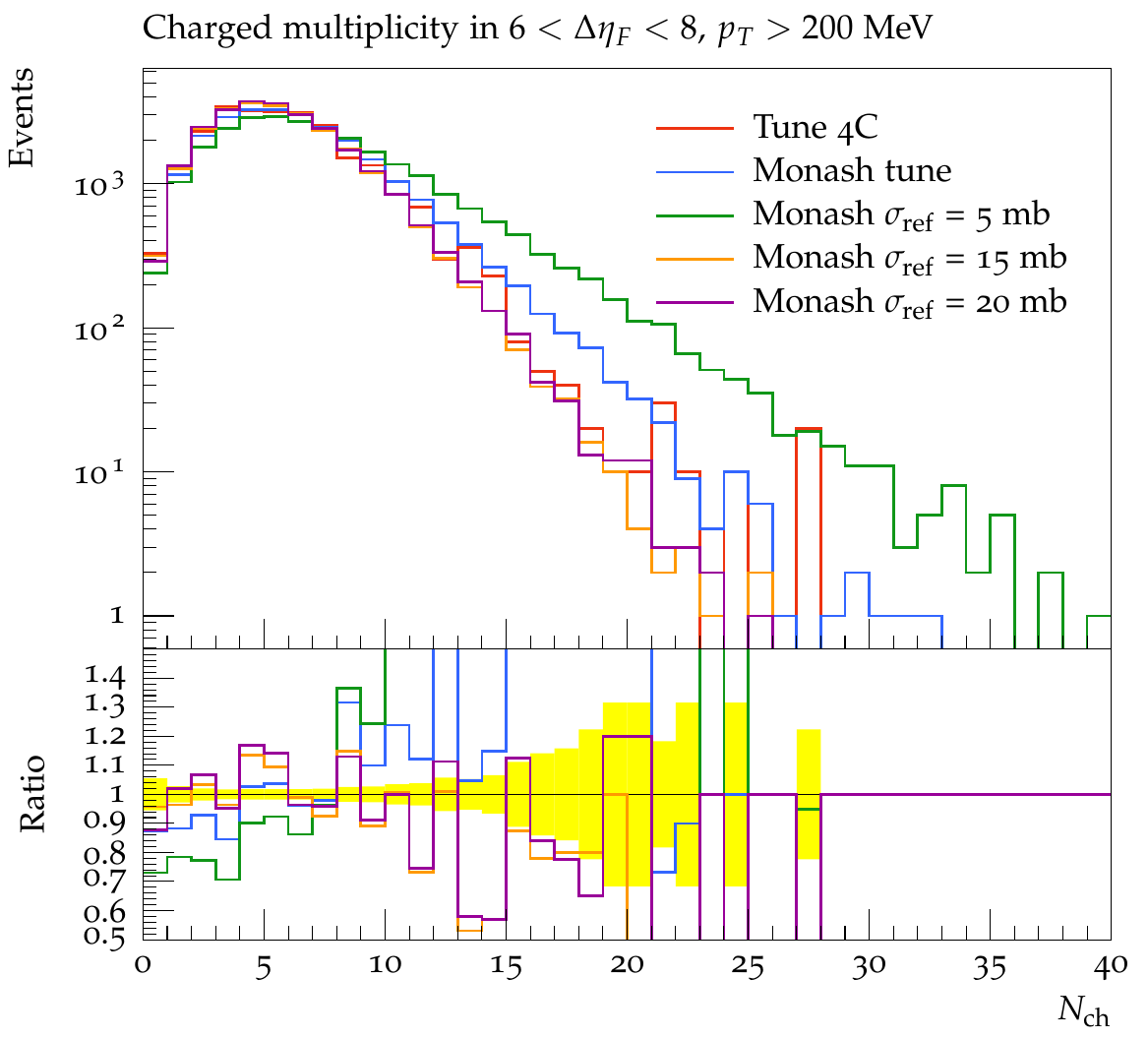}\\
(d)
\end{minipage}
\caption{\label{Fig:sigRef} The effects of changing the reference
$\Pom\p$ cross section on the charged multiplicity distribution using
the SaS model at 7 TeV in the four gap ranges: 0 to 2 (a), 2 to 4 (b), 
4 to 6 (c) and 6 to 8 (d).}
\end{figure}

\begin{figure}[ht!]
\begin{minipage}[c]{0.475\linewidth}
\centering
\includegraphics[width=\linewidth]{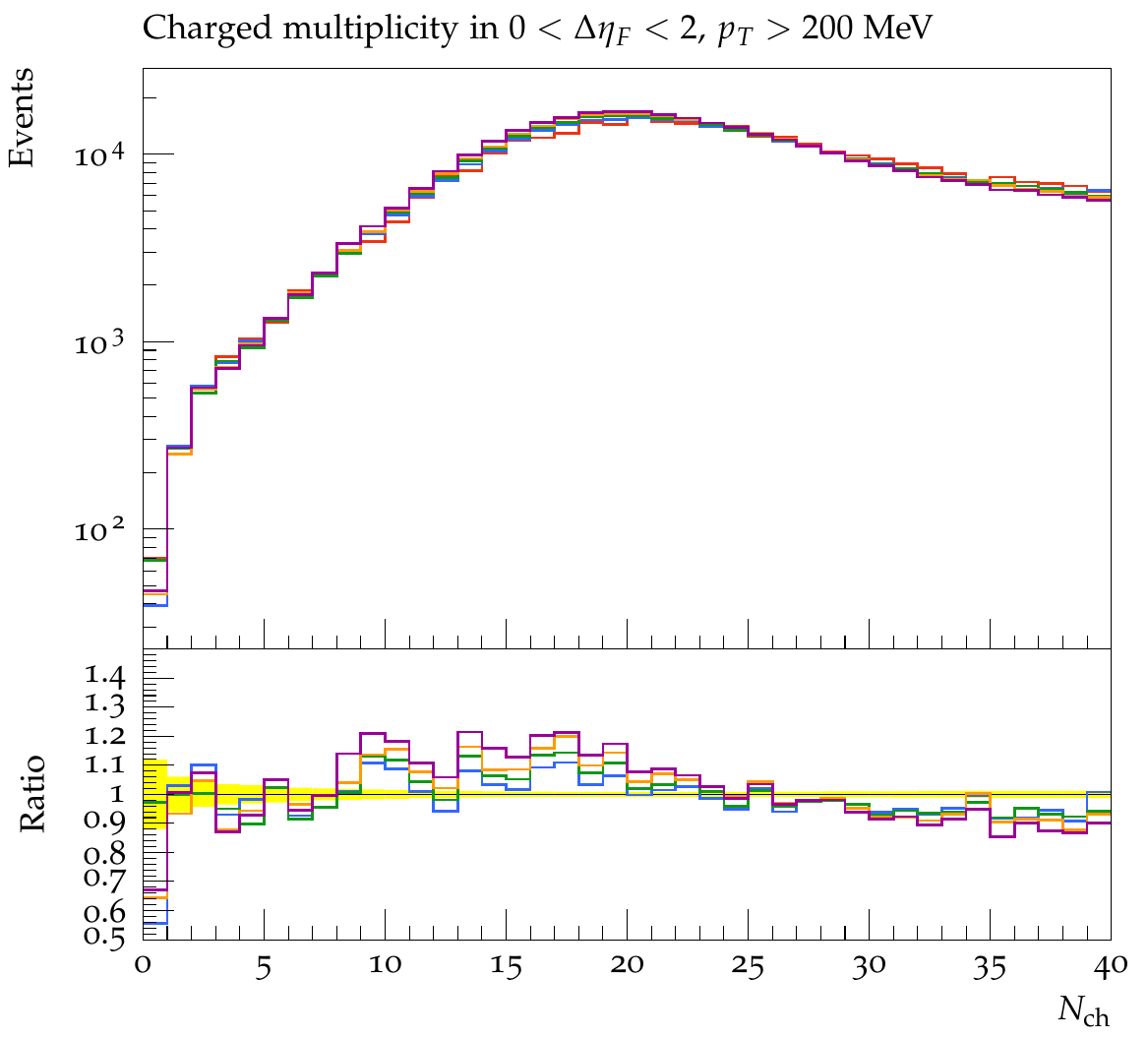}\\
(a)
\end{minipage}
\hfill
\begin{minipage}[c]{0.475\linewidth}
\centering
\includegraphics[width=\linewidth]{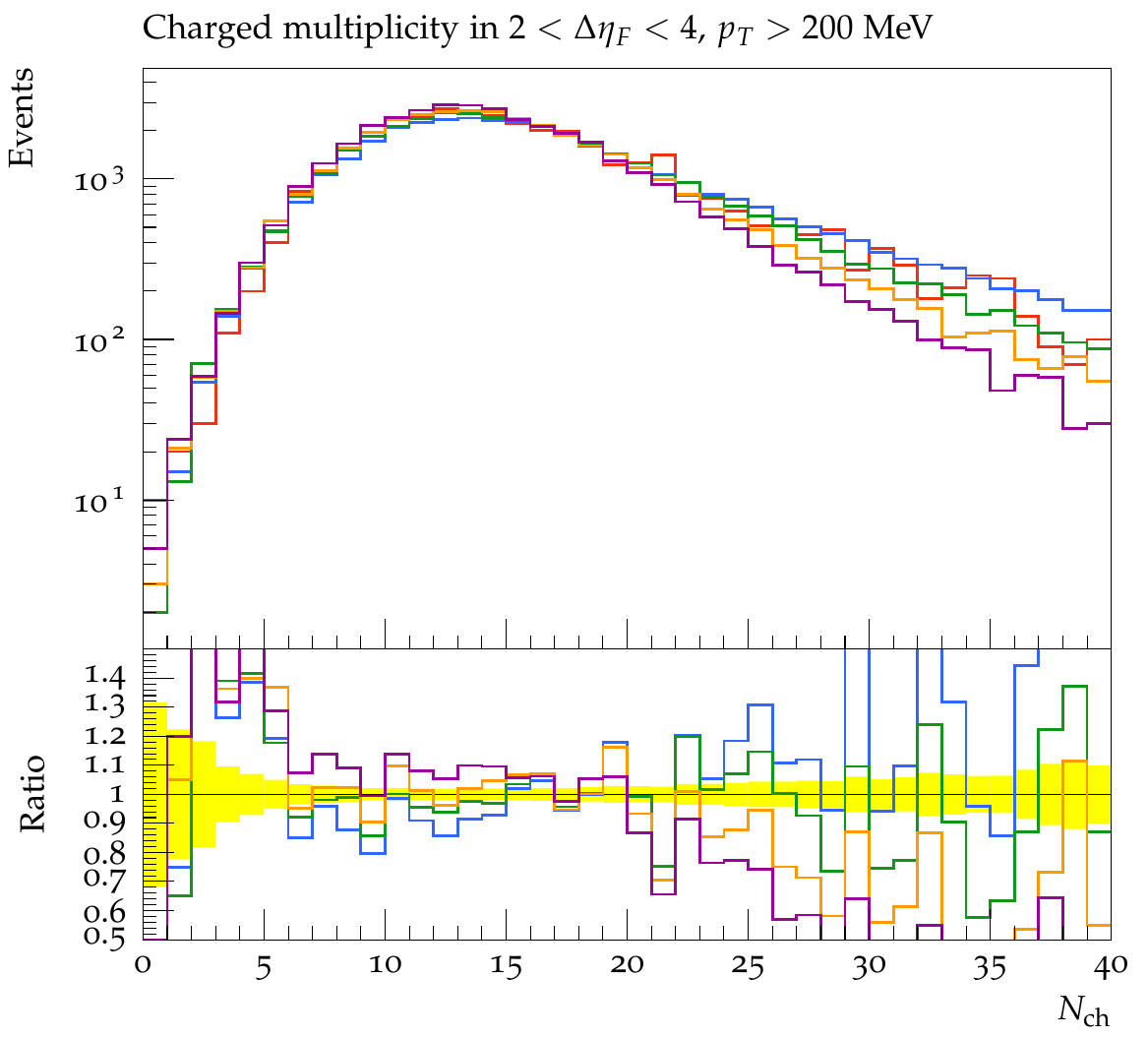}\\
(b)
\end{minipage}
\begin{minipage}[c]{0.475\linewidth}
\centering
\includegraphics[width=\linewidth]{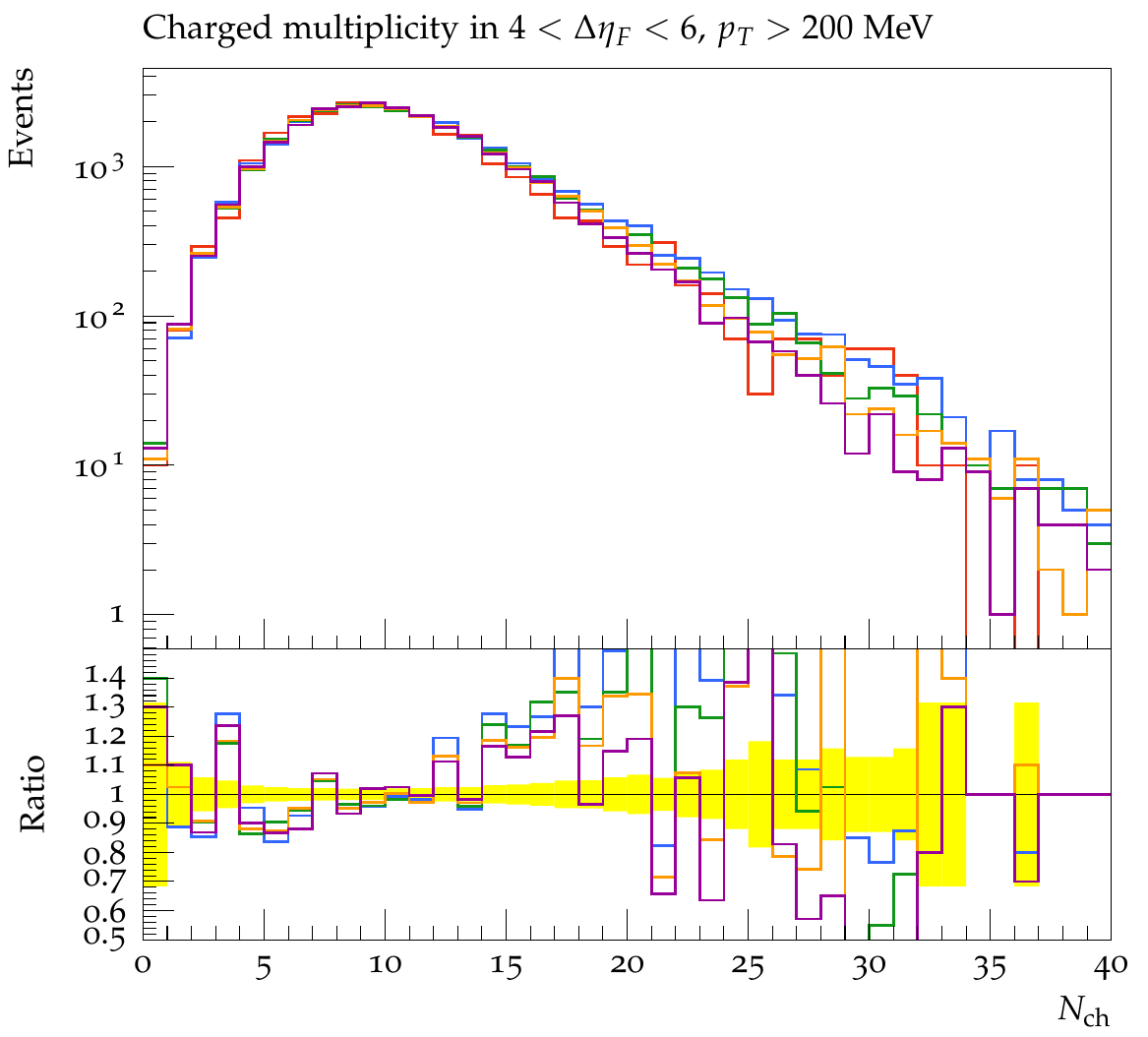}\\
(c)
\end{minipage}
\hfill
\begin{minipage}[c]{0.475\linewidth}
\centering
\includegraphics[width=\linewidth]{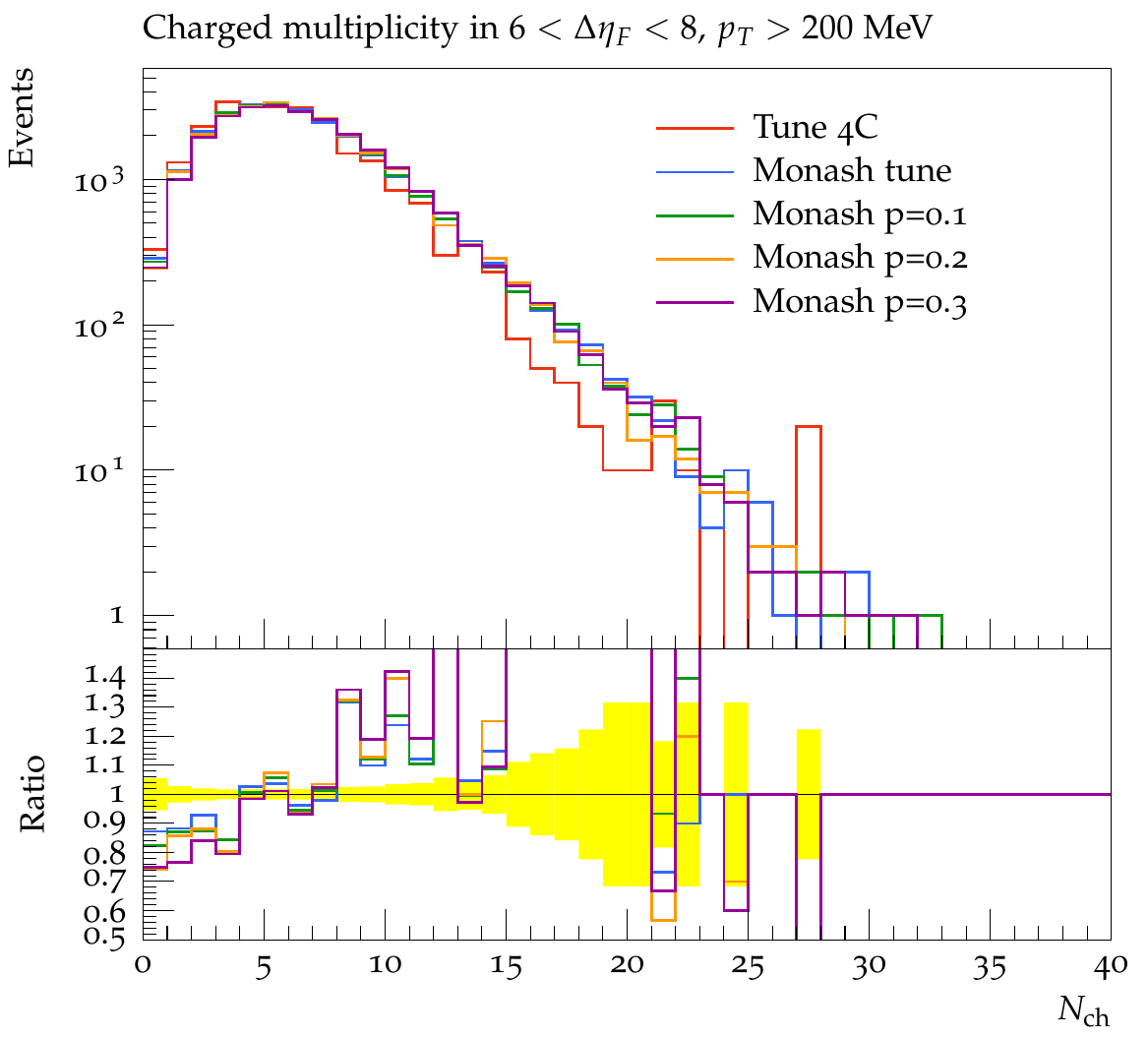}\\
(d)
\end{minipage}
\caption{\label{Fig:mPowPomP}The effects of changing the power of the
mass dependence in the $\Pom\p$ cross section on the charged 
multiplicity distribution using the SaS model at 7 TeV in the four gap 
ranges: 0 to 2 (a), 2 to 4 (b), 4 to 6 (c) and 6 to 8 (d).}
\end{figure}

None of the datasets available in the Rivet framework are able to
constrain the parameters related to the hadronic event properties.
This includes both the low-to-high-mass transition probability
parameters as well as the parameters of the non-perturbative and 
perturbative description of the evolution of the diffractive system.
In particular, the non-perturbative description is left as is in this
study, while the effects of changing the $\Pom\p$ cross section is
shown in figs.~\ref{Fig:sigRef} and \ref{Fig:mPowPomP}. This cross
section determines the amount of multiparton interactions activity 
in a high-mass diffractive event, and thereby e.g.\ the charged 
multiplicity distribution. It is interesting because of discrepancies 
between uncorrected ATLAS data and the Pythia 4C tune (figs.~3a-d in 
\cite{Aad:2012pw}). A direct comparison cannot be made, since 
the ATLAS distributions show the number of electromagnetic clusters
rather than that of charged particles, but the two clearly are related.

Figs.~\ref{Fig:sigRef} and \ref{Fig:mPowPomP} show the effects of 
changing the $\Pom\p$ cross section on the charged particle 
distributions in the different $\Delta\eta_F$ bins compared to 
the 4C tune. In \cite{Aad:2012pw} the 4C tune generally was seen to 
undershoot the low cluster-multiplicities, while overshooting the 
mid to high cluster multiplicities. In the highest $\Delta\eta_F$ bin,
dominated by the diffractive events, Tune 4C undershoots both the 
low- and high-multiplicity activity. Reducing the $\Pom\p$ cross
increases the multiplicity, and vice versa. Thus, to describe the
high-multiplicity events, a smaller $\Pom\p$ cross section would be
preferred. This could be compensated by allowing the perturbative
description to go below $M_X=m_{\mrm{min}}=10$ GeV, thus allowing
slightly more activity in low-mass systems, possibly increasing the
number of low-multiplicity events. The effects of including a 
mass dependence in the $\Pom\p$ cross section is seen in 
fig.~\ref{Fig:mPowPomP}. A parametrization has been chosen as
$\sigma_{\Pom\p}^{\mrm{eff}}(M_X) = \sigma_{\Pom\p}^{\mrm{ref}} \,
(M_X / M_{\mrm{ref}})^p$, with $M_{\mrm{ref}} = 100$~GeV.
Here, an increase of $p$ slightly decreases
the high-multiplicity region, albeit more subtly than with an increase
of the $\Pom\p$ cross section. Recall that the mass of the diffractive 
system is related to the collision energy, such that a value of 
$p\sim 0.2-0.3$ is not unreasonable, corresponding to a rise of the 
cross section with energy of $s^{0.1}-s^{0.15}$. 

\begin{figure}[ht!]
\begin{minipage}[c]{0.475\linewidth}
\centering
\includegraphics[width=\linewidth]{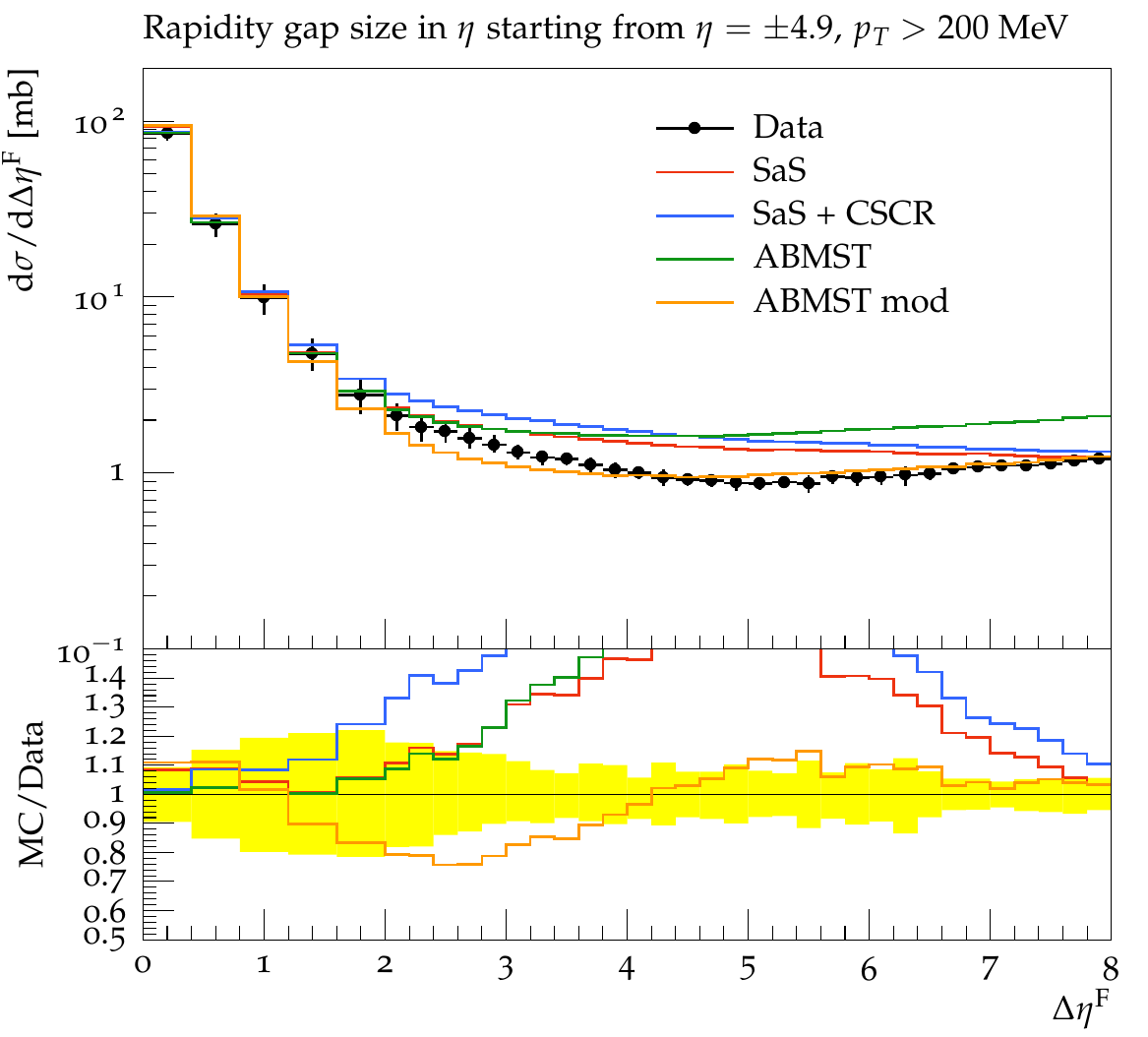}\\
(a)
\end{minipage}
\hfill
\begin{minipage}[c]{0.475\linewidth}
\centering
\includegraphics[width=\linewidth]{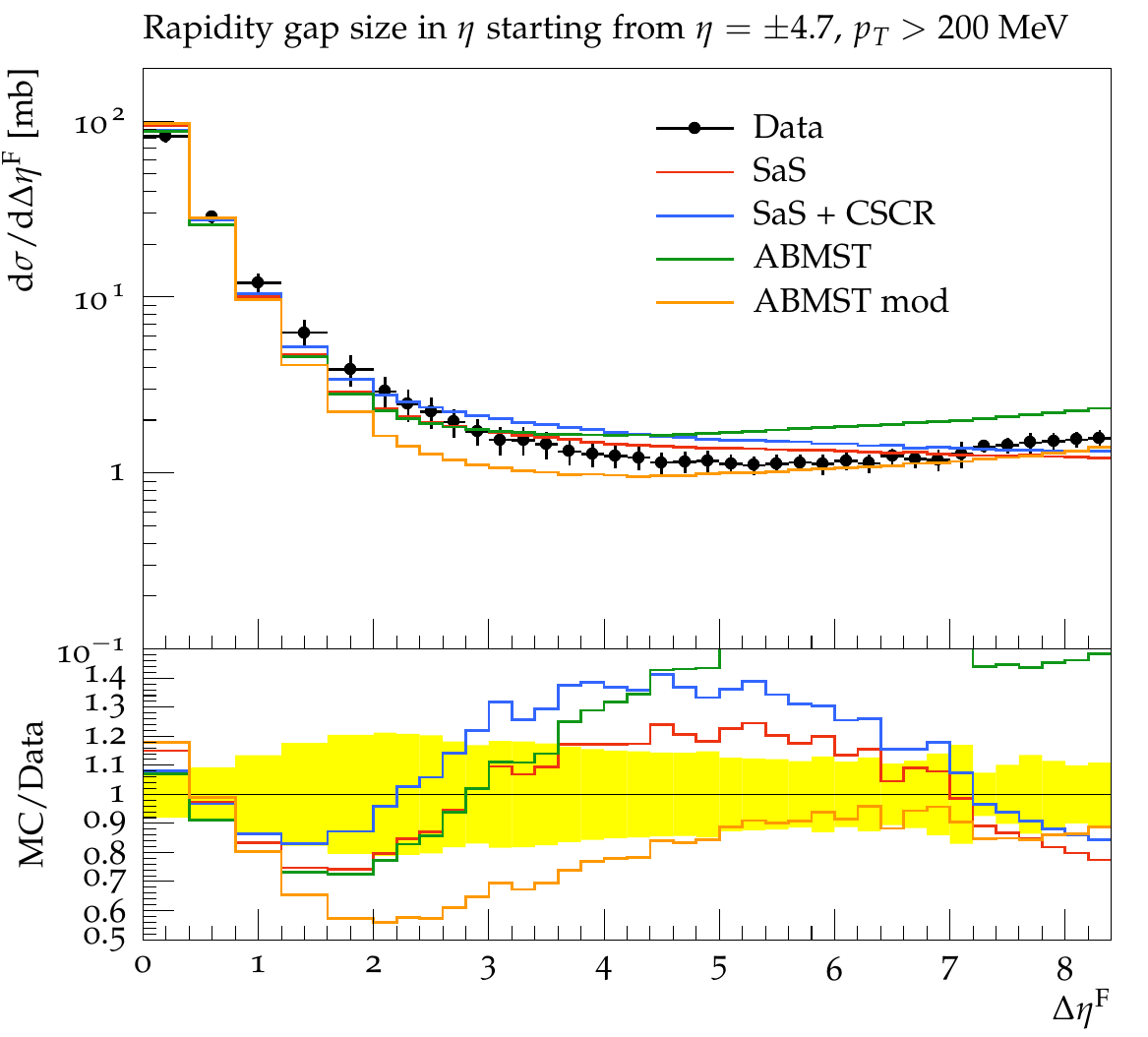}\\
(b)
\end{minipage}
\begin{minipage}[c]{0.475\linewidth}
\centering
\includegraphics[width=\linewidth]{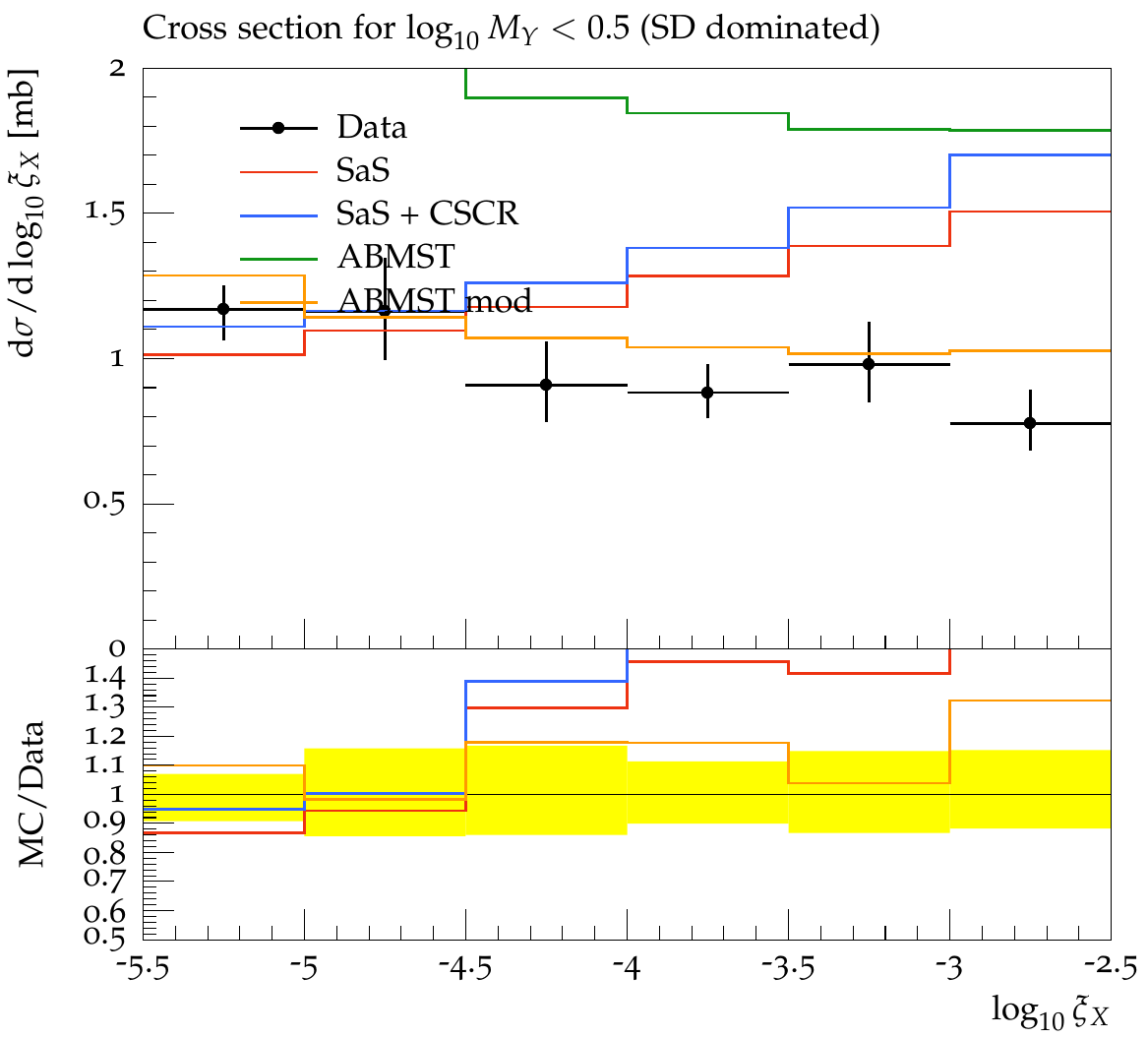}\\
(c)
\end{minipage}
\hfill
\begin{minipage}[c]{0.475\linewidth}
\centering
\includegraphics[width=\linewidth]{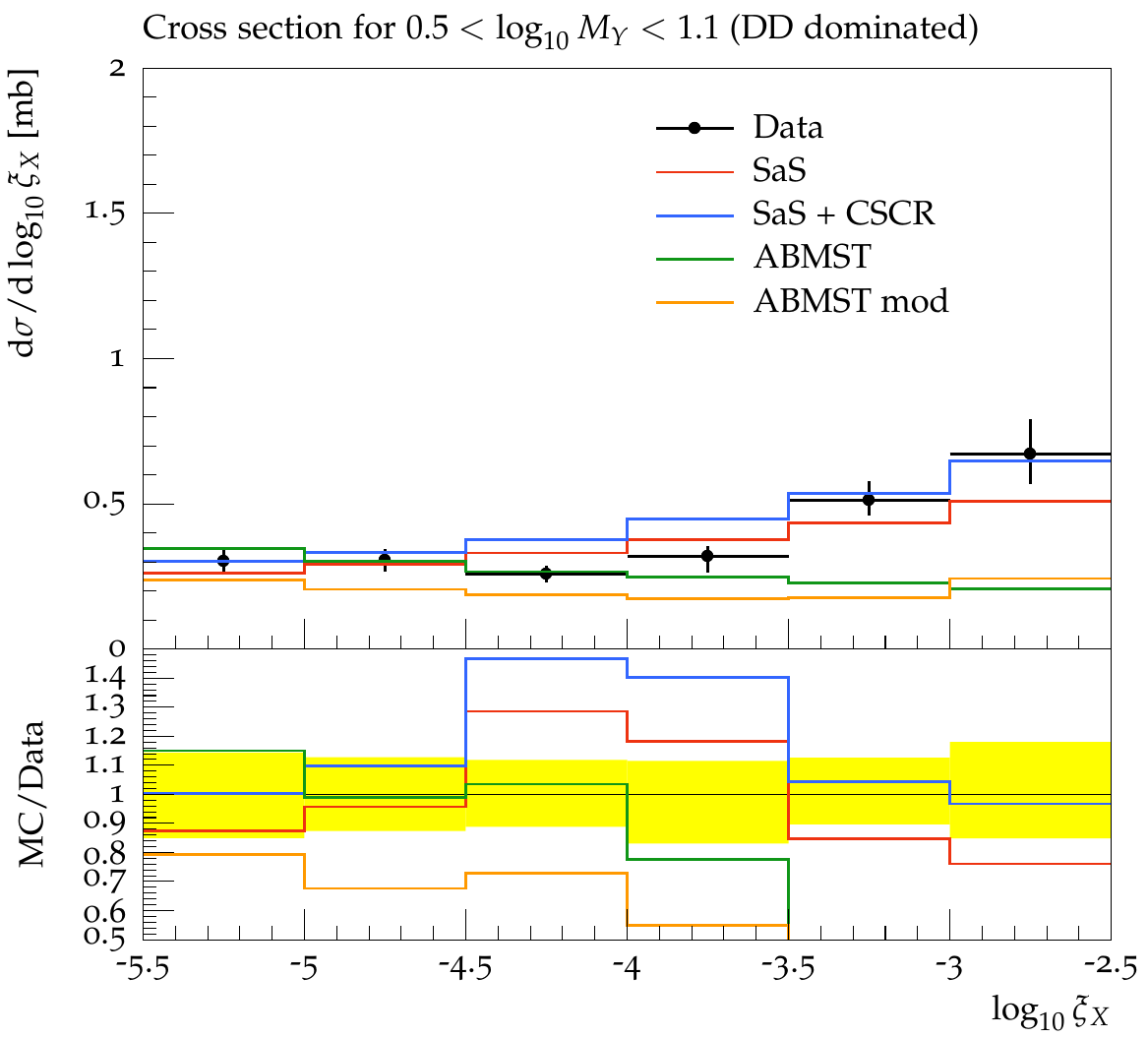}\\
(d)
\end{minipage}
\caption{\label{Fig:DefaultComparison} The cross section as a function of
gap size for the default SaS model, the SaS+CSCR model and the untuned
ABMST models compared to ATLAS \cite{Aad:2012pw} (a) and CMS 
\cite{Khachatryan:2015gka} (b) data. The cross section as a function of 
$\mrm{log}_{10}\xi_X$ in a single-diffraction dominated region (c) and 
double diffraction dominated region (d) compared to CMS 
\cite{Khachatryan:2015gka} data.}
\end{figure}

A full study of particle production in diffractive events with 
\textsc{Pythia}~8, \textsc{Herwig}~7 \cite{Bahr:2008pv,Bellm:2015jjp}, 
\textsc{Sherpa} \cite{Gleisberg:2008ta,Martin:2012nm} and
\textsc{Phojet} \cite{Bopp:1998rc} could provide further
valuable information on the hadronic event properties of diffractive 
systems, as the generators differ in how they describe such
production. The effects of colour reconnection in a diffractive system
is also of interest, as the amount of ``accidental'' gaps could be
constrained in these systems, if one assumes that the CR scheme
is the same in both diffractive and non-diffractive systems. At
present we leave the $\Pom\p$ cross section as is at 10 mb, and
show the results with the models presented so far in
fig.~\ref{Fig:DefaultComparison}.  

The bulk of the cross section arises from nondiffractive events.
These tend to only give rise to small rapidity gaps, as the 
phase space is more or less evenly filled by multiparton interactions. 
Gaps of intermediate or large size can occur, however, e.g.\ by 
colour reconnection between the partons \cite{Edin:1995gi,Edin:1996mw}. 
The default \textsc{Pythia} CR framework has been designed to avoid 
accidental gaps, so as to keep a clean separation between diffractive
and nondiffractive topologies. In other models, e.g.\ the CSCR one
\cite{Christiansen:2015yqa}, the colour reshuffling tends to give 
somewhat larger probability for intermediate gaps. A combination of 
the CSCR model and the default SaS diffractive setup then results in 
too large a cross section in the intermediate-gap range,
cf. figs.~\ref{Fig:DefaultComparison}a,b.
 
Diffractive events are more likely to give rise to
intermediate to large gaps. Hence, depending on colour-reconnection
model used, they will dominate from gap sizes of approximately two 
and larger. The size of the gap is closely connected to the mass of 
the diffractive system. Thus a model with a $\d M_X^2/M_X^2$ ansatz, 
like the SaS one (modulo some corrections), will give an approximately 
flat distribution of measured gap sizes. This can be modified by the 
recent inclusion of the mass-correction factor $\epsilon_{\mrm{SaS}}$,
which introduces an additional 
$1/M_X^{2\epsilon_{\mrm{SaS}}}$ factor to the differential model. 
Depending on the sign of $\epsilon_{\mrm{SaS}}$, it will either increase 
or decrease the high-mass cross section. In both the ATLAS and CMS 
datasets an increase of the large-gap cross section is seen. Thus we 
expect a positive sign for $\epsilon_{\mrm{SaS}}$, as this will enhance 
the activity at low masses. For simplicity, adding the mass correction 
will not affect the integrated diffractive cross section. 

The ABMST models show slightly better agreement with the shape of the
rapidity gap distributions, although the original ABMST model overshoots
both datasets. This was to be expected, as the model had trouble with the
increase of the single diffractive cross section at LHC energies. The
modified version of the ABMST model shows very nice agreement with
both datasets, except for an undershoot of the high-mass region of the
double-diffractive-dominated region in fig.~\ref{Fig:DefaultComparison}d.
This behaviour closely correlates with the flatness of the 
$\xi\d\sigma/\d\xi$-spectrum, fig.~\ref{Fig:DD}b. Both the ABMST 
models have a mass spectrum shape comparable to data in the 
single-diffraction-dominated region, unlike the SaS model, which 
overshoots the high-mass systems. 

\subsection{The tuned models}

The tunes provided here are performed with the Professor framework 
\cite{Buckley:2009bj}, varying the high-mass diffractive parameters 
given in table \ref{Tab:TuneParameters}. All non-diffractive parameters 
are left at their default values, as given by the Monash tune 
\cite{Skands:2014pea}, except for the CSCR-specific changes in that 
setup. The H1 leading order Pomeron PDF \cite{Aktas:2006hy} is used 
in all the tunes.

\begin{table}[tbp]
\centering
\begin{tabular}{|l|c|c|c|c|c|c|}
\hline
 & $\epsilon$ & $\alpha'$ & $\sigma_{\mrm{SD}}^{\mrm{max}}$ &
 $\sigma_{\mrm{DD}}^{\mrm{max}}$ & $\sigma_{\mrm{CD}}^{\mrm{max}}$ &
 $\epsilon_{\mrm{SaS}}$ \\
\hline
SaS & 0.06 & 0.4 & 22.31 & 39.83 & 0. & 0. \\
SaS+CSCR & 0.15 & 0.26 & 20.81 & 13.13 & 0. & 0. \\
SaS+$\epsilon_{\mrm{SaS}}$ & 0.04 & 0.30 & 24.78 & 52.49 & 0. & 0.08 \\
\hline
 & $k_{\mrm{SD}}$ & $k_{\mrm{DD}}$ & $k_{\mrm{CD}}$ & $p_{\mrm{SD}}$ &
 $p_{\mrm{DD}}$ & $p_{\mrm{CD}}$\\
\hline
ABMST & 0.58 & 2.45 & 1.0 & 0. & 0.05 & 0.03 \\
ABMST modified & 0.92 & 1.72 & 1.38 & 0. & 0.1 & 0.04 \\
\hline
\end{tabular}
\caption{\label{Tab:TuneParameters} The parameters used in the tunes
for the different models.}
\end{table}

\begin{figure}[ht!]
\begin{minipage}[c]{0.475\linewidth}
\centering
\includegraphics[width=\linewidth]{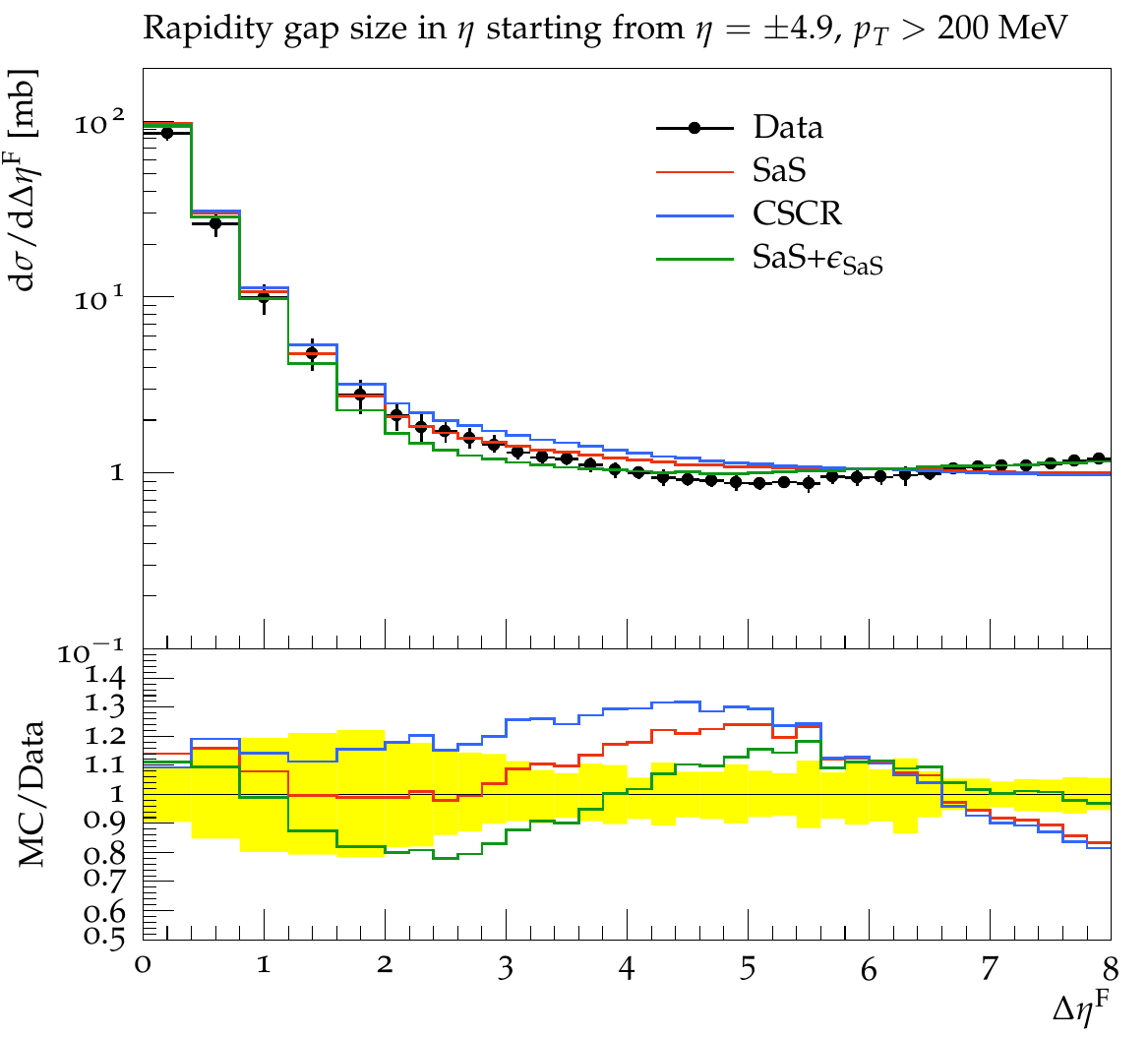}\\
(a)
\end{minipage}
\hfill
\begin{minipage}[c]{0.475\linewidth}
\centering
\includegraphics[width=\linewidth]{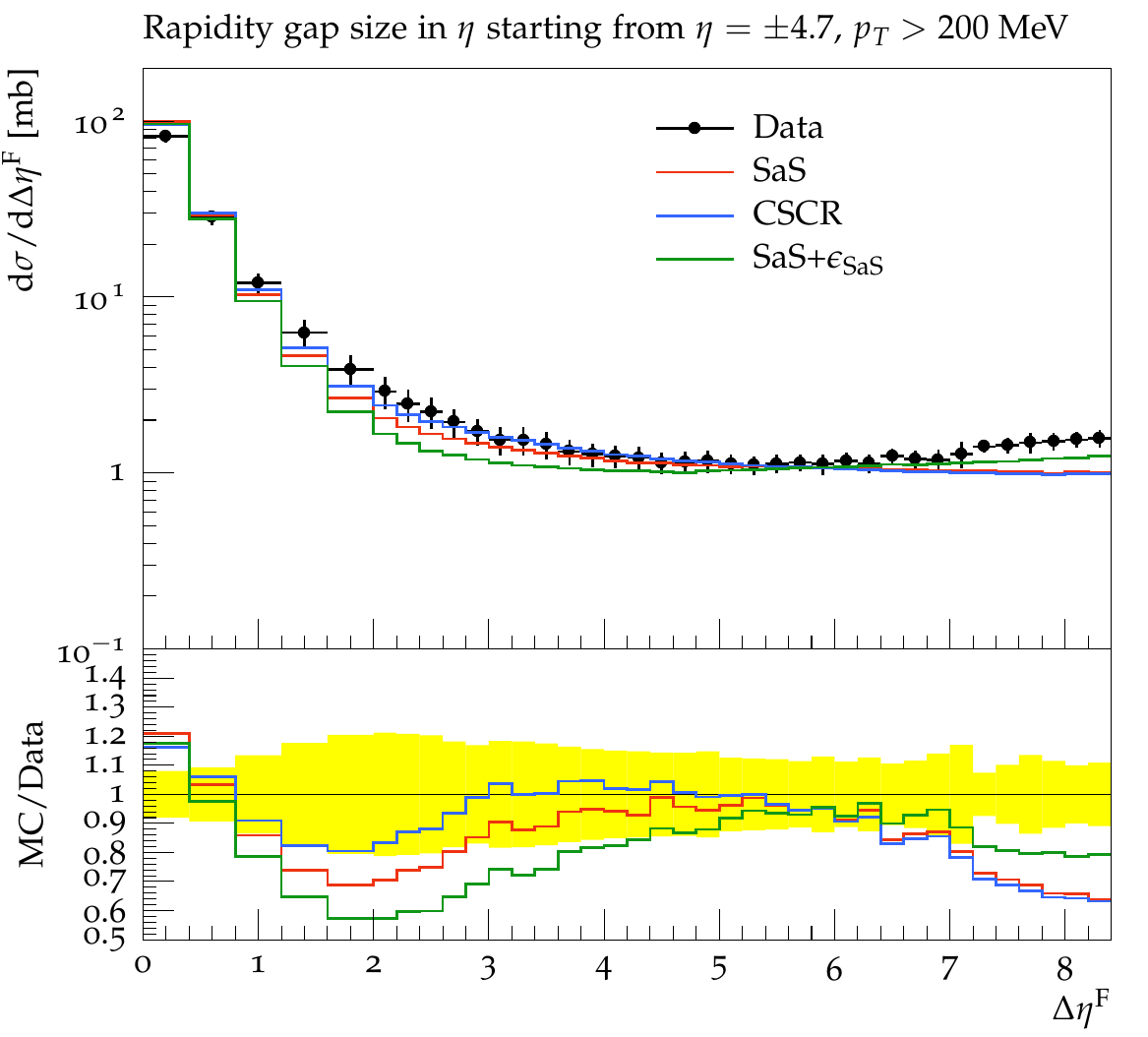}\\
(b)
\end{minipage}
\begin{minipage}[c]{0.475\linewidth}
\centering
\includegraphics[width=\linewidth]{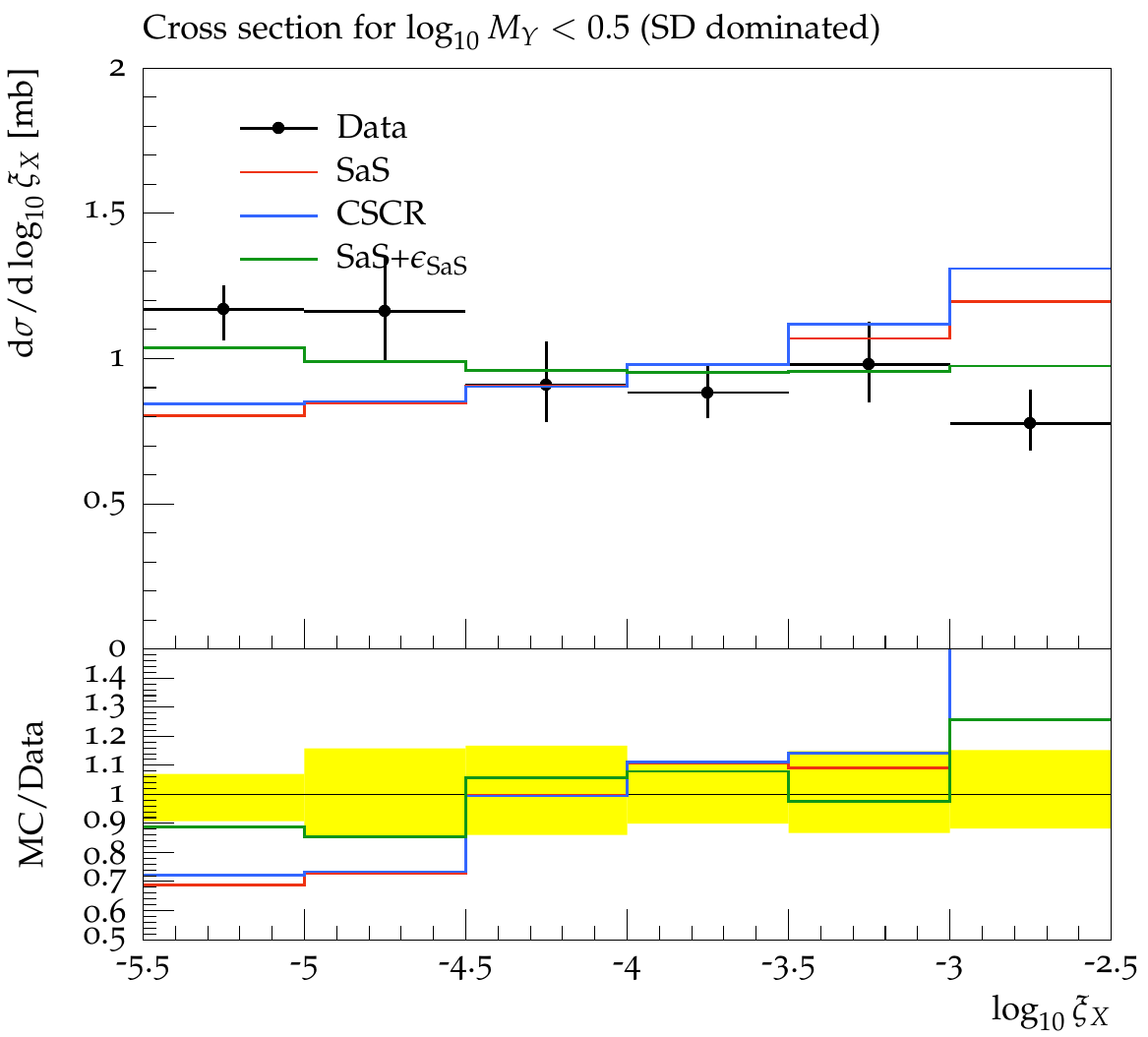}\\
(c)
\end{minipage}
\hfill
\begin{minipage}[c]{0.475\linewidth}
\centering
\includegraphics[width=\linewidth]{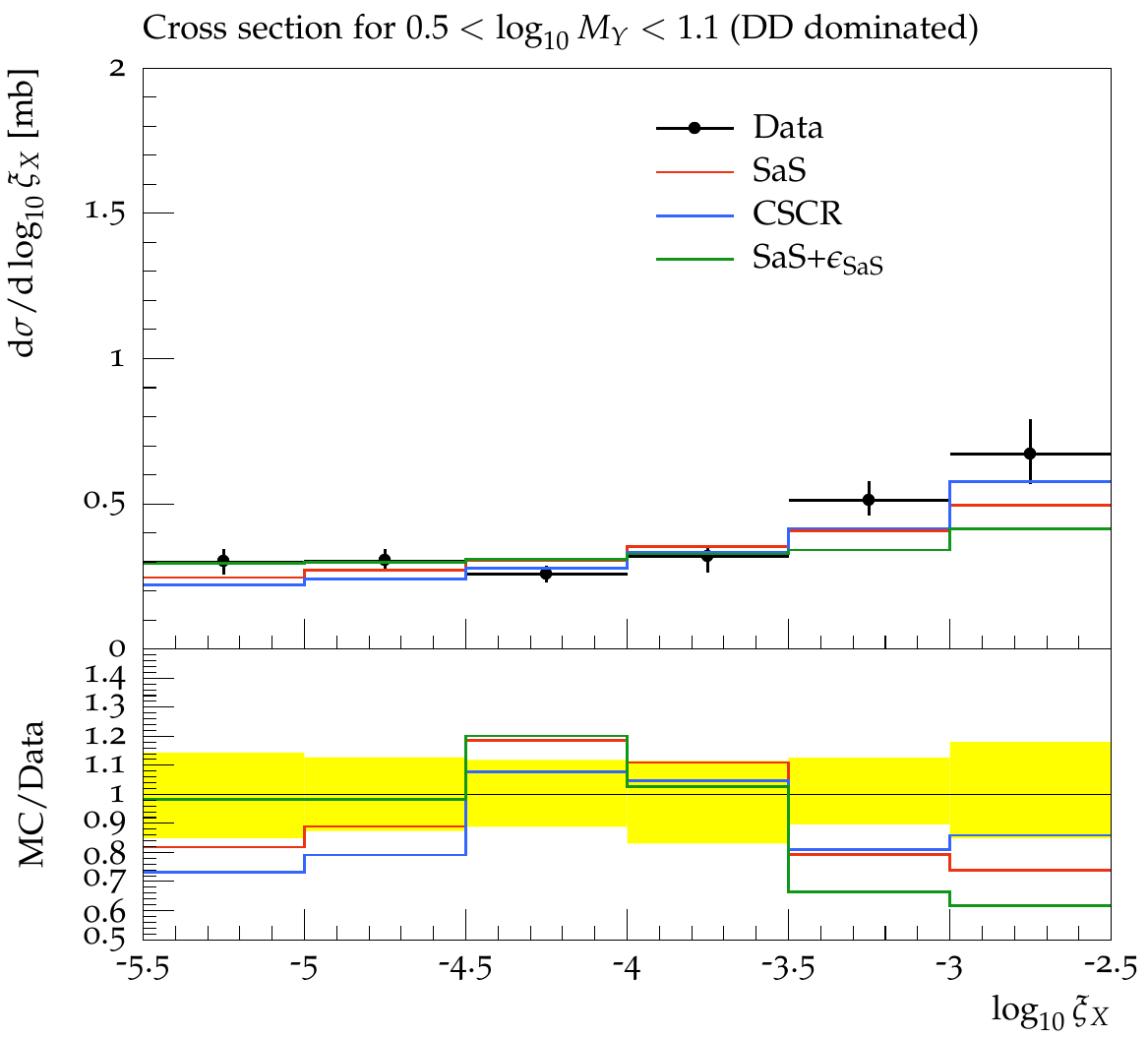}\\
(d)
\end{minipage}
\caption{\label{Fig:SaScomparison} The cross section as a function of
gap size for the three SaS-based models compared to ATLAS 
\cite{Aad:2012pw} (a) and CMS \cite{Khachatryan:2015gka} (b) data. 
The cross section as a function of $\mrm{log}_{10}\xi_X$ in
a single-diffraction dominated region (c) and double diffractive
dominated region (d) compared to CMS \cite{Khachatryan:2015gka} data.}
\end{figure}

Fig.~\ref{Fig:SaScomparison} shows the three SaS-based models tuned to 
the above-mentioned data. Neither of the three models are able to describe
the shape of the gap data perfectly, figs.~\ref{Fig:SaScomparison}a,b. 
The tune has decreased the amount of activity in the mid- to large-gap 
region by a decrease of the $\sigma_i^{\mrm{max}}$ values used in 
eq.~(\ref{Eq:SaSmod}). The inclusion of $\epsilon_{\mrm{SaS}}$ has shifted 
some of the activity from intermediate-gaps to larger ones, while keeping 
the integrated cross section fixed. Unfortunately this is at the expense 
of an undershoot in the transition region $\Delta\eta^{\mrm{F}} \sim 2$
between diffractive and nondiffractive topologies. This is the region 
where CSCR does better, so a combination of CSCR with an 
$\epsilon_{\mrm{SaS}} > 0$ could provide a flatter MC/data distribution
in fig.~\ref{Fig:SaScomparison}a,b.

For the mass spectra measured by CMS, fig.~\ref{Fig:SaScomparison}c,d, 
evidently only the SaS+$\epsilon_{\mrm{SaS}}$ model is able to 
describe the single-diffraction-dominated mass spectrum, whereas it 
undershoots the high-mass double diffraction region since, relative to 
the original SaS model, it has shifted some of the high-mass activity 
to lower masses. 

\begin{figure}[ht!]
\begin{minipage}[c]{0.475\linewidth}
\centering
\includegraphics[width=\linewidth]{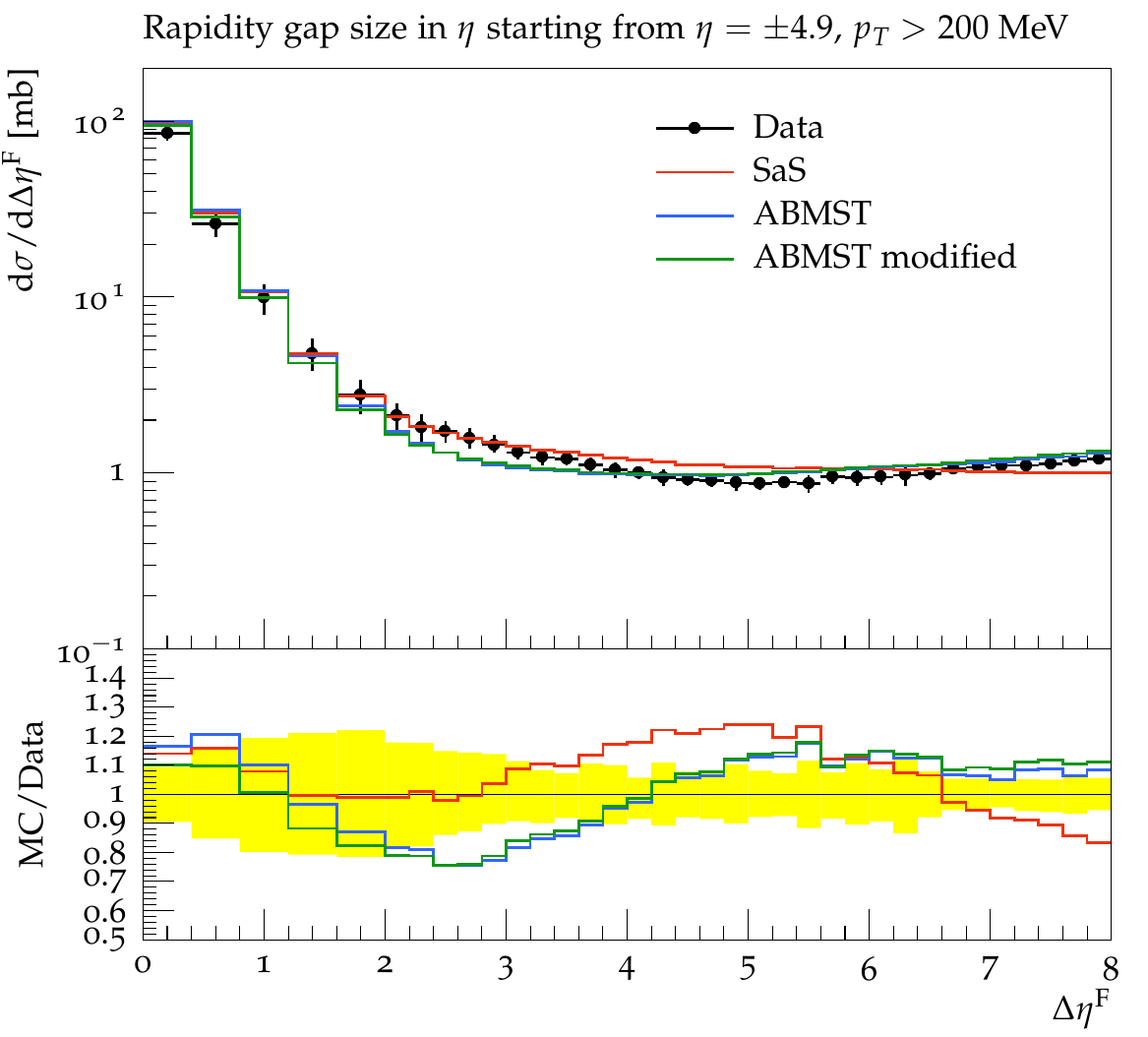}\\
(a)
\end{minipage}
\hfill
\begin{minipage}[c]{0.475\linewidth}
\centering
\includegraphics[width=\linewidth]{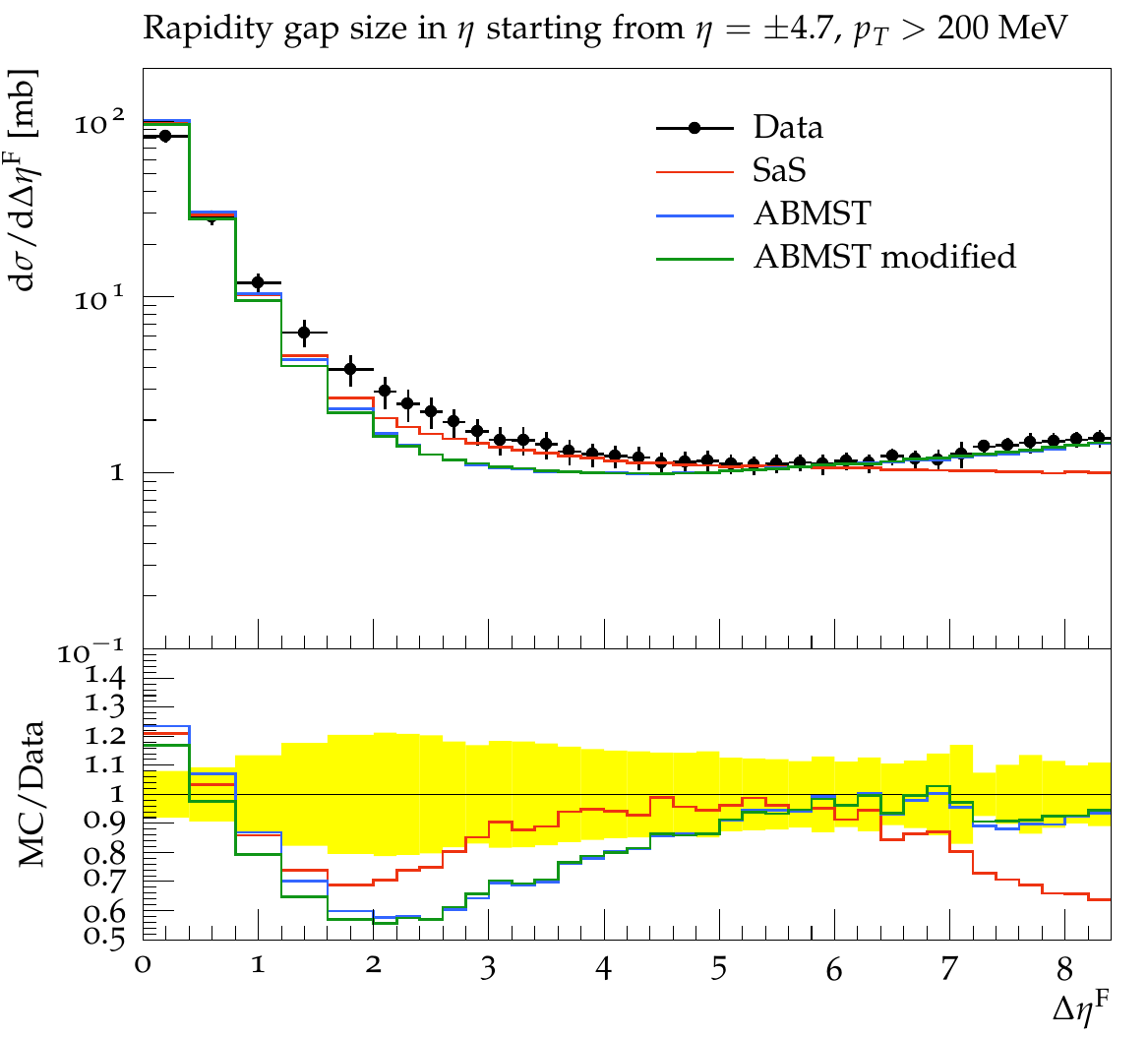}\\
(b)
\end{minipage}
\begin{minipage}[c]{0.475\linewidth}
\centering
\includegraphics[width=\linewidth]{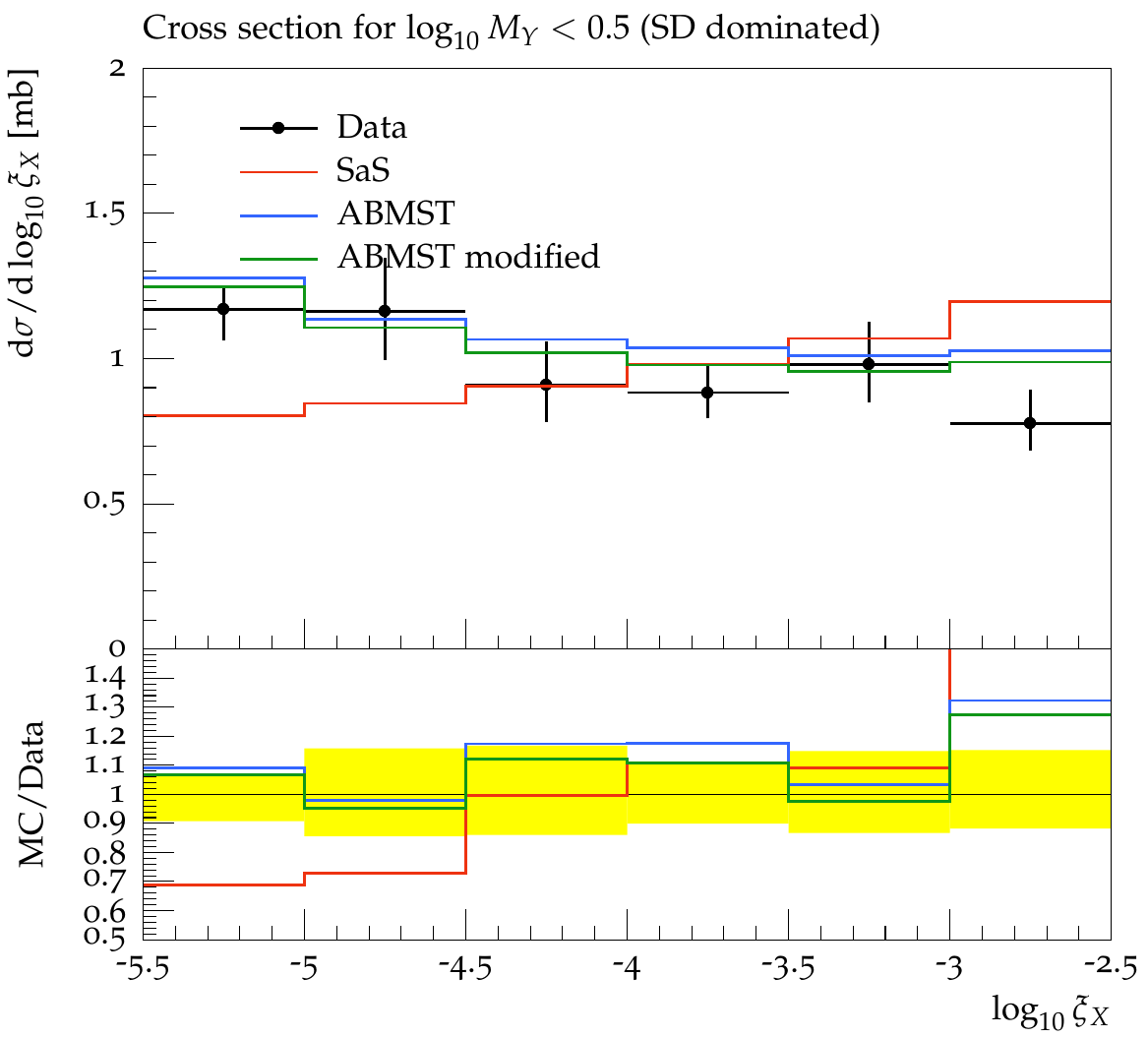}\\
(c)
\end{minipage}
\hfill
\begin{minipage}[c]{0.475\linewidth}
\centering
\includegraphics[width=\linewidth]{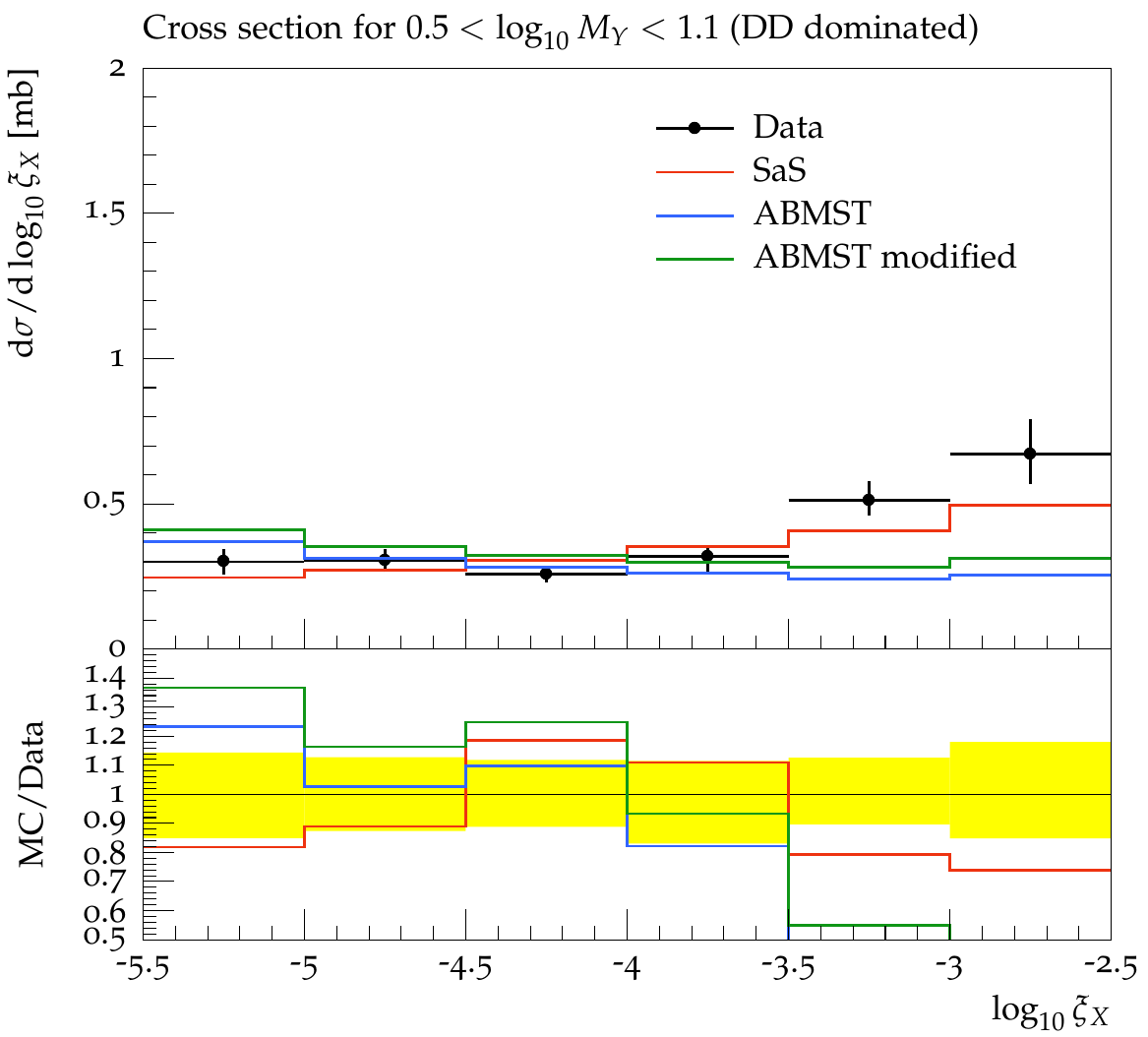}\\
(d)
\end{minipage}
\caption{\label{Fig:ABMSTcomparison} The cross section as a function of
gap size for the two ABMST-based models compared to ATLAS \cite{Aad:2012pw} 
(a) and CMS \cite{Khachatryan:2015gka} (b) data. The cross section as a 
function of $\mrm{log}_{10}\xi_X$ in a single-diffraction dominated region 
(c) and double diffractive dominated region (d) compared to CMS 
\cite{Khachatryan:2015gka} data. For reference the tuned SaS model
is also shown.}
\end{figure}

Fig.~\ref{Fig:ABMSTcomparison} shows the tuned ABMST models. The
tune has a hard time improving the modified ABMST model, as this gave a
good agreement with data already to begin with. The original ABMST model,
however, is significantly improved by rescaling, and is now
very similar to the modified ABMST model developed in this paper. Note 
that the mass spectrum of the single-diffraction-dominated region,
fig.~\ref{Fig:ABMSTcomparison}c, shows the proper shape, while the
double-diffraction-dominated one, fig.~\ref{Fig:ABMSTcomparison}d, 
seems to overestimate the low-mass region and underestimate the high-mass 
one. This is a result of the reduction of the vanishing-gap topologies 
of double-diffractive systems, that has been kept unchanged in this tune. 
Combining the ABMST models with the CSCR model has potential also
here, as the ABMST models underestimate data in the intermediate
gap range, cf. figs.~\ref{Fig:ABMSTcomparison}a,b. 

\begin{figure}[ht!]
\begin{minipage}[c]{0.475\linewidth}
\centering
\includegraphics[width=\linewidth]{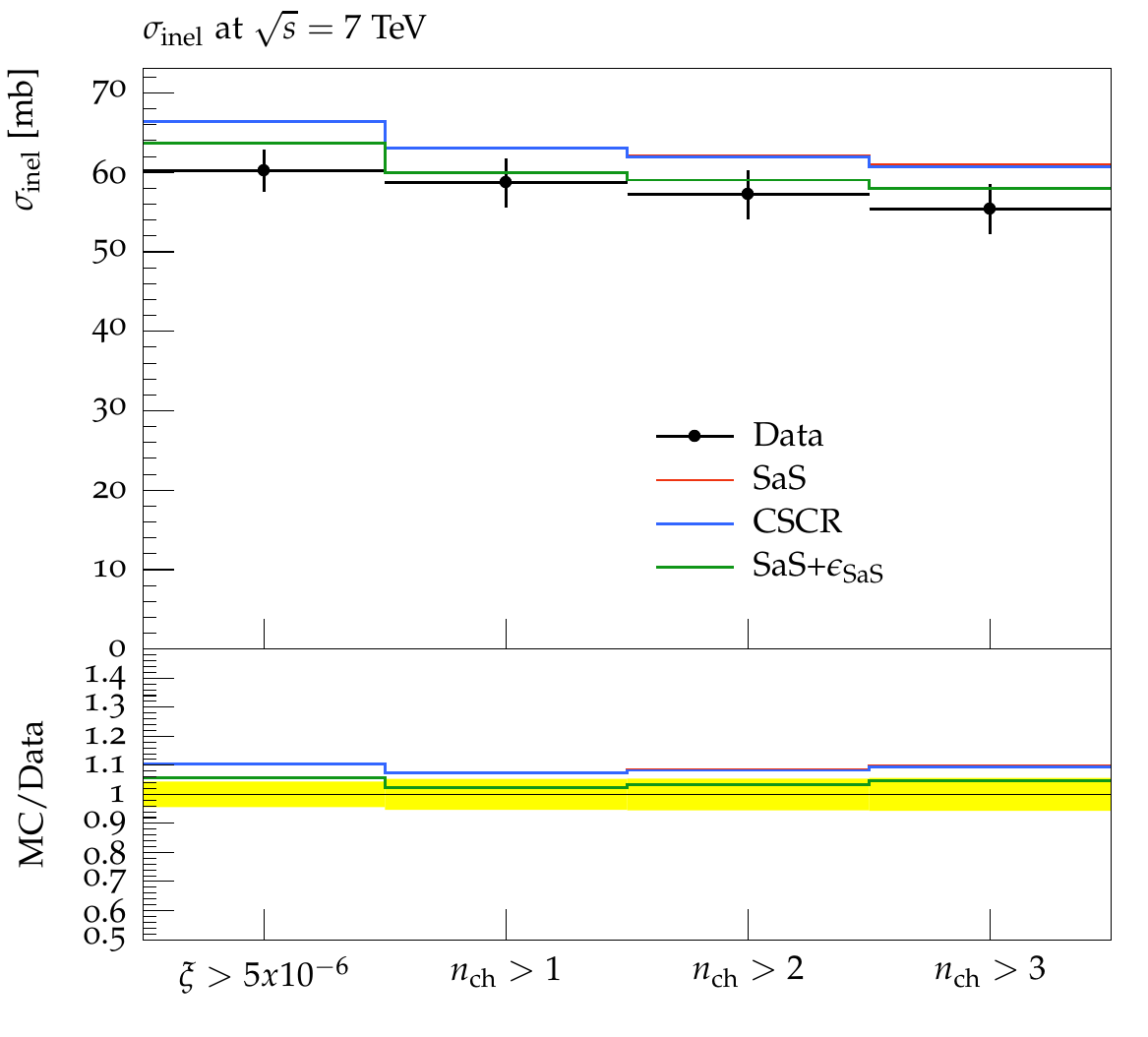}\\
(a)
\end{minipage}
\hfill
\begin{minipage}[c]{0.475\linewidth}
\centering
\includegraphics[width=\linewidth]{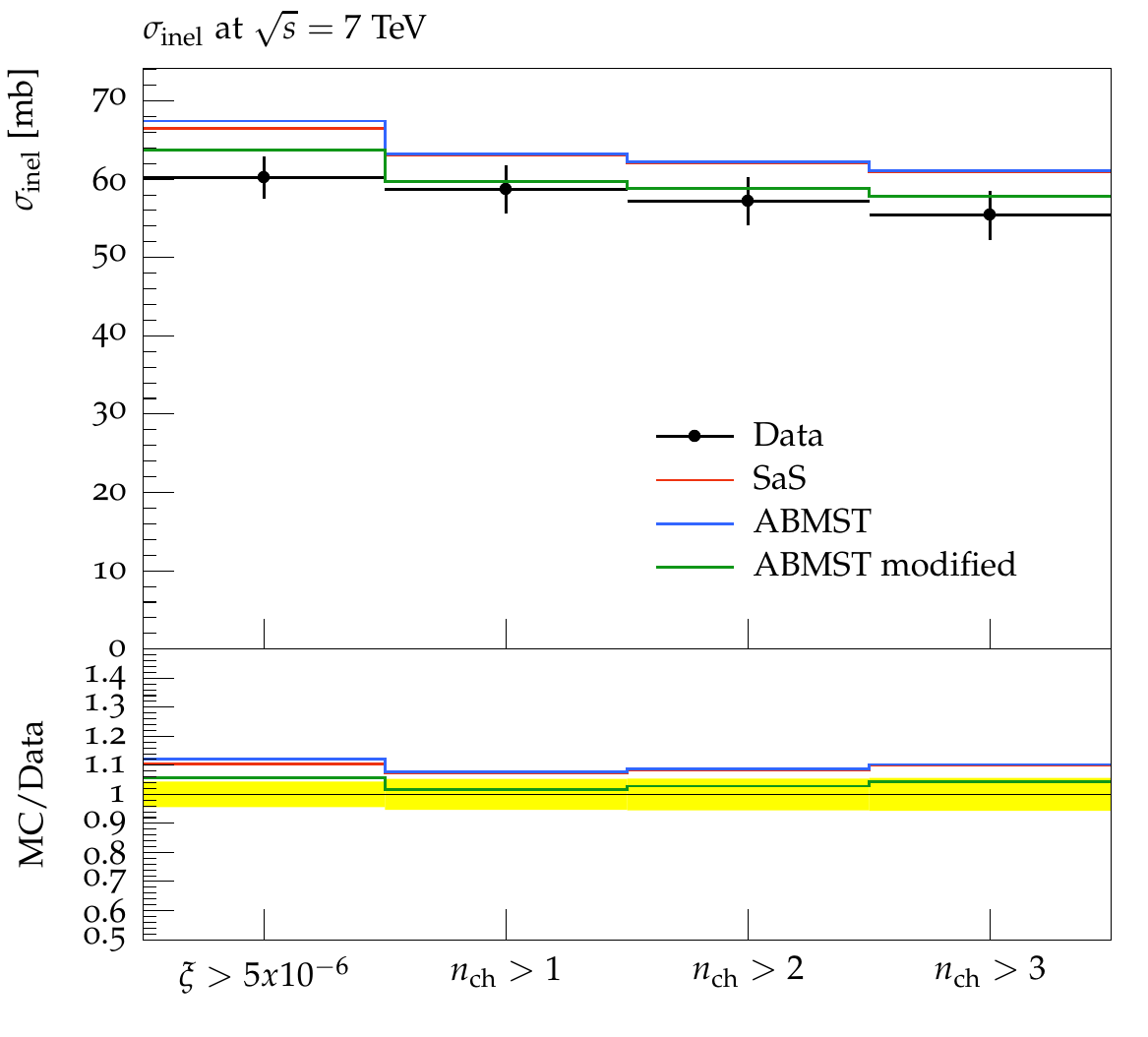}\\
(b)
\end{minipage}
\caption{\label{Fig:SigmaInelCMS} The inelastic cross section as a 
function of method described in the text compared to CMS 
\cite{Chatrchyan:2012nj} data.}
\end{figure}

Fig.~\ref{Fig:SigmaInelCMS} shows the CMS inelastic cross section obtained
with two different approaches. One uses forward calorimetry 
($3 < |\eta| < 5$), to measure protons with fractional momentum loss 
greater than  $\xi > 5\cdot10^{-6}$, corresponding to everything but 
low-mass diffractive systems ($M_X>16$ GeV). The other uses the central 
tracker, requiring either one, two or
three tracks. The SaS+$\epsilon_{\mrm{SaS}}$ and the modified ABMST models 
perform better than the others, with a maximum 5\% deviation 
from CMS data. The SaS and the CSCR models has the same model for
diffractive systems, and hence it is not expected that these differ in the
measured inelastic cross section. With the SaS+$\epsilon_{\mrm{SaS}}$
model, however, some of the activity has been shifted to lower
diffractive masses, resulting in a lower inelastic cross section.
For ABMST, the reduction of the high-mass systems in the modified model 
results in a reduction of the inelastic cross section relative to the
original one.

Table \ref{Tab:SaSparameters} shows the integrated cross sections 
obtained with the ALICE and ATLAS 2011 analyses mentioned above. The
ALICE results have been obtained for $M_X<200$ GeV ($\xi < 0.0008$) for
single diffraction, for gap sizes larger than $\Delta\eta>3$ for double
diffraction, and with a van der Meer scan using diffractive events
adjusted to data for the inelastic cross section. In the Rivet analysis,
this corresponds to at least two tracks in the final state, i.e.\
effectively without any experimental cuts and hence returning the
generator-level cross section. The SaS+$\epsilon_{\mrm{SaS}}$ model gives a 
better prediction for the single diffractive data, because of the increased 
low-mass cross section. The CSCR model predicts a larger double diffractive 
cross section, because of the larger probability for ``accidental'' gaps. 
The inelastic cross section, however, is the same for all three SaS-based 
models when compared with the ALICE data, as all have the 
same generator-level integrated cross section. In the ATLAS measurement 
of the inelastic cross section (for $\xi>5\cdot10^{-6}$) the 
SaS+$\epsilon_{\mrm{SaS}}$ model predicts a lower inelastic cross 
section, again because of the larger low-mass cross section.

Both the ABMST models give larger single-diffractive cross sections 
than SaS, having improved in the low-mass region. But both 
underestimate the double-diffractive cross section as they both 
underestimate the medium-sized gaps, compared with SaS and data. An 
addition of the CSCR model would be likely to improve this prediction. 
The ABMST models predict the same inelastic cross section for ALICE,
since the generator-level inelastic cross section is the same for the
two models. They differ for the ATLAS analysis, again because of the 
reduced high-mass systems of the modified ABMST model. 

\begin{table}[tbp]
\centering

\begin{tabular}{|c|c|c|c|c|}
\hline
 & $\sigma_{\mrm{SD}}$ (mb) & $\sigma_{\mrm{DD}}$ (mb) 
    & $\sigma_{\mrm{inel}}$ (mb) & $\sigma_{\mrm{inel}}$ (mb) \\
 & (ALICE) & (ALICE) & (ALICE) &  (ATLAS) \\ \hline
data & 14.9 $\pm$ 5.90 & 9.00 $\pm$ 2.60  
    & 73.20 $\pm$ 5.28 & 60.33 $\pm$ 2.10 \\
SaS & 6.13 $\pm$ 0.01 &  5.72 $\pm$ 0.01  
    & 71.06 $\pm$ 0.02 & 66.48 $\pm$ 0.02 \\ 
SaS + CSCR &  6.15 $\pm$ 0.01 & 6.19 $\pm$ 0.01 
    & 71.06 $\pm$ 0.02 & 66.43 $\pm$ 0.02 \\  
SaS + $\epsilon_{\mrm{SaS}}$ & 7.98 $\pm$ 0.01 & 5.62 $\pm$ 0.01 
    & 71.06 $\pm$ 0.02 & 63.69 $\pm$ 0.02 \\
ABMST & 7.24 $\pm$ 0.01 & 4.69 $\pm$ 0.01
    & 71.62 $\pm$ 0.02 & 67.44 $\pm$ 0.02 \\
ABMST mod & 9.41 $\pm$ 0.01 & 5.09 $\pm$ 0.01
    & 71.62 $\pm$ 0.02 & 63.72 $\pm$ 0.02 \\
\hline
\end{tabular}
\caption{\label{Tab:SaSparameters} The integrated cross section obtained 
with the three aforementioned Rivet analyses for the tuned models. For
ALICE \cite{Abelev:2012sea}, the SD cross section is for $M_X < 200$ GeV, 
the double diffractive for gaps larger than 3, the inelastic using a van 
der Meer scan using diffractive events adjusted to data. The ATLAS 
\cite{Aad:2011eu} inelastic cross section is for $\xi > 5\cdot10^{-6}$.}
\end{table}

In general, however, all models fail to describe the measured integrated
cross sections, although some of the more sophisticated models do improve 
in some respects. Similarly, it seems that neither of the models describe 
well the transition from a non-diffractive-dominated region to a
diffraction-dominated one. Including a colour-reconnection model that
allows for larger gaps in the non-diffractive events, like CSCR,
is likely to improve the description in the mid-sized-gap range, if
combined with a model that predicts a lower diffractive cross section 
there, like the ABMST models and SaS+$\epsilon_{\mrm{SaS}}$. The
overall question of how to combine the descriptions of non-diffractive and
diffractive topologies, however, will still exist even if the
CR model ``accidentally'' (i.e.\ by ``accidental'' gaps) improves 
the description of data. All this highlights our still limited 
understanding of nonperturbative QCD, which forces us to work with
models e.g.\ rooted in Regge theory. This may be good enough for an
overall understanding, but still not for a precise reproduction of all
relevant data.

\section{Conclusions}

In this paper we provide an updated description of the cross sections
and hadronic event shapes in the event generator \textsc{Pythia}~8. The
update has been required since the first results appeared from the LHC
experiments, showing significant discrepancies between the models
provided by Donnachie and Landshoff for the total cross section, as well 
as the elastic and diffractive cross sections by Schuler and Sj\"ostrand. 
By chance the DL undershooting of the total cross section and the SaS 
undershooting of the elastic cross section partly cancel in the  
inelastic cross section. Further to that, the SaS overshooting of the 
diffractive cross sections gave rise to a reasonable agreement between 
\textsc{Pythia}~8 and LHC measurements on the observable non-diffractive 
cross section, which is the relevant one for many of the measurements 
performed at the LHC. Thus, in spite of these shortfalls, the default 
\textsc{Pythia}~8 cross sections usually were good enough, notably when
diffractive cross sections had been reduced somewhat (eq.~\ref{Eq:SaSmod}).

The discrepancies became largely evident with the precision measurements
of the elastic and total cross sections performed by both TOTEM and 
ATLAS+ALFA. Here the exponential shape of the $t$ spectrum in
\textsc{Pythia}~8 is too simplistic, and other models have to be used 
for comparisons. Some of these models have now been implemented into
\textsc{Pythia}~8, thereby providing a more sophisticated framework for 
elastic scattering and total cross sections. 

For diffractive topologies the precision is less. The studies 
are marred by non-diffractive events mimicking diffractive ones,
and vice versa, making the explicit distinction between the various 
diffractive and non-diffractive event topologies hard. The possibility 
of tagging the elastically scattered protons would greatly improve the 
separation of the samples, but so far no analyses on diffraction with 
tagged protons have appeared from CMS+TOTEM or ATLAS+ALFA. Thus we are 
left with measurements only using the central general-purpose detectors. 
Unfortunately these do not give fully consistent answers. Notably the 
CMS and ATLAS rapidity-gap measurements disagree in the 
diffraction-dominated region, making it hard to compare models with data. 
Lacking any further guidance, we have here aimed for a middle ground
between the two data sets.

The situation is even worse for of hadronic event shapes. Single
diffractive data is available for very low energies, most of which goes
into the ABMST model, but rather little for higher energies.
This means that, even if integrated cross sections were provided for 
diffractive topologies from the LHC experiments, no constraints are put 
on the internal structure of diffractive systems.
The ansatz of \textsc{Pythia}~8, that the diffractive system properties 
are similar to those of non-diffractive events, could be wrong. A future 
study of these event shapes, and of the different strategies underlying
commonly used event generators, would help provide a guideline
what would be interesting distributions to see measured at the LHC.

In conclusion, we provide an updated and extended framework for elastic
and diffractive topologies, as well as an update for all parts of the
total cross section. We rely on previous work provided by several other 
authors, but have corrected and extended the models where need be.
Each of the models have been tuned to available data, thus providing an
upgrade of the already present models in \textsc{Pythia}~8. We have
discussed some of the consequences of different approaches for creating
rapidity gaps, such as the CSCR model, and how this affects the
predictions for LHC. Still, the lack of data or the discrepancies of
present data, leaves us with imperfect descriptions and predictions,
in particular for diffraction. The situation may be ``good enough'' for 
current needs, but will hopefully improve with new data in the future.
At present we are not able to decide which model is ``the better
one'' for diffraction, but in the case of total and elastic cross
section the new models, COMPAS and ABMST, offer an improved
description as compared to the SaS model. As the COMPAS model offers
no description of diffraction, we propose to use the ABMST model for
total and elastic cross section and the modified ABMST model for
diffraction, with the tuned parameters as provided in this paper. We
expect to change the default behaviour in the next \textsc{Pythia}
release.

Foreseeable further work could include a low-mass description 
for central diffractive topologies, possibly modelling the resonances 
present there. Other work would be an extensive study of the diffractive 
event shapes as discussed above. A study on eikonalisation aspects,
e.g.\ of events with both diffractive and nondiffractive Pomeron 
exchanges, could also provide more insight on both cross sections and 
event topologies. Finally, the diffractive framework could be extended
also to other processes, such as $\gamma\p$ and $\gamma\gamma$
collisions.

\section*{Acknowledgements}

We would like to thank Vladimir Ezhela from the COMPAS group for
assistance with their model. Similarly we would like to thank Rob
Appleby, James Molson and Sandy Donnachie for their huge effort in
explaining various aspects of their model, as well as providing all of
their data used in their fit.
 
Work supported in part by the Swedish Research Council, contracts number
621-2013-4287 and 2016-05996, and in part by Marie Curie Initial Training 
Networks, FP7 MCnetITN (grant agreement PITN-GA-2012-315877) and H2020 
MCnetITN3 (grant agreement 722104). 
This project has also received funding from the European Research
Council (ERC) under the European Union's Horizon 2020 research
and innovation programme (grant agreement No 668679).


\end{document}